\documentclass[aps, superscriptaddress, showpacs, twocolumn, longbibliography, prx]{revtex4-2}
\usepackage{units}
\usepackage{amsmath}
\usepackage{amssymb}
\usepackage{graphicx}
\usepackage{bm}
\usepackage{multirow,color,relsize,ulem,microtype,mathtools}
\usepackage{appendix}

\usepackage{soul} %for \hl and \st
\soulregister\ref{7}
\soulregister\eqref{7}
\soulregister\cite{7}
\soulregister\onlinecite{7}
\usepackage[colorlinks=true, allcolors=blue]{hyperref}
\renewcommand{\st}[1]{}

\newcommand{\be}{\begin{equation}}
\newcommand{\ee}{\end{equation}}

\begin{document}
\title{Theory of Stimulated Brillouin Scattering in Fibers for Highly Multimode 
Excitations}

\author{Kabish Wisal}
\affiliation{Department of Physics, Yale University, New Haven, CT 06520, USA}

\author{Stephen C. Warren-Smith}
\affiliation{Future Industries Institute, University of South Australia, Mawson Lakes, Australia}
\author{Chun-Wei Chen}
\affiliation{Department of Applied Physics, Yale University, New Haven, CT 06520, USA}
\author{Hui Cao}
\author{A. Douglas Stone}
\email{douglas.stone@yale.edu}
\affiliation{Department of Applied Physics, Yale University, New Haven, CT 06520, USA}
%\affiliation{Yale Quantum Institute, Yale University, New Haven, CT 06520, USA}
\date{\today}

\begin{abstract}
Stimulated Brillouin scattering (SBS) is an important nonlinear optical effect which can both enable and impede optical processes in guided wave systems. Highly multi-mode excitation of fibers has been proposed as a novel route towards efficient suppression of SBS in both active and passive fibers. To study the effects of multimode excitation generally, we develop a theory of SBS for arbitrary input excitations, fiber cross section geometries and refractive index profiles. We derive appropriate nonlinear coupled mode equations for the signal and Stokes modal amplitudes starting from vector optical and tensor acoustic equations. Using applicable approximations, we find an analytical formula for the SBS (Stokes) gain susceptibility, which takes into account the vector nature of both optical and acoustic modes exactly. We show that upon multimode excitation, the SBS power in each Stokes mode grows exponentially with a growth rate that depends parametrically on the distribution of power in the signal modes. Specializing to isotropic fibers we are able to define and calculate an effective SBS gain spectrum for any choice of multimode excitation. The peak value of this gain spectrum determines the SBS threshold, the maximum SBS-limited power that can be sent through the fiber. We show theoretically that peak SBS gain is greatly reduced by highly multimode excitation due to gain broadening and relatively weaker intermodal SBS gain. We demonstrate that equal excitation of the 160 modes of a commercially available, highly multimode circular step index fiber raises the SBS threshold by a factor of 6.5, and find comparable suppression of SBS in similar fibers with a D-shaped cross-section.

\end{abstract}


\maketitle

%%%%%%%%%%%%%%%%%%%%%%%%%%  body  %%%%%%%%%%%%%%%%%%%%%%%%%%
\section{Introduction}

Stimulated Brillouin Scattering (SBS) is the nonlinear scattering of light by acoustic phonons generated by optical forces~\cite{boyd2020nonlinear,agrawal2000nonlinear,kobyakov2010stimulated,bai2018stimulated,wolff2021brillouin}. It was first theoretically predicted by Brillouin in 1922~\cite{brillouin1922diffusion} and experimentally demonstrated in liquids in 1964~\cite{chiao1964stimulated} and in optical fibers in 1972~\cite{ippen1972stimulated}. SBS has been utilized for diverse applications such as slow light~\cite{thevenaz2008slow}, non-reciprocal light storage~\cite{kim2015non}, strain and temperature sensing~\cite{horiguchi1995development}, Brillouin microscopy~\cite{ballmann2015stimulated}, and integrated photonics~\cite{rakich2012giant,eggleton2013inducing, shin2013tailorable, otterstrom2018silicon,  eggleton2019brillouin}. On the other hand, SBS is often a highly undesirable effect in both active and passive systems. Of particular interest is its role
 in limiting the power capacity of narrow linewidth high-power fiber lasers and high-power delivery fibers~\cite{richardson2010high,zervas2014high, kobyakov2010stimulated, fu2017review,pannell1993stimulated}. As input power is increased, SBS can cause almost complete reflection above a certain threshold power, rendering both active and passive fibers inoperable above that power level~\cite{ippen1972stimulated,agrawal2000nonlinear,kobyakov2010stimulated,panbhiharwala2018investigation}. Significant research efforts have therefore been devoted to suppressing SBS efficiently. Due to concerns about maintaining high output beam quality almost all of the experimental and theoretical work has focused on exciting a single fundamental mode (FM) of the fiber (whether or not the fiber itself is nominally single-mode or multimode fiber (MMF)). Recently highly multi-mode excitation of fibers has been proposed as a novel route towards efficient suppression of SBS in both active and passive fibers~\cite{wisal2022generalized,chen2022suppressingSBS,SBSexpinPrep}. In the current work we explore theoretically the effect on SBS of highly multimode excitation of MMFs. While a number of previous works have introduced elements of our current theory \cite{ke2014stimulated,poulton2013acoustic,dong2010formulation}, none have developed a quantitative formalism for computing SBS under highly multimode excitation, nor has any explored and identified the physical effects that arise when controlled, highly multimode signals are imposed.

 As noted, most of the previous efforts to suppress SBS were in single-mode fibers and employ one of the following approaches: broadening the Brillouin spectrum by dynamic seed modulation~\cite{supradeepa2013stimulated,coles2010bandwidth,liu2009research} or applying temperature and strain gradients along the fiber~\cite{yoshizawa1993stimulated,liu2007suppressing}, tailoring the fiber acoustic properties to reduce acousto-optic overlap~\cite{kobyakov2005design,dragic2005optical,li2007ge,poulton2013acoustic,hawkins2021kilowatt}, and altering the geometry (shape and composition) of the fiber cross-section~\cite{shiraki1995suppression,robin2011acoustically}. Although all of these efforts have had some success, they suffer from a number of drawbacks: for seed modulation, a large linewidth broadening~\cite{supradeepa2013stimulated,coles2010bandwidth,liu2009research} makes the resultant beams unsuitable for coherent beam combining and other narrowband applications~\cite{augst2007beam,loftus2007spectrally,buikema2019narrow}; for acoustic mode tailoring, the need for precise fiber design and fabrication to control both the acoustic-index and optical-index profiles~\cite{kobyakov2005design,dragic2005optical,li2007ge,poulton2013acoustic,hawkins2021kilowatt}; and for increased core size, the emergence of other undesirable effects such as transverse mode instability~\cite{ward2012origin}.

The current investigation of a highly multimode approach was motivated by recent developments in the field of wavefront shaping, which have shown that: 1) It is possible to control nonlinear effects by manipulating the input excitation in multimode fibers~\cite{tzang2018adaptive,deliancourt2019wavefront,shutova2019coherent,teugin2020controlling,chen2022suppressing}. 2) It is possible, even for multimode excitation, to obtain a high-quality output beam by wavefront shaping, since for a narrow input linewidth, the light in various fiber modes remains mutually coherent throughout the fiber~\cite{ploschner2015seeing,xiong2016spatiotemporal,florentin2017shaping,gomes2022near}. Thus, multimode excitation combined with appropriate wavefront shaping can in principle provide a novel method of SBS suppression, while maintaining good beam quality. 

Previous studies, both experimental and theoretical, of SBS in few-mode fibers appear to support the viability of increasing the SBS threshold by using multimode excitation~\cite{tei2001critical,kovalev2002waveguide,sjoberg2003dependence,minardo2014experimental,wang2020mode}. Moreover several such studies indicate that intermodal SBS gain is weaker than intramodal 
gain~\cite{song2013characterization,srinivasan2021role,ke2014stimulated,lu2015theoretical}.  This suggested that division of power in many modes may reduce the effective SBS gain within a given fiber; the efficacy of this approach is a key result of the theory and computational examples in the current work.  Finally, assuming that the fiber is in the phase-matched limit, previous work~\cite{ke2014stimulated,lu2015theoretical}, and our findings below, show that the SBS threshold only depends on the input power in various modes, leaving the input phases of various modes as free parameters. Using wavefront shaping, these phases can in principle be carefully selected to control and focus the output beam profile via modal interference \cite{florentin2017shaping,SBSexpinPrep}. In order to understand the potential of power division and wavefront shaping for suppressing SBS in MMFs, we have developed a quantitative and computationally tractable theory of the effective SBS spectrum under highly multimode excitation.  This theory is general enough to be applicable to other questions in the theory of multimode SBS, beyond maximizing the threshold.

Most previous SBS theories assume single transverse mode operation for both the signal (which acts as the pump for Brillouin scattering) and Stokes field~\cite{cotter1983stimulated,agrawal2000nonlinear,kobyakov2010stimulated,dragic2010accurate,dong2010formulation}. There have been efforts to model SBS in MMFs in the study of phase conjugation and beam cleaning using MMFs~\cite{suni1986theory,hu1989theoretical,lombard2006beam}. However, these approaches focus only on identifying the selection rules for non-zero mode couplings, and do not model the individual couplings accurately. In particular, the guided nature of acoustic modes is generally ignored resulting in inaccurate SBS thresholds. There is also extensive literature on SBS theory in nano-waveguides~\cite{rakich2012giant,qiu2013stimulated,wolff2015stimulated,wolff2021brillouin,shi2017invited,poulton2013acoustic}, although only a few studies focus on intermodal gain and none of them study multimode excitation. The most comprehensive theory to date of SBS in MMFs was done by Ke et al.~\cite{ke2014stimulated}. They showed that within the phase-matched regime, the power in each Stokes mode grows exponentially at a rate proportional to the signal power in various modes multiplied by the multimode Brillouin gain spectra, determined by the overlap of the optical and acoustic modes involved. The analysis by Ke et al. is, however, limited in several ways: It treats both acoustic and optical fields as scalar quantities, the phase-matching assumption is made at the outset in calculating the SBS gain, and, importantly, they did not undertake any explicit calculations of Brillouin gain spectra for a highly multimode excitation of a MMF. Hence they did not study the major physics questions addressed in the current work, nor did they provide an accurate enough computational framework for doing so.

 In this paper, we formulate a theory of SBS in MMFs starting from the full vector and tensor equations for electric and acoustic fields~\cite{shi2017invited}. These equations have been used extensively to model the SBS in the nano-waveguides~\cite{rakich2012giant,wolff2015stimulated,shi2017invited}. The optical field equations are vector Helmholtz equations~\cite{jackson1999classical}, sourced by a nonlinear polarization due to the photoelastic effect~\cite{nelson1971theory}. The acoustic field equations~\cite{auld1973acoustic,landau1986theory} are those of general continuum elasticity with the stiffness and viscosity tensors describing the restoring force and damping, respectively and the driving source arising from optical forces generated due to electrostriction~\cite{feldman1975relations}. We demonstrate that for accurate calculations of SBS coupling, it is important to take into account the vector nature of both the optical and acoustic modes, especially for highly multimode fibers with large numerical apertures (NA). Even when we specialize the theory to the case of elastically isotropic fibers~\cite{auld1973acoustic}, with only two independent parameters for each tensor, we find corrections to the acoustic modes due to shear--longitudinal coupling~\cite{dong2010formulation,dragic2010accurate,waldron1969some}, neglected in the widely used scalar acoustic equations. Our formalism is general enough to permit the calculation of the SBS threshold for an arbitrary input excitation in MMFs, with any fiber cross-section, refractive index profile, and fiber length. In particular, our theory is well suited to study the optimization of the SBS threshold with respect to degrees of freedom in MMFs. We use our theory to calculate the SBS gain for all the $\sim 10^4$ mode pairs for an example corresponding to a 
 commercially available highly multimode circular step-index fiber. We show that when all the modes are equally excited, a roughly 6.5$\times$-higher SBS threshold can be obtained, compared to exciting only the fundamental mode. 

The paper is organized as follows:

In Section~\ref{sec:II}, we start by deriving the coupled mode equations for Stokes and signal amplitudes, assuming only the translational invariance in $z$, the slowly varying envelope approximation, and highly damped phonons~\cite{agrawal2000nonlinear}. All of these approximations are valid for most multimode fibers. We do not assume  phase matching {\it a priori}. Then we apply the undepleted-signal approximation~\cite{boyd2020nonlinear}, valid below SBS threshold, to obtain linear growth equations for each Stokes amplitude. This allows us to identify a linearized SBS coupling which accurately captures the vector nature of all the fields involved.

In Section~\ref{sec:III}, we systematically simplify the SBS coupling under a series of applicable approximations. We show that the length of the fiber compared to a phase-mismatch length scale determines whether phase-mismatched terms need to be kept or can be discarded. In the phase-matched limit, the equations for various Stokes modes decouple, leading to independent exponential growth in the backward direction. The rate of growth for power in each Stokes mode depends on the modal content of the signal, and is given by a sum of signal power in each mode weighted by the corresponding Brillouin gain spectra (BGS), which now generalizes to a matrix of pairwise modal spectra. The BGS for a Stokes--signal pair is given by a sum of Lorentzians, for each acoustic mode, with peak values proportional to the overlap integrals of the Stokes, signal, and acoustic modes involved.  We derive a reasonably accurate simplified form of the overlap integrals for elastically isotropic fibers with small shear acoustic velocities, where the dot product of the two vector optical modes appears in the integrand. This structure is critical for accurately evaluating the intermodal gain, especially for large numerical aperture (NA)  fibers, when the scalar linearly polarized (LP) modes are not good approximations of exact vector fiber modes. Both intramodal and intermodal terms contribute to the growth rate of a given mode $m$ in a manner determined by the signal power distribution. Hence we introduce the important concept of an effective gain for each mode, $\Tilde{g}_m (\Omega)$, which captures the effect of a particular power distribution on the gain experienced by a Stokes mode, $m$. This allows us to define a generalized, signal-dependent formula for the SBS threshold for MMFs within the phase-matched theory. 

In Section~\ref{sec:IV}, we explicitly calculate the BGS for highly  multimode step-index fibers with circular and D-shaped cross sections. We show that the SBS threshold is increased significantly if the input power is optimally distributed in multiple fiber modes, initially using a two-mode example.  This increase arises from two effects. First, the intermodal gain is typically weaker than the intramodal gain, due to the mismatch in different optical modal profiles. Second, the peaks of the BGS for different mode pairs are shifted relative to each other since they are mediated by different acoustic modes; thus power division among many modes tends to broaden the gain spectrum, leading to further SBS suppression~\cite{kovalev2002waveguide}.  As an example, we calculate that when all 160 modes of a circular step-index fiber are equally excited, a 6.5-$\times$-higher SBS threshold is obtained, compared to exciting just the fundamental mode. Due to the distinct polarization properties of D-shaped fiber, we find that the increase of the SBS threshold in this case is dependent on the input polarization and is highest for input polarized at 45 degrees with respect to the axis of symmetry, an effect which is explained by our theory. 
%The origin of this effect This effect has a similar origin to a known result in single-mode polarization-maintaining fibers~\cite{van1994polarization}. It is due to the vanishing intermodal interaction between orthogonally polarized ($x$ and $y$) modes of the fiber. 
Finally, in Section~\ref{Sec:V} we provide a discussion of experimental validation, applicability and future directions for our work.
\section{Multimode SBS model} \label{sec:II}

\subsection{Generalized SBS theory}

SBS is a result of optical scattering by acoustic phonons generated by electrostriction~\cite{boyd2020nonlinear,agrawal2000nonlinear,kobyakov2010stimulated}. A schematic of the SBS in a fiber is shown in Fig.~\ref{Fig:fiber}. The forward-going signal wave at frequency $\omega_1$ interferes with a backward-going and Stokes shifted wave at frequency $\omega_2= \omega_1-\Omega$, generating a moving intensity grating. This intensity pattern results in optical forces through electrostriction, which generate acoustic phonons with frequency $\Omega$. These phonons in turn reflect light in the backward direction, leading to exponential growth in the Stokes power. To model SBS, we solve the optical and acoustic wave equations coupled through nonlinear source terms due to electrostriction and the photoelastic effect. The total electric field $\vec{E}$ satisfies the vector Helmholtz equation~\cite{shi2017invited,wolff2015stimulated}:
\begin{equation}
    \left[\nabla^2  - \frac{n^2}{c^2}\frac{\partial}{\partial t^2} \right]\vec{E}= \mu_0  \frac{\partial}{\partial t^2}\left(\stackrel{\leftrightarrow}{\chi}_{\tiny N} \cdot \vec{E}\right).
    \label{Eq:Efield}
\end{equation}
\noindent
Here, $n$ is the linear refractive index of the fiber, $\mu_0$ is the free space permeability, $c$ is the speed of light in vacuum and $\stackrel{\leftrightarrow}{\chi}_{\tiny N}$ is the photoelastic susceptibility tensor~\cite{nelson1971theory} and is given by: 

\begin{figure*}[t!]
		\centering\includegraphics[width=0.7\textwidth]{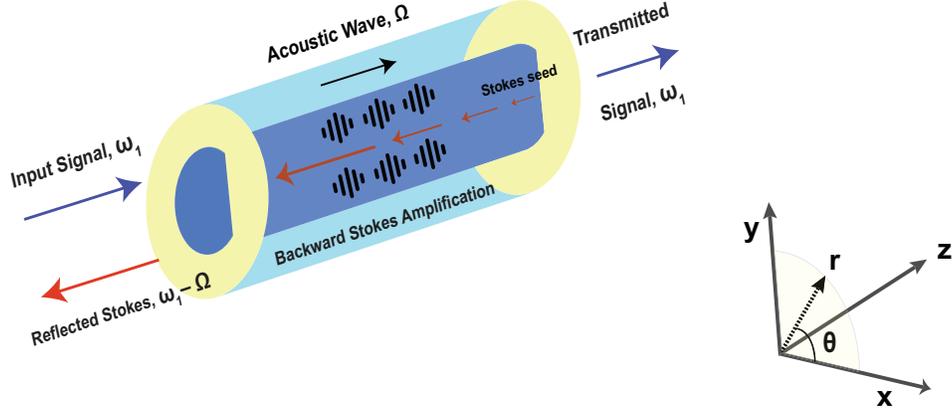}
		\caption{{\bf Schematic of SBS in a multimode fiber with arbitrary core shape (here, D-shaped).} Stokes shifted backward traveling light (seeded by spontaneous Brillouin scattering) experiences amplification due to the scattering of the forward-going signal by the acoustic phonons, which are generated by electrostriction. This process can take away significant power from the signal and limits the transmitted power.}
        \label{Fig:fiber}
\end{figure*}
\begin{equation}
    {\stackrel{\leftrightarrow}
    {\chi}_{\tiny N}}={\stackrel{\leftrightarrow}{\pi}}:\nabla \otimes \vec{u}={\stackrel{\leftrightarrow}{\pi}}:{\stackrel{\leftrightarrow}{S}}.
    \label{Eq:chi}
\end{equation}
\noindent
Here $\otimes $ denotes the outer product of two vectors and 
\linebreak
$(\nabla \otimes \vec{u})_{ij} = \partial_i u_j\equiv {\stackrel{\leftrightarrow}{S}}_{ij}$ is the second-rank strain tensor. ${\stackrel{\leftrightarrow}{\pi}}:{\stackrel{\leftrightarrow}{S}}= \sum_{kl} \pi_{ijkl}S_{kl}$ denotes the contraction of the strain tensor with the fourth-rank photoelastic tensor, $\stackrel{\leftrightarrow}{\pi}$, to yield the second-rank susceptibility tensor, $\stackrel{\leftrightarrow}{\chi}_{N}$. The photoelastic tensor generalizes the electrostriction constant, $\gamma_e$, often used in scalar SBS formulations~\cite{ke2014stimulated,dong2010formulation}. 
 We consider an appropriately scaled  $\stackrel{\leftrightarrow}{\pi}$ such that $\gamma_e=\pi_{1122}$ (see Appendix~\ref{appendixA}). Similarly, the strain tensor generalizes the scalar density fluctuations in the drive term such that $\delta \rho = \rho_0\, {\rm Tr}[\stackrel{\leftrightarrow}{S}]$, where, $\rho_0$ is the average material density. Going forward, for clarity, we will continue to suppress the Cartesian vector and tensor indices using the notation of Eq.~ (\ref{Eq:chi}) and display integer modal indices, which are typically most relevant.
%A point of notation: we have chosen to write  vector dot products and tensor products in their compact form rather than expanding in terms of Cartesian indices. To clarify the multimode nature of the equations, we explicitly write the modal indices. The choice to keep various products in their compact form helps avoid the confusion between modal indices and vector (tensor) indices. 

\begin{figure*}[t!]
		\centering\includegraphics[width=6.8 in]{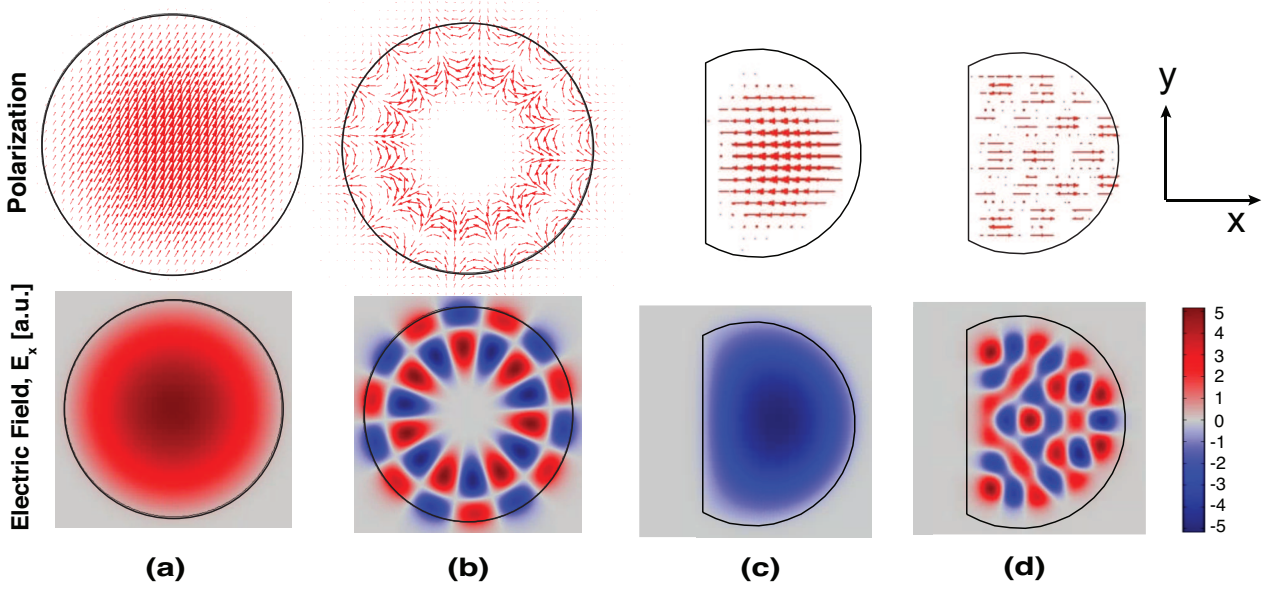}
		\caption{	 Polarization (upper panel) and x-component of the electric field, $E_x$ (lower panel) profiles for fundamental (a,c) and higher order (b,d) optical modes of step-index multimode fibers with circular (a,b) and D-shaped (c,d) core cross-sections.  Note the more ergodic behavior of the D fiber modes compared to the circular case.  Modes of the fiber with D-shaped cross section have polarization either along the axis of symmetry (x-axis) or orthogonal to it (y-axis). Modes of the circular cross-section fiber have spatially varying polarization depending on the mode.}
        \label{Fig:optmodes}
\end{figure*}

The photoelastic susceptibility $\stackrel{\leftrightarrow}{\chi}_N$ depends on the displacement field $\vec{u}$, which follows a linear elastic wave equation with nonlinear source terms given by the optical force due to electrostriction~\cite{shi2017invited,wolff2015stimulated},

\begin{equation}
\left[\nabla\cdot\left(\stackrel{\leftrightarrow}{C}-i\frac{\partial}{\partial t}\stackrel{\leftrightarrow}{\eta}\right):\nabla \otimes - \rho_0\frac{\partial}{\partial t^2}\right] \vec{u}=\vec{F},
\label{Eq:ufield}
\end{equation} 
\noindent
where $\stackrel{\leftrightarrow}{C}$ and $\stackrel{\leftrightarrow}{\eta}$ are fourth-rank elasticity and viscosity tensors, respectively~\cite{auld1973acoustic,landau1986theory}. The elasticity tensor generalizes various elasticity moduli which determine the acoustic velocities in the fiber. The viscosity tensor plays the role of generalized phonon loss in the fiber.  The source term $\vec{F}$ is the optical force, given by~\cite{shi2017invited}:

\begin{equation}
   \vec{F} =-\frac{1}{2} \nabla \cdot \left[\stackrel{\leftrightarrow}{\pi}: \vec{E}\otimes \vec{E}\right].
   \label{Eq:force}
\end{equation}
\noindent
The optical force is given by the divergence of the contraction between the fourth-rank photoelastic tensor $\stackrel{\leftrightarrow}{\pi}$ and the second-rank tensor formed by the tensor product of electric field $\vec{E}$ with itself. Here, we have included only the electrostriction term in the optical force and neglected the force generated due to the deformation of the fiber boundary~\cite{rakich2012giant,wolff2015stimulated,shi2017invited}. This is justified if the core size of the fiber is much larger than the wavelength of light, which is typically the case for multimode fibers. To study the Stokes growth, we split the total electric field, $\vec{E}=\vec{E}_1+\vec{E}_2$, into a forward-going signal wave $\vec{E_1}$ and a backward-going Stokes wave $\vec{E_2}$. Assuming the refractive index is translationally invariant along the fiber axis, we can decompose both $\vec{E_1}$ and $\vec{E_2}$ in terms of relevant vector fiber modes~\cite{ndagano2015fiber,okamoto2021fundamentals,snyder1978modes} with slowly varying amplitudes~\cite{agrawal2000nonlinear}:

\begin{equation}
\begin{aligned}     
    E_1=\sum_{m}A_m(z,t)\vec{f}^{(1)}_{m}(r,\theta)e^{i (\omega_1 t-\beta_mz)}+{\rm c.c.}\\
    E_2=\sum_{m}B_m(z,t)\vec{f}^{(2)}_{m}(r,\theta)e^{i (\omega_2 t+\gamma_m z)}+{\rm c.c.}
    \label{Eq:Eansatz}
\end{aligned}
\end{equation}
\noindent
Here, $\beta_m$ $(\gamma_m)$ and $\vec{f}^{(1)}_{m}(r,\theta)$ [$\vec{f}^{(2)}_{m}(r,\theta)$] denote the propagation constant in the $z$ direction and the transverse mode profile for the $m^{\rm th}$ signal (Stokes) mode, respectively. $\omega_1$ is the signal frequency, $\omega_2$ is the Stokes frequency, and the difference of the two frequencies, $\Omega= \omega_1-\omega_2$, represents the Stokes shift. The transverse modes satisfy the vector fiber modal equations given by~\cite{okamoto2021fundamentals,snyder1978modes}:

\begin{equation}
\begin{aligned}
    \left[\nabla_T^2+\left(\frac{n^2\omega_{1}^2}{c^2}-{\beta_m}^2\right)\right]\vec{f}^{(1)}_{m}(r,\theta)=0 \\
    \left[\nabla_T^2+\left(\frac{n^2\omega_{2}^2}{c^2}-{\gamma_m}^2\right)\right]\vec{f}^{(2)}_{m}(r,\theta)=0.
    \label{Eq:Emode}
\end{aligned}
\end{equation}
\noindent
The solution for the mode profiles and the corresponding propagation constants can be obtained analytically in the case of a circular step-index geometry~\cite{okamoto2021fundamentals,snyder1978modes} or numerically for an arbitrary fiber cross-section and refractive index profile. In Fig.~\ref{Fig:optmodes}, we have plotted the polarization and the $x$-component of the electric field for the fundamental and a higher-order mode in step-index fibers with circular and D-shaped cross sections. The fields were determined using the finite element method in the wave optics module of COMSOL Multiphysics (version 5.5)~\cite{comsol}. In general, both the electric field amplitude and the polarization vary spatially for the different modes. For the circular fiber, the fundamental mode (Fig.~\ref{Fig:optmodes}a) is approximately uniformly polarized while higher-order modes (Fig.~\ref{Fig:optmodes}b) have spatially varying polarization with azimuthal and radial components. For the D-shaped fibers, modes are approximately uniformly polarized either along the axis of symmetry ($x$-polarized) or orthogonal to it ($y$-polarized) for both the fundamental mode (Fig.~\ref{Fig:optmodes}c) and the higher order modes (Fig.~\ref{Fig:optmodes}d). The choice to study D-shaped core fibers in particular is motivated by chaotic ray dynamics in D-shaped cavities~\cite{bittner2018suppressing}, leading to more ergodic field profiles, especially for significantly higher order modes.

We normalize both the signal and Stokes modes such that the power in the $m^{\rm th}$ signal and Stokes modes is proportional to $|A_m|^2$ and $|B_m|^2$ respectively. The interference between the signal and Stokes fields gives rise to an acoustic source term for each signal-Stokes pair $\{i,j\}$ which oscillates at frequency $\Omega$ and has propagation constant $ q_{ij} =\beta_i+\gamma_j$ in the $z$ direction. Therefore, we expand the displacement field into a series of acoustic modes $\{k\}$ for each source term $\{i,j\}$:

\begin{equation}
     \vec{u} =\sum_{i,j}\sum_{k}c_{k}^{ij}(z,t)\vec{u}^{\:ij}_{k}(r,\theta)e^{i(\Omega t-q_{ij} z)}+{\rm c.c.}
     \label{Eq:uansatz}
\end{equation}
\noindent
Here, $c_k^{ij}(z,t)$ is the slowly varying amplitude for $ k^{\rm th}$ acoustic mode corresponding to optical pair, $\{i,j\}$. The associated transverse mode profile is given by $\vec{u}^{\:ij}_{k}(r,\theta)$, which satisfies the following modal equation~\cite{wolff2015stimulated}:

\begin{equation}
(\nabla_T-i q_{ij}\hat{z}).[\stackrel{\leftrightarrow}{C}]:(\nabla_T-i q_{ij}\hat{z}) \otimes \vec{u}^{\:ij}_{k}+\rho_0{\Omega^2_{ijk}}\vec{u}^{\:ij}_{k}=0.
\label{Eq:umode}
\end{equation}
\noindent
Here, $\nabla_T$ is the transverse gradient operator, and $\hat{z}$ is the unit vector along the fiber axis. We have chosen to not include the acoustic loss given by viscosity tensor $\stackrel{\leftrightarrow}{\eta}$ explicitly in the modal equation; it will be accounted for through the coefficients, $c_{ij}^k$. This choice has an advantage that it allows the operator of the modal equation to be Hermitian, which leads to useful properties such as orthogonality and completeness of the modal basis~\cite{jackson1999classical}.
Also, note that $\stackrel{\leftrightarrow}{C}$ is in general a non-diagonal tensor which leads to cross derivatives such as $\partial_x \partial_y$ to be part of the modal equation as opposed to the presence of only $\nabla_T^2$ in the scalar acoustic wave equation~\cite{ke2014stimulated}. 
\newline
\newline
\noindent
To obtain the solution for the displacement field, $\vec{u}$, we evaluate source terms in Eq.~\ref{Eq:ufield} using Eq.~\ref{Eq:force} and Eq.~\ref{Eq:Eansatz}. The left-hand side of Eq.~\ref{Eq:ufield} can be simplified by substituting the ansatz for $\vec{u}$ from Eq.~\ref{Eq:uansatz}. We then take the dot product with acoustic mode profile $\vec{u}^{\:ij*}_k$ and utilize orthogonality of acoustic modes to isolate the equation for a single acoustic amplitude. Upon applying the slowly varying approximation, where we ignore the second-order z-derivatives of $c_k^{ij}$ and using the modal equations, Eq.~\ref{Eq:umode},  we get the following equation:

\begin{equation}
\begin{aligned}
    \left[2i q_{ij} \frac{\partial}{\partial z}+2i\Omega\frac{\partial}{\partial t}+(\Omega^2_{ijk}-\Omega^2+  i\Omega\Gamma_{ijk})\right]&c_{k}^{ij}\\
    &=\frac{O_1}{2\rho_0}A_i B^{*}_j,
    \label{Eq:usimplify}
    % I guess this should be lower case c
\end{aligned}
\end{equation}
\noindent
where $O_1$ denotes the overlap integral of the source term with the relevant acoustic mode and is given by:
\begin{equation}
   O_1= -\langle(\vec{q}\cdot[\stackrel{\leftrightarrow}{\pi}: \vec{f}^{(1)}_i\otimes \vec{f}^{(2)^*}_j])\cdot \vec{u}^{\:ij*}_{k} \rangle
   \label{Eq:overlap1}
\end{equation}
\noindent
The angular brackets $\langle.\rangle$ denote integration over the entire fiber cross-section, and $\vec{q}$ is the gradient operator with $z$ component set equal to $-i q_{ij} \hat{z}$. The derivative of the tensor product of optical mode profiles, $\vec{f}^{(1)}_i$ and $\vec{f}^{(2)}_j$, contracted with the photoelastic tensor, $\stackrel{\leftrightarrow}{\pi}$, represents the optical force. This is then projected onto the relevant acoustic mode profile $\vec{u}^{\:ij}_{k}$ and integrated over the fiber cross-section to give the relevant source integral.   We have also introduced an effective phonon loss rate, $\Gamma_{ijk}$, which is given by projecting the acoustic loss term in Eq.~\ref{Eq:ufield} onto the acoustic modal basis:
\begin{equation}
    \Gamma_{ijk}=\frac{1}{\rho_0} \langle \vec{u}^{\:ij*}_k\cdot(\vec{q} \cdot [\stackrel{\leftrightarrow}{\eta}:\vec{q} \otimes \vec{u}^{\:ij}_{k}]) \rangle.
    \label{Eq:uloss}
\end{equation}
\noindent
To study the steady-state behaviour, we set the time derivative in Eq.~\ref{Eq:usimplify} to zero. Also, the typical phonon propagation length in fibers is much smaller than the fiber length~\cite{boyd2020nonlinear,agrawal2000nonlinear}. Thus, SBS is effectively mediated by localized phonons, which allows us to drop the $z$ derivative term in Eq.~\ref{Eq:usimplify} as well. In that case, the acoustic modal coefficients are simplified to:
\begin{equation}
c_{k}^{ij}=\frac{1}{2\rho_0}\frac{O_1}{\Omega^2_{ijk}-\Omega^2+ i\Omega\Gamma_{ijk}}A_i B^{*}_j.
\label{Eq:ucoeff}
\end{equation}
\noindent
This completes a formal solution for the displacement field in terms of amplitudes of the signal and Stokes fields and the relevant overlap integrals, assuming acoustic modes have been solved either numerically or analytically (using Eq.~\ref{Eq:umode}). For special cases, such as elastically isotropic, circular, step-index fibers, it is possible to obtain acoustic modes semi-analytically~\cite{dong2010formulation,waldron1969some}. However, in general for arbitrary fiber geometries and refractive index profiles, they can be obtained using a numerical solver such as COMSOL~\cite{comsol}. 
\subsection{Coupled modal equations}
We can now use this displacement field to determine the source terms in the optical equation Eq.~\ref{Eq:Efield} and obtain coupled modal equations for the optical fields. The left-hand side of Eq.~\ref{Eq:Efield} can be simplified by substituting the modal decomposition of the electric field as given in Eq.~\ref{Eq:Eansatz} and using the optical modal equations, Eq.~\ref{Eq:Emode}. The equation for a particular mode $m$ can be isolated by taking a dot product with $\vec{f}_m^*$ and integrating over the fiber cross-section. This leads to following coupled amplitude equations (for more detail, see Appendix~\ref{appendixC}):
\begin{equation}
    -\frac{d B_m(\Omega)}{dz}=\sum_{i,j,l}Y_{mlij}(\Omega) A_l A_i^* B_j e^{i (\beta_i+\gamma_j-\beta_l-\gamma_m)z}
    \label{Eq:CAEs1}
\end{equation}
\begin{equation}
    \frac{d A_l(\Omega)}{dz}=\sum_{i,j,m}X_{mlij}(\Omega) B_m B_j^* A_ie^{-i (\beta_i+\gamma_j-\beta_l-\gamma_m)z}.
    \label{Eq:CAEs}
\end{equation}
\noindent
These are scalar, one-dimensional, ordinary differential equations in $z$, which accurately describe the acousto-optic interaction in a guided, translationally invariant in $z$, material system. The growth in each Stokes amplitude $B_m$ is proportional to the sum of the products of two signal amplitudes, $A_l$ and $A_i^*$, and a Stokes amplitude, $B_j$. The strength of each contribution depends on both a phase-mismatch factor and a coupling coefficient, $Y_{ijlm} (\Omega)$, which has a resonant frequency dependence and quantifies how efficiently the optical and acoustic modes in the source terms overlap with each other. The transverse dependence and vector nature of the interaction are contained within the coupling coefficients $Y_{ijlm}$ and $X_{ijlm}$, given by:
\begin{equation}
\begin{aligned}
Y_{mlij}=\sum_{k}\frac{{\mu_0\omega_2}}{8\rho_0}\frac{O_1^*O_2}{-i(\Omega^2_{ijk}-\Omega^2)+\Omega\Gamma_{ijk}}\\
X_{mlij}=\sum_{k}\frac{{\mu_0\omega_1}}{8\rho_0}\frac{O_1O_2^*}{-i(\Omega^2_{ijk}-\Omega^2)-\Omega\Gamma_{ijk}}.
\label{Eq:coupcoeff}
\end{aligned}
\end{equation}
\noindent
The coupling coefficient for a particular four-wave-mixing term $\{m,l,i,j\}$ is a sum of Lorentzians for each acoustic mode, $k$, with a center frequency given by the eigenfrequency of the acoustic mode, $\Omega_{ijk}$, and the linewidth given by the effective acoustic loss, $\Gamma_{ijk}$. The peak value of the curves is proportional to the overlap integrals, $O_1$ and $O_2$, where $O_1$ is given in Eq.~\ref{Eq:overlap1}. The second overlap integral $O_2$ is the projection of the optical source term onto the modal basis:
\begin{equation}
    O_2={\langle}((\stackrel{\leftrightarrow}{\pi}:\vec{q}^{\:*} \otimes{\vec{u}^{\:ml*}_{k})}.{\vec{f}^{(1)}_l}).{\vec{f}^{(2)*}_m}{\rangle}.
    \label{Eq:overlap2}
\end{equation}
\noindent
The derivative of the acoustic mode profile, $\vec{u}^{\:ml*}_{k}$, contracted with the photoelastic tensor, $\stackrel{\leftrightarrow}{\pi}$, gives the nonlinear susceptibility, which, when multiplied with optical mode profile, $\vec{f}_l^{(1)}$, gives the acousto-optic polarization field. The projection of this polarization on the particular Stokes mode profile, $\vec{f}_{m}^{(2)}$, gives the optical scattering strength. It can be shown by using integration by parts that the two overlap integrals $O_1$ and  $O_2$ are equal to each other for the same optical and acoustic mode indices.

The only assumptions we have used so far are translational invariance, neglecting the moving boundary terms~\cite{rakich2012giant}, slowly varying approximation~\cite{boyd2020nonlinear}, and damped phonon approximation~\cite{boyd2020nonlinear}. To the best of our knowledge, this is the first time these equations have been derived at this level of generality, for acousto-optic interactions in multimode, translationally invariant systems. These equations accurately capture the acousto-optic interaction at any length scale for arbitrary input excitations in multimode fibers, with any cross-section geometry.  
The equations can be integrated numerically much more efficiently than the original 3D nonlinear coupled optical and acoustic wave equations. This is because here the transverse degrees of freedom only need to be accounted for once (in the modal equations), due to the translational invariance in the longitudinal direction. 
In particular, the equations have made no phase-matching assumptions, and can be useful for studying SBS over length scales smaller than the phase-mismatch length scales, which can be quite substantial for systems like graded-index fibers~\cite{okamoto2021fundamentals,snyder1978modes}. In general, it should be remembered that phase matching is not a fundamental physical principle of nonlinear optics, but only becomes a good approximation when the system is long enough that phase-mismatched terms self-average to zero. For shorter systems the ``particle conservation approaches" to describe SBS~\cite{rakich2012giant,qiu2013stimulated}, can become inaccurate, since the interference effects between the quasiparticles describing the nonlinear interaction become significant. 
Note that the more fundamental Manley-Rowe relations~\cite{manley1959general}, which are a consequence of adiabatic invariance~\cite{brizard1995local}, still remain valid even without imposing the phase matching. Their validity is a consequence of permutation symmetry of the coupling strengths, which results from a fundamental symmetry in the acousto-optic interaction term in the Lagrangian~\cite{wolff2015stimulated}.
%Is the statement about adiabatic invariance one we still believe?

\subsection{Undepleted Signal}

Despite the simplifications due to translational invariance imposed above, the general coupled mode equations for Stokes and signal amplitudes derived above are nonlinear equations and are difficult to solve analytically. However, for analyzing the question of the threshold for significant SBS loss, we can assume that the Stokes amplitudes are much smaller than the signal amplitudes. The SBS threshold is typically defined as when the power of the backward-going Stokes waves reaches a few percent of the signal power~\cite{kobyakov2010stimulated,agrawal2000nonlinear}. In this limit, the decay in signal power, {\it due to SBS}, is negligible; with this undepleted signal approximation, the signal modal amplitudes can be assumed to be constants, determined by the input. The equations for the Stokes amplitudes then become a set of {\it linear} ordinary differential equations and can be rewritten in the following matrix representation: 
\begin{equation}
\frac{d B(\Omega)}{dz}=M(z,\Omega)B(\Omega)
\label{Eq:matODE}
\end{equation}
Here, $B(\Omega)$ is an $\rm N \times 1$ column vector with $m^{\rm th}$ entry equal to the Stokes amplitude, $B_m(\Omega)$. $\rm N$ is the total number of optical modes in the fiber. The coupling matrix $M(z,\Omega)$ is a $\rm N \times \rm N$ matrix whose entries are given by:
\begin{equation}
    M_{mj} = \sum_{il} Y_{mlij} A_lA_i^*e^{i (\beta_i+\gamma_j-\beta_l-\gamma_m)z}.
    \label{Eq:coupmat}
\end{equation}
\noindent
The above equation is a coupled linear system of first-order, homogeneous ordinary differential equations; its solution is given by:
\begin{equation}
    B(z)= \mathcal{P}\exp{\left[\int_{L}^{z} M(z') dz'\right]} B(L),
    \label{Eq:matgrowth}
\end{equation}
\noindent
where the Stokes amplitude vector $B$ at any point $z$ is given by the path-ordered exponential~\cite{giscard2015exact} of the mode-coupling matrix, $M$, times the Stokes amplitude vector at the output end of the fiber, $B(L)$, which is typically seeded by the spontaneous Brillouin scattering. At this level of approximation the Stokes amplitudes in various modes remain coupled, so that the growth of each $B_m$ is affected by all the others. The signal amplitudes and the fiber properties act as parameters in the Stokes growth through the coupling matrix $M(z,\Omega)$. 

\section{Phase Matching Limit} \label{sec:III}
\subsection{Phase-matched Stokes Power Growth}
The coupling matrix $M_{mj}$ dictates the effect of Stokes amplitude $B_j$ on the growth of Stokes amplitude $B_m$. Most of the elements of $M$ have complex phases, which vary on the length scale given by the mismatch of longitudinal wavevectors (propagation constants) for the modes involved. When the total fiber length is much greater than these length scales, the terms with complex phases oscillate between fixed values rather than growing exponentially. At the same time the phase-matched terms~\cite{kobyakov2010stimulated,agrawal2000nonlinear}, which have no net mismatch in their propagation constants, grow exponentially. In this limit, the effect of phase-mismatched terms becomes negligible compared to phase-matched terms and thus can be neglected. We shall focus on this limit for the remainder of this work.  The condition for phase-matching of terms can be obtained directly from Eq.~\ref{Eq:coupmat}:
\begin{equation}
    \beta_i+\gamma_j-\beta_l-\gamma_m=0.
    \label{Eq:phasematching}
\end{equation}
\noindent
A straightforward solution to this condition is $i=l,j=m$. This solution corresponds to transfer of power from signal mode $l$ to Stokes mode  $m$. Note that there can be alternate solutions to the phase-matching equation. For instance, when there are exactly degenerate modes in the fiber, i.e., two different modes have the same propagation constant, there are more solutions to Eq.~\ref{Eq:phasematching}.  We call them ``non-trivially phase-matched terms". Since exact degeneracies are usually lifted in realistic fibers due to fabrication imperfections, these alternate phase matching solutions are typically absent. When the exact degeneracies are present, all the solutions need to be included for maintaining consistency in the theory. For an ideal circular step-index fiber, the near degeneracy of the vector modes within the same group~\cite{okamoto2021fundamentals} gives rise to the non-trivially phase-matched terms over relevant length scales, which provides a connection between the vector and scalar SBS theories (see Appendix~\ref{appendixB} for more details). 
\noindent
For the rest of this work, we will assume that in real materials any exact symmetries are broken and thus there are no exactly degenerate solutions for the propagation constants leading to the uniqueness of the $i=l,j=m$ solution. This leads to a dramatic simplification of the Stokes growth equations. The coupling matrix becomes diagonal ($m=j$) leading to independent growth for each of the Stokes amplitudes:
\begin{equation}
    \frac{d B_m(\Omega,z)}{dz}=\left[\sum_l Y_{mlml}(\Omega) {|A_l|}^2 \right]\:B_m(\Omega,z)  
    \label{Eq:ampgrowth}
\end{equation}
The growth rate of Stokes amplitude $m$ is proportional to the signal power in various modes, ${l}$, weighted by effective coupling, $Y_{mlml}$. The Stokes amplitude growth equations can be converted to the growth equations for Stokes power by multiplying with the complex conjugate of the Stokes amplitude on both sides and adding the complex conjugate term, leading to: 

\begin{equation}
\begin{aligned}
    \frac{d P^s_m(\Omega,z)}{dz}&=-\left[\sum_l g^{(m,l)}_B(\Omega) \Tilde{P}_l\right]P_0\:P^s_m(\Omega,z)\\& \equiv -\Tilde{g}_m(\Omega) P_0 P^s_m(\Omega,z),
    \label{Eq:powergrowth}
\end{aligned}
\end{equation}
where $P^s_m(\Omega,z)$ is the Stokes power in mode $m$, $\Tilde{P}_l$ is the fraction of signal power in mode $l$ ($\sum_l \Tilde{P}_l=1$), $P_0$ is the total signal power, $\Omega$ is the Stokes frequency shift, and $g^{(m,l)}_B(\Omega)$ is the BGS for Stokes--signal mode pair $(m,l)$. The Stokes power in each mode grows independently in the backwards direction. Here we have defined for each Stokes mode $m$, an effective BGS, $\Tilde{g}_m(\Omega)$, which is equal to the weighted sum of pairwise BGS $g^{(m,l)}_B(\Omega)$ with weights equal to the fractional signal power $\Tilde{P}_l$ in various modes. Both the pairwise and effective BGS have units of [$\rm W^{-1}m^{-1}$]. We shall see below that the effective BGS, $\Tilde{g}_m(\Omega)$, is physically meaningful; it is the engineered Brillouin spectrum for the mode $m$ induced by our choice of input power distribution into signal modes, and can be used to understand the increase in the SBS threshold due to multimode excitation.

The pairwise BGS $g^{(m,l)}_B$ depends on the fiber properties and is equal to twice the real part of coupling coefficient $Y_{mllm}$, thus it can be calculated using Eq.~\ref{Eq:coupcoeff}:
\begin{equation}
    g^{(m,l)}_B(\Omega) = \alpha\sum_{k}{|O_{mlk}|}^2\frac{\frac{\Gamma_{mlk}}{2}}{(\Omega_{mlk}-\Omega)^2+{(\frac{\Gamma_{mlk}}{2})}^2}.
    \label{Eq:Brillouin}
\end{equation}
%Is the approximation sign here due to phase-matching assumption; if so it should be dropped.
\noindent
Here, $\alpha$ is a constant including various material and optical constants. The BGS for mode pairs $(m,l)$ is a sum of Lorentzian curves for each acoustic mode $k$ with a center frequency equal to the acoustic eigenfrequency $\Omega_{mlk}$, and the linewidth is equal to effective acoustic loss $\Gamma_{mlk}$. Each acoustic mode contribution is weighted by ${|O_{mlk}|}^2$, the corresponding overlap integral of the optical and acoustic modes involved, given by:

\begin{equation}
       O_{mlk}= \langle(\vec{q}\cdot[\stackrel{\leftrightarrow}{\pi}: \vec{f}^{(1)}_l\otimes \vec{f}^{(2)^*}_m])\cdot \vec{u}^{\:ml*}_{k} \rangle.
       \label{Eq:overlap}
\end{equation}
\noindent
The overlap integral takes into account the full tensorial nature of the source terms in both optical and acoustic equations as well as the vector nature of both the optical and the acoustic modes. Finally, we can solve Eq.~\ref{Eq:powergrowth}, to derive the exponential growth in Stokes power in various modes:
\begin{equation}
    P^s_m(\Omega,0)= P^s_m(\Omega,L)e^{\Tilde{g}_m(\Omega) P_0 L}
    \label{Eq:powersolution}
\end{equation}
\noindent
The Stokes power in each mode grows exponentially in the backward direction with a growth rate equal to the effective BGS $\Tilde{g}_m$ multiplied by the total signal power $P_0$.

\subsection{Elastically Isotropic Fibers}

Up to this point the theory has been quite general, in terms of the material system and also the fiber geometry in that arbitrary transverse cross-section and breaking of geometric isotropy, (i.e. non-circular or transversely patterned geometries), obey the above equations when the appropriate optical and acoustic modes are used in calculating the overlap integrals. The photoelastic tensor of each fiber material depends on the atomic scale geometry and whether the system is crystalline, amorphous or has any other anisotropy on that scale.  Commonly used optical fibers, usually made from glass, are elastically isotropic, in which case the full photoelastic tensor can be described by just two independent constants~\cite{nelson1971theory,feldman1975relations,biegelsen1974photoelastic}. This results in a substantial simplification of the overlap integral in Eq.~\ref{Eq:overlap}. A detailed discussion and derivation of the relevant overlap integrals in this case is given in Appendix~\ref{appendixA}.1). We will henceforth focus on the elastically isotropic case, where the integrals simplify to:

\begin{equation}
O_{mlk} \approx  \gamma_e\langle\vec{f}^{(1)}_l\cdot \vec{f}^{(2)^*}_m\vec{\nabla}\cdot \vec{u}^{\:ml*}_{k}\rangle.
\label{Eq:overlap-simple}
\end{equation}
\noindent
The overlap integral is proportional to the electrostriction constant, $\gamma_e$, which is equal to one of the independent components of the photoelastic tensor, $\gamma_e=\pi_{1122}$. The overlap integral for optical modes $m$ and $n$ with acoustic mode $k$ involves the dot product of optical mode profiles $\vec{f}^{(1)}_l$ and $\vec{f}^{(2)}_m$ multiplied with the strain field for the acoustic mode $\vec{\nabla}\cdot \vec{u}^{\:ml*}_{k}$, integrated over the fiber cross section. 

To calculate the strain field, the vector acoustic modal equation, Eq.~$\ref{Eq:umode}$ need to be solved to obtain the displacement field, $u^{ml*}_k$, followed by taking its divergence. For isotropic fibers this results in a contribution from both longitudinal acoustic velocity and the shear acoustic velocity to the acoustic mode profiles ~\cite{dong2010formulation,waldron1969some,auld1973acoustic} (for a detailed discussion, see Appendix~\ref{appendixA}.2).
Typically in silica fibers the shear acoustic velocity is significantly lower than the longitudinal acoustic velocity, resulting in predominantly longitudinal acoustic modes, with small perturbations due to the shear velocity (see  Fig.~\ref{Fig:acmodes}). In Appendix~\ref{appendixA}.2, we compare results for solving the isotropic version of Eq.~$\ref{Eq:umode}$ with and without the shear term. We find that, for the case of silica fibers, neglecting the shear velocity contribution leads to only a small error ($1-5 \%$) in the overlap integrals, and the strain field can be directly calculated by solving the scalar acoustic wave modal equation Eq.~\ref{Eq:scalaraceq}. Thus, to calculate the overlap integral in Eq.~\ref{Eq:overlap-simple}, we make the substitution $\frac{1}{q_{ml}}\vec{\nabla} \cdot \vec{u}^{\: ml}_{k}=\delta\rho^{\: ml}_{k}$,
where $\delta \rho^{\: lm}_{k}$ is normalized $k^{\rm th}$ eigenmode (for optical mode pair $\{l,m\}$) of the scalar density fluctuation equation. The overlap integral simplifies to:

\begin{equation}
O_{mlk} \approx  \gamma_e q_{ml}\langle\vec{f}^{(1)}_l\cdot \vec{f}^{(2)^*}_m \delta\rho^{ml}_{k}\rangle.
\label{Eq:overlap-ac-scalar}
\end{equation}
We use this form of the overlap integral to calculate the Brillouin gain spectra in the rest of the paper. The presence of the dot product between the optical mode profiles is a key  difference from standard scalar SBS theories~\cite{ke2014stimulated}. For single mode fibers ($l=m$) the dot product simply reduces to scalar multiplication of modes, making scalar SBS theories appropriate. For multimode fibers the polarization for different modes can be significantly different and spatially varying (see Fig.\ref{Fig:optmodes}). Therefore, for accurate calculation of intermodal gain ($l\neq m$), correctly evaluating this dot product is necessary. Neglect of this feature results generically in significant inaccuracies in scalar theories (for a more detailed discussion, see Appendix~\ref{appendixB}).

\subsection{Multimode SBS Threshold}
\noindent
The SBS threshold is typically defined as the output signal power at which the backward reflected Stokes power becomes a non-negligible fraction (typically $ > 1 \% $) of the signal power~\cite{agrawal2000nonlinear,kobyakov2010stimulated,boyd2020nonlinear,lu2015theoretical}. We can use the Stokes growth equation derived in Sec.\ref{sec:II}.A (Eq.~\ref{Eq:powersolution}) to obtain the formula for the multimode SBS threshold. We assume SBS is seeded by spontaneous Brillouin scattering at the far end of the fiber, which leads to average photon density of one photon per mode in all the modes of the fiber. We set the threshold to be when the ratio of exponentially amplified Stokes power over the total signal power is equal to $\xi = 1 \%$. The seed power from spontaneous Brillouin scattering is typically multiple orders of magnitude smaller than the signal power, thus the amplification factor required to reach SBS threshold is very high. The mode with the highest Stokes growth rate thus exponentially dominates the Stokes power and can be used to approximate the total reflected power. Therefore the threshold condition becomes:
\begin{equation}
   P^s={P
_{\tiny \rm \:N}}\:\:e^{P_{\rm th}\:L\: g_B}=\xi P_{\rm th}, 
    \label{Eq:SBSth0}
\end{equation}
which can be rearranged as:
\begin{equation}
    P_{\rm th}\:L\: g_B = \log(\frac{\xi P_{\rm th}}{P
_{\tiny \rm \:N}}), 
    \label{Eq:SBSth}
\end{equation}
\noindent
where, $P_{\rm th}$ is the SBS threshold, $L$ is the length of the fiber, and ${P
_{\tiny \rm \:N}}$ is the Stokes noise power seeded by the spontaneous Brillouin scattering. We have introduced an overall Brillouin gain coefficient $g_B$ which is equal to:

\begin{equation}
    g_B \approx \max_{\Omega, m} \Tilde{g}_m(\Omega) = \max_{\Omega, m} \sum_l g_B^{(m,l)}(\Omega)\Tilde{P}_l
    \label{Eq:GB}
\end{equation}

The SBS threshold is inversely proportional to the length of the fiber and the overall Brillouin gain coefficient $g_B$. It also weakly (logarithmically) depends on the output power level, seed power and the fraction ($\xi$) at which the threshold is set. Additionally, we have approximated the final Stokes power by the Stokes power in the mode with the highest growth rate, which is justified due to the exponential nature of the growth. In case there are multiple modes (say $M_s$) with similar growth rates, the Stokes power will be $M_s$ times higher than our estimation, which will lead to a $\log(M_s)$ correction to the effective SBS gain and threshold, which we found to be quite small.  It can be verified that the multimode threshold formula in Eq.~\ref{Eq:SBSth} reduces to the formula for single-mode fiber when only the fundamental mode is present. More generally, however, the SBS threshold depends on both the fiber properties (through $g_B^{(m,l)}(\Omega)$) and the distribution of power in various signal modes, $\{\Tilde{P}_l\}$. Our formalism allows an efficient calculation of the SBS threshold for different input multimode excitations for highly multimode fibers with any cross-section. In addition, this formalism allows the investigation of SBS suppression by using highly multimode excitation to vary the power distribution in the signal modes at the input, a restricted form of wavefront shaping. In the next section, we show that by distributing power in multiple modes it is possible to substantially reduce the effective SBS gain, since $g_B^{(m,l)}$ strongly depends on mode indices $(l,m)$.  This leads to a significant increase in SBS threshold with the number of excited modes when all the modes are equally excited.

%caption previously  (b) Effective Brillouin gain spectra for different input excitations. When only the fundamental mode is excited (FM-only), effective BGS (equal to the intramodal BGS for FM  $g_B^{(1,1)}$, shown by dotted red curve) has maximum peak value, leading to the lowest SBS threshold. Upon excitation of only the higher-order mode (HOM-only) effective BGS (equal to the intramodal BGS for HOM $g_B^{(120,120)}$, shown by blue dotted curve) has slightly lower peak value, resulting in a higher SBS threshold. When an optimal combination of FM and HOM is excited, the effective BGS (equal to the weighted sum of intramodal and intermodal BGS) in both the modes $\Tilde{g}_{1}$ and $\Tilde{g}_{120}$ have lower peak value than the previous two cases, leading to the highest SBS threshold. The optimal power division ($\Tilde{P}_1\sim40 \%$, $\Tilde{P}_{120}\sim60 \%$) results in an SBS Threshold roughly two times higher. (Note that for optimal combination $\Tilde{g}_1(\Omega),\Tilde{g}_{120}(\Omega)$ are not identical, but their peak values are equal). Further power division among many modes will lead to even higher thresholds and significant broadening of the effective BGS.
\section{Numerical Results} \label{sec:IV}

\begin{figure*}[t!]
		\centering\includegraphics[width=\textwidth]{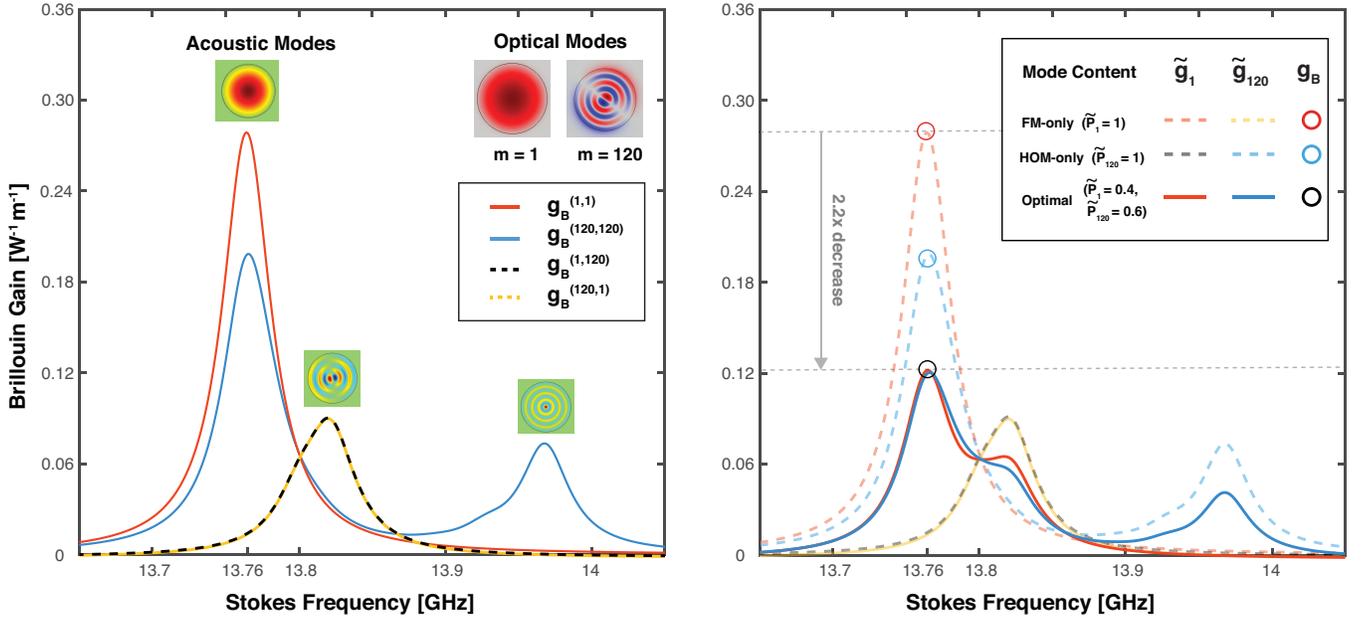}
		\caption{	{\bf Illustration of SBS suppression with multimode excitation using a two-mode example} (a) Brillouin gain spectra (BGS) for a circular step index fiber (fiber \textit{A}) (summed over all possible acoustic interactions) for the fundamental mode (FM) ($m=1$) and a higher order mode (HOM) ($m=120$). The intermodal BGS (black and yellow curves, remains identical under index interchange) have significantly lower peak values and relatively higher Brillouin frequency (frequency of the peak) compared to both intramodal BGS (red and blue curves). Acoustic modes with dominant contributions are shown corresponding to each peak in the BGS. The intramodal BGS have dominant contribution from the fundamental acoustic mode whereas the intermodal BGS has dominant contribution from higher-order acoustic mode. As shown, intramodal BGS  for HOMs can have multiple peaks. \textbf{(b) Effective BGS in the FM and HOM Stokes modes for three different input excitations.} For single mode input excitations (FM-only or HOM-only) the effective BGS in each mode $\Tilde{g}_m$ is given by corresponding intramodal or intermodal BGS (reproduced in figure (b) as dotted curves). For a multimode excitation the effective BGS in each Stokes mode is a weighted sum of intramodal and intermodal BGS (shown as solid curves for optimal two-mode excitation). The overall Brillouin gain coefficient $g_B$ for a particular input excitation is given by the maximum value of effective BGS across all the Stokes modes and frequencies (Eq.~\ref{Eq:GB}). $g_B$ is highest for FM-only excitation (red circle), is 1.4 times lower for HOM-only excitation (blue circle) and is 2.2 times lower for optimal combination (black circle) of FM and HOM ($\Tilde{P}_1\sim0.4$, $\Tilde{P}_{120}\sim0.6$). Thus, SBS threshold (which is inversely proportional to $g_B$) is 2.2 times higher for optimal two-mode excitation compared to FM-only excitation. Further power division among many modes will lead to even higher SBS thresholds.}
        \label{Fig:BGS}
\end{figure*}
To summarize, we have shown that in a MMF the Stokes power in each mode grows exponentially with a growth rate equal to the total signal power multiplied with effective Brillouin gain spectra (BGS)(Eq.~\ref{Eq:powersolution}).
Therefore, effective BGS and the SBS threshold (Eq.~\ref{Eq:SBSth}) in a MMF depends on both the input signal power distribution $\{\Tilde{P}_l\}$  and the pairwise BGS $g_B^{(m,l)}(\Omega)$. For a given Stokes mode $m$ and signal mode $l$, the BGS can be calculated by the formula given in Eq.~\ref{Eq:Brillouin}.
The BGS for each mode pair is a sum of Lorentzians corresponding to individual acoustic modes, weighted by the overlap of the optical and acoustic modes. For elastically isotropic fibers, we identified a simplified form of the overlap integral, which is reasonably accurate for calculating BGS, involving a dot product of the vector optical modes multiplied with a scalar acoustic eigenmode (Eq.~\ref{Eq:overlap-ac-scalar}). These equations (Eq.~\ref{Eq:Brillouin}, Eq.~\ref{Eq:powersolution}, Eq.~\ref{Eq:SBSth}, and Eq.~\ref{Eq:overlap-ac-scalar}) are sufficient to determine the SBS threshold for a given multimode fiber for any input multimode excitation. 

\begin{table}[b]
    \centering
    \begin{tabular}{|c|c|}
 \hline
 Parameter & fiber \textit{A} \\ [0.5ex] 
 \hline\hline
 Core Shape & Circular \\
 \hline
 Core Diameter [$\mu m$] & 10 \\ 
 \hline
 Cladding Diameter [$\mu m$] & 50\\
 \hline
 Core Refractive Index & 1.4803 \\
 \hline
Cladding Refractive Index & 1.4496   \\
 \hline
Signal Wavelength [$\mu m$] & 1.064 \\
\hline
Number of Optical Modes & 160\\
  \hline
Core Acoustic Velocity, $v_L$ [m/s] & 4946  \\
  \hline
Core Acoustic Velocity(shear), $v_s$ [m/s] & 3189  \\
  \hline
Cladding Acoustic Velocity, $v_L$  [m/s]& 5944 \\ 
\hline
Cladding Acoustic Velocity(shear), $v_s$  [m/s]& 3749 \\ 
\hline
Number of Acoustic Modes & $\geq 1000$\\ [1ex] 
\hline
    \end{tabular}
    \caption{Detailed parameters for fiber \textit{A}.}
    \label{tab:my_label}
\end{table}

The extent to which the input power division influences the SBS threshold is determined by the fiber properties through the pairwise BGS, $g_B^{(m,l)}$. To study the properties of the pairwise BGS, we calculate the BGS for all the mode pairs for a highly multimode circular step-index fiber. We consider a commercially available fiber (fiber \textit{A}) with germanium-doped silica core and pure silica cladding. The core radius of the fiber is 10 $\mu$m and with a numerical aperture of 0.30, supporting 160 optical modes. The detailed parameters are given in Table~\ref{tab:my_label}.

To illustrate key properties of multimode BGS, we plot the BGS for the intramodal gain of the (FM) (mode number, $m=1$) and a higher order mode (HOM) (mode number, $m=120$), as well as the intermodal gain between them, in Fig.~\ref{Fig:BGS}a. The intramodal gain for the FM (red curve) has the maximum peak value. The intramodal gain for the HOM (blue curve) has a slightly lower peak value, due to a larger effective acousto-optic area~\cite{kobyakov2010stimulated,mermelstein2007sbs} and also shows secondary peaks corresponding to interactions with higher order acoustic modes. Interestingly, the intermodal gain (black and yellow curves) between the FM and the HOM has a substantially lower peak value than both intramodal curves. This is a result of inefficient acousto-optic overlap between the FM and the HOM due to significant variations in the polarization and intensity profile (shown in inset). Further, the intermodal BGS (black and yellow) peak at a higher Brillouin frequency than the intramodal BGS (red and blue). This is because lower-order radially symmetric acoustic modes facilitate intramodal gain, whereas higher-order acoustic modes (with higher eigenfrequecy) are responsible for intermodal gain. We have shown the profile of the acoustic modes with dominant contribution corresponding to each of the peaks in the BGS in Fig.~\ref{Fig:BGS}a. 
\begin{figure*}[t!]
		\centering\includegraphics[width=\textwidth]{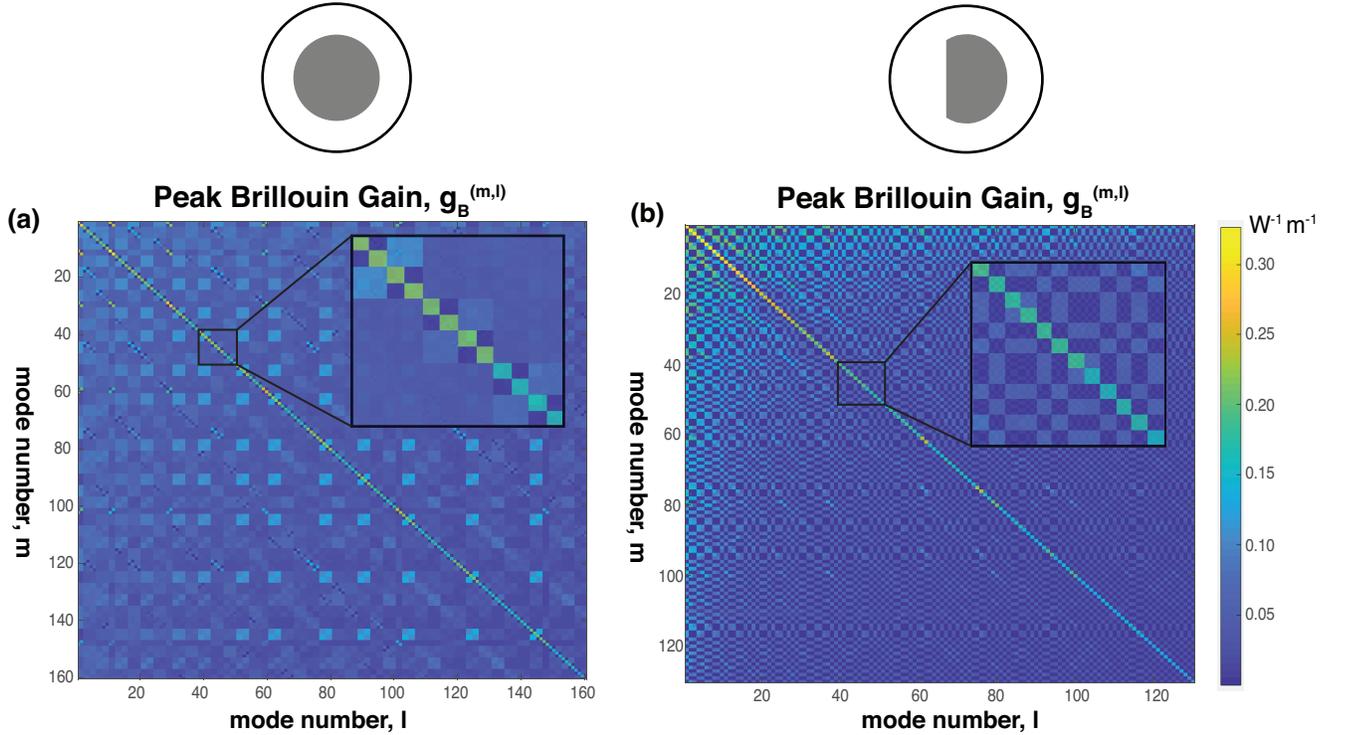}
		\caption{	Matrix of peak values of pairwise BGS for all possible Stokes-signal mode pairs for highly multimode step-index fibers with (a) circular (fiber \textit{A}) and (b) D-shaped (fiber \textit{B}) cross-sections. The insets show zoomed-in views of sections of the matrices in each case. The modes are ordered according to their effective refractive indices.  The peak values of intermodal BGS (off-diagonal elements) are typically lower than the intramodal BGS (diagonal elements) for both fibers A and B. The matrix for fiber \textit{B} has a checkerboard structure since modes are either completely x or y polarized. Fiber \textit{A} does not exhibit this structure due to the  spatially varying polarization patterns of the optical modes.}
        \label{Fig:peak}
\end{figure*}

\begin{figure*}[t!]
		\centering\includegraphics[width=0.919\textwidth]{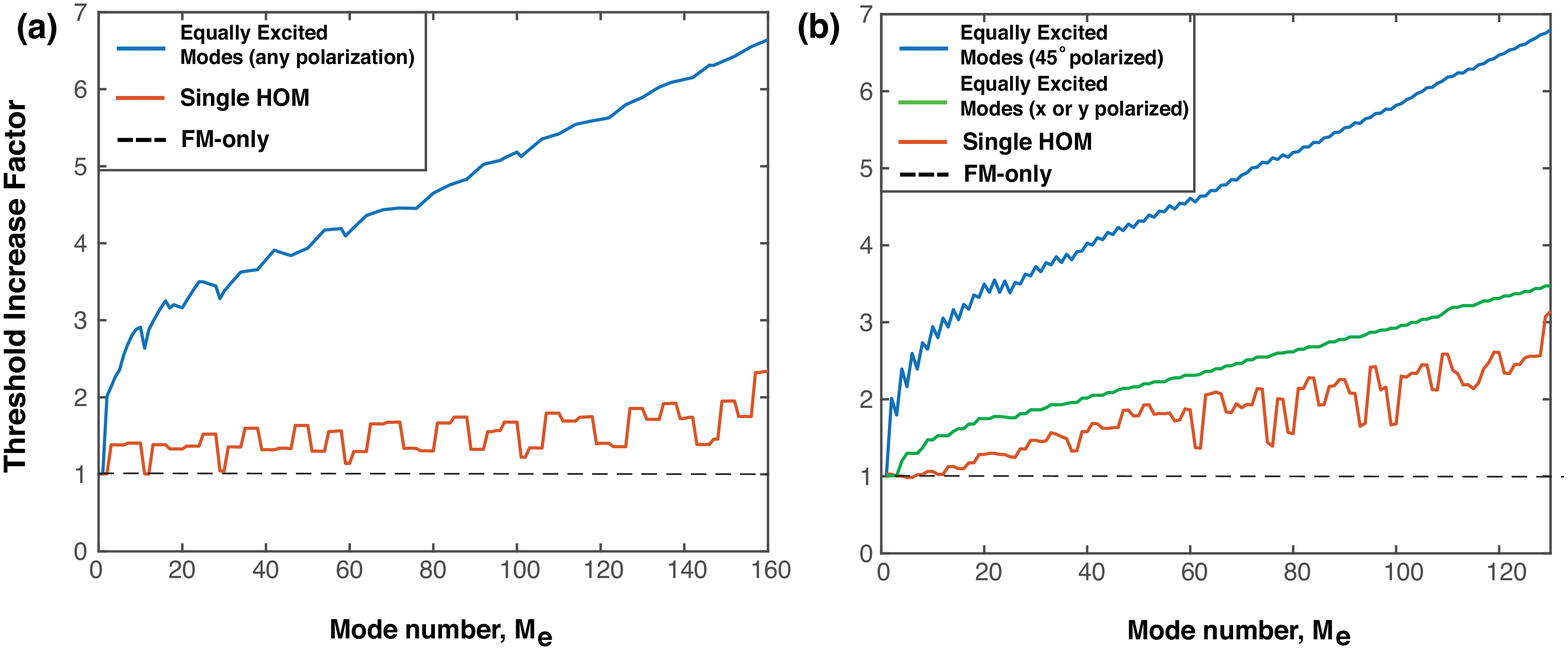}
		\caption{Scaling of the SBS threshold with the number of modes for different input excitations in multimode step index fibers with (a) circular (fiber \textit{A}) and (b) D-shaped (fiber \textit{B}) cross-sections. For reference, the fundamental mode (FM)-only excitation is set to be 1, shown as a black dashed line. In (a) the blue curve represents the case when all the modes are equally excited up to mode number $M_e$ with any polarization. When $M_e=160$, a 6.5-times-higher SBS threshold is obtained for equal mode excitation, substantially higher than exciting a single HOM (mode number $=M_e$) shown by the red curve. In (b), the SBS threshold for equal mode excitation increases more slowly for $x$- or $y$-polarized input (green) as compared to input polarized at $45^{\circ}$ (blue) with respect to the axis of symmetry ($x$-axis). A 6.7 times higher SBS threshold is obtained when $M_e=130$ for equal mode excitation and $45^{\circ}$ polarization, significantly higher than exciting a single HOM (mode number $=M_e$) shown by red curve.}
        \label{Fig:scaling}
\end{figure*}

The relatively lower peak value of intermodal BGS along with the shifted spectrum suggests that multimode excitation can lead to SBS suppression by lowering the effective Brillouin gain (Eq.~\ref{Eq:powergrowth}), leading to a higher SBS threshold. To show this explicitly, we consider three different input excitations: (1) all of the signal power is in the FM $(m=1)$, this will be used as a reference since this is by default the case in SMFs, (2) all of the signal power is in a single HOM $(m=120)$, and (3) the signal power is divided between the FM and HOM. The SBS threshold is inversely proportional to the overall Brillouin gain coefficient $g_B$ (Eq.~\ref{Eq:SBSth}), which is given by the maximum value of the effective BGS across all the Stokes modes and frequencies (Eq.~\ref{Eq:GB}). Recalling that the effective Brillouin gain is a weighted combination (depending on the input signal mode content) of pairwise BGS (see. Eq.~\ref{Eq:powergrowth}), we can compare the three cases.  When all of the power is in a single mode, as in cases (1) and (2) the effective BGS is simply equal to the intramodal BGS for that mode, $g_B^{m,m}$, and the intermodal BGS $g_B^{n,m}$ for modes $n\neq m$. Typically, we find that intramodal BGS have higher peak value than intermodal BGS. Hence for single mode excitations $g_B$ is simply the peak value of respective intramodal BGS shown in Fig.~\ref{Fig:BGS}a, which we reproduce in Fig.~\ref{Fig:BGS}b (dashed red and blue curves). Since the FM-FM BGS has the highest peak value (red  circle in Fig.~\ref{Fig:BGS}b), FM-only excitation will result in the lowest SBS threshold (for reference we denote it by $P_{\rm th}^0$). The HOM-only curve has a slightly lower peak (blue  circle in Fig.~\ref{Fig:BGS}b) and will lead to a higher SBS threshold (1.4$P_{\rm th}^0$). In case (3) (two-mode excitation) the effective BGS in each Stokes mode is a weighted sum of intramodal and intermodal BGS (shown as solid curves in Fig.~\ref{Fig:BGS}b). Maximum peak value $g_B$ in this case can be 2.2 times lower(blue circle in Fig.~\ref{Fig:BGS}b) for optimal combination of FM and HOM ($\Tilde{P}_1\sim0.4$, $\Tilde{P}_{120}\sim0.6$) leading to a 2.2 higher SBS threshold compared to FM-only excitation. Note that for this two-mode case the optimal power distribution corresponds to matching the two peak values of the effective BGS for the FM and HOM.

The SBS suppression (increase in SBS threshold) illustrated with two modes ($m=1$ and $m=120$) in Fig.~\ref{Fig:BGS} is due to generic properties of BGS such as relatively weaker intermodal gain and shifted BGS peaks, and hence generalizes to many-mode excitation. In Fig.~\ref{Fig:peak}a, we show in a color scale the matrix of peak values of BGS for all possible mode pairs for the circular step-index fiber described before (Fiber \textit{A}). This gives a $160 \times 160$ matrix of positive entries, where each element $(m,n)$ describes the SBS interaction between stokes mode $m$ and signal mode $l$. A zoomed-in view of a section of the matrix is shown in the inset. It can be clearly seen that the intermodal gain (off-diagonal elements) is generically smaller than the intramodal gain (diagonal elements). For generality, we also consider another fiber (fiber \textit{B}), which has all the same material properties as fiber \textit{A} but with a D-shaped cross-section (see. Fig.~\ref{Fig:optmodes}). A motivation to study this fiber is the ray chaotic nature of the D-shaped cavities, leading to more ergodic modal profiles. This fiber supports 130 optical modes with polarization roughly aligning with either the $x$ axis (axis of symmetry) or $y$ axis (perpendicular to $x$ axis), unlike the circular cross-section fiber (see. Fig.~\ref{Fig:optmodes}). The matrix of peak values of the BGS for fiber \textit{B} is shown in Fig.~\ref{Fig:peak}b. A zoomed-in view of a section of the matrix is shown in the inset. The intermodal gain is again generically weaker than the intramodal gain, similar to fiber \textit{A}. In addition, there is a checkerboard pattern in the matrix which is a result of complete decoupling of modes with two orthogonal polarizations ($x$ and $y$). The contrast between intermodal and intramodal gain (for modes with the same polarization) is actually found to be lower than for fiber \textit{A}. 
\begin{figure}[t!]
		\centering\includegraphics[width=0.45\textwidth]{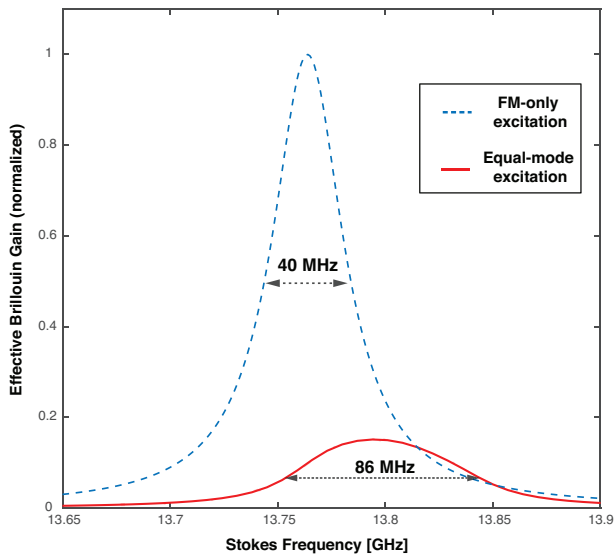}
		\caption{Effective Brillouin gain spectra in fundamental Stokes mode for FM-only excitation (dotted blue curve) and 160-equal-mode excitation (red curve) in a circular step index fiber (fiber \textit{A}). The spectra are normalized such that peak value of gain for FM-only excitation is equal to one. Brillouin gain for equal mode excitation is significantly broadened leading to a full width at half maximum (FWHM) of 86 MHz, which is more than twice the FWHM for Brillouin gain for FM-only excitation (40 MHz). The broadening is a result of addition of multiple intermodal and intramodal BGS which peak at different frequencies, upon equal mode excitation.  As a result of this broadening and weak intermodal coupling, the peak value of Brillouin gain is 6.5 times lower for equal mode excitation, leading to an equivalent increase in SBS threshold.}
        \label{Fig:bgseq}
\end{figure}

The presence of relatively weaker intermodal gain in both circular and D-shaped fiber suggests that exciting multiple modes instead of exciting only the FM or a single HOM, can lead to substantially higher SBS threshold. To show this, we consider equal division of signal power in all of the modes up to mode number $M_e$, and calculate the SBS threshold as $M_e$ is varied.  The results for fiber \textit{A} are shown in Fig.~\ref{Fig:scaling}a. For reference, we compare all the results to the SBS threshold for FM-only excitation (dotted black line) by defining a threshold increase factor as the ratio of the SBS threshold for a given excitation divided by that with FM-only excitation. The SBS threshold increases almost monotonically for equal mode excitation (blue curve) as $M_e$ is increased and reaches 6.5 times when $M_e=160$. As illustrated with the two-mode example above, the increase in SBS threshold upon multimode excitation results from relatively small intermodal gain and not simply from increasing the acousto-optic effective area, due to increasing mode order.  To illustrate this further, we also show in the plot the SBS threshold when the highest single HOM (mode number = $M_e$ ) is excited (red curve). For the best HOM ($M_e=160$), the threshold increase is 2.1 times, significantly lower than for equal mode excitation. 
As we noted above, power division tends to broaden the effective SBS gain spectrum, which also reduces the peak value. This is seen dramatically in the case of equal power division among all $160$ modes of this circular fiber.  The effective gain spectrum is shown in Fig.~\ref{Fig:bgseq}, compared to that of FM-only excitation. We find a more than doubling of the gain bandwidth under equal power division, along with the peak value of the spectrum decreasing greatly, leading to a factor of 6.5 increase in the SBS threshold.

The threshold behavior for fiber \textit{B}, shown in Fig.~\ref{Fig:scaling}b, has an interesting new feature. For the D-shaped fiber core, the SBS threshold strongly depends on the input polarization. When the polarization of light is either along $x$ or $y$ axis, and power is equally divided in the modes (green curve), the maximum SBS threshold obtained is 3.5 times higher than the FM-only threshold for $M_e=130$. However, with polarization at an angle of $45^{\circ}$ to the $x$ axis, the maximum SBS threshold is 6.7 times higher than FM-only excitation, when all 130 modes are equally excited (blue curve). A strong dependence of SBS threshold on input polarization has been previously observed in single-mode birefringent fibers~\cite{van1994polarization}. Similar dependence has been observed for the threshold of transverse mode instability~\cite{jauregui2021mitigation}, which is a thermo-optic nonlinear effect that results from a four-wave mixing type interaction, similar to SBS. In our model, it is easy to understand this dependence through the checkerboard structure of the BGS matrix (see Fig.~\ref{Fig:peak}b). When the light is launched at $45^{\circ}$, power is launched equally in $x$- and $y$-polarized modes, which have zero intermodal interaction with each other, qualitatively reducing the intermodal gain, leading to roughly a factor of two reduction in the total effective SBS gain, and almost a two-fold increase in SBS threshold. Note that this polarization dependence is not present in fiber \textit{A} (circular core), since there is no preferred axis of symmetry in an ideal circular fiber. Overall results in both the circular and D-shaped fiber show that a significant enhancement ($\sim 6.5$) in SBS threshold can be obtained upon equal excitation of many modes. 
%Mention possible difference in non-symm chaotic fiber

\section{Discussion and Conclusion}\label{Sec:V}
In this work we have presented and implemented the first accurate theoretical model for predicting the SBS threshold under arbitrary multimode excitation. The theory elucidates the physics of intermodal gain and explains why it is generically weaker than intramodal gain, hence favoring highly multimode excitation to achieve a substantially higher SBS threshold.  
The theory presented above can be used, after the applicable approximations (primarily, the assumptions of undepleted signal and phase matching), to calculate the SBS spectrum and threshold for highly multimode fibers with any refractive index profile and cross-sectional geometry, taking into account fully the vector and tensor nature of the optical and acoustic fields and forces. The signal modal power distribution enters the theory as a set of control parameters, determining the effective spectrum and the threshold.
The generality of our linearized equations, with correct treatment of vector modes and tensor interaction coefficients, makes them suitable for micro and nanowaveguides, although in the latter case boundary terms~\cite{rakich2012giant,shi2017invited,poulton2013acoustic} will need to be added to the sources, which will modify the overlap integrals.

As demonstrated, these equations provide a realistic computational framework for calculating the intramodal and intermodal gain spectra for all pairs of modes in a highly multimode fiber. The power in each Stokes mode grows exponentially, with a modal growth rate independent of the power in the other Stokes modes.  Typically the SBS threshold will be determined by the mode with the highest growth rate and each growth rate depends strongly on the modal power distribution of the signal excitation.  We have shown that dividing the input power among modes generically decreases the maximal Stokes gain, increasing the SBS threshold, due to the relative weakness of the intermodal gain, and the broadening of the effective SBS spectrum.

 In parallel with this theoretical work, experiments have been performed on SBS in passive multimode fibers and have confirmed the basic physical principle implied by our theory: that exciting a multimode fiber with many modes substantially increases the SBS threshold compared to single-mode excitation of the same fiber. The experiments have also validated an important property not studied here: the feasibility of refocusing the speckled multimode beam to a reasonable focal spot in the far field.  Further applications of the theory to experimental data under different excitation conditions are made in that work \cite{SBSexpinPrep}. One important prediction of the current theory is that the increase of the SBS threshold is independent of the relative phases of the signal modal amplitudes, and only depends on the power in each mode.  This suggests that these relative phases can be controlled (e.g., using an SLM) so as to refocus the beam at the output, while maintaining an increased SBS threshold due to multimode excitation. 
%(Practically speaking the refocusing imposes some restrictions on the modal amplitude distribution, but is consistent with highly multimode excitation of the fiber).

 In the current work we have only presented results for single-mode and equal mode excitation of a fiber. Beyond this, our theoretical framework allows us to pose the maximization of the SBS threshold as a convex optimization problem for a given matrix of Brillouin gain coefficients.  In future work we will explore this approach and expect that an even larger suppression of SBS can be achieved with a highly multimode, but non-uniform input signal power distribution \cite{SBSoptinPrep}.

Suppression of SBS is especially important in fiber amplifiers, since it will allow power scaling in narrow-linewidth high-power fiber lasers~\cite{zervas2014high}. In this work we have not explicitly included the signal gain of the active fiber.  This is trivial in the absence of linear mode-dependent gain and loss~\cite{ho2012exact}, but to model realistic fiber amplifiers these effects will need to be included.  This can be modelled and/or measured and included without significant complication of the analytic and computational framework.  A further quite important effect is that of gain saturation of the signal~\cite{paschotta1997ytterbium}. This is a space-dependent nonlinear effect which cannot be neglected if one wants to describe such systems quantitatively. However we can still neglect the depletion of the signal to the Stokes mode when calculating the saturated signal field; moreover, there are iterative self-consistent approaches to include this in a semi-analytic framework. The qualitative physics which makes intermodal gain weaker than intermodal gain and favors highly multimode excitation is not changed by gain saturation.

%% applications  - Kabish write this paragraph.
An important application of SBS in fibers has been in developing distributed sensors for temperature and strain. Most studies focus on single-mode fibers but it has been posited that SBS in multimode fibers can lead to better performance in sensing applications \cite{iezzi2011stimulated}. In MMFs different Stokes modes have distinct Brillouin frequency shifts, which also depend on external parameters such as temperature and strain.  Thus, sensing platforms based on SBS in MMFs can possibly be utilized to extract more information about the fiber environment compared to SMFs. Our multimode SBS theory will be quite useful in providing a comprehensive framework for any such future studies.

% talked with Brandon Redding (Hui's former postdoc, working in naval research lab, interested in multimode SBS).

Finally, our general approach is applicable to model other non-linear effects for which wavefront shaping and modal control can be applied to affect their manifestation.  Already we have applied this theoretical approach to the study of transverse modal instability (TMI), and uncovered new physical effects due to the thermal origin of the instability \cite{chen2022suppressing}.  In this case multimode excitation is predicted to be even more effective in suppressing the instability.  A similar approach seems possible for controlling the effect of Kerr nonlinearity in multimode fibers \cite{liu2016kerr}. In our view, wavefront shaping in multimode fibers has the potential to become a standard tool to control nonlinear effects in fibers, and possibly in waveguides, of great practical utility.

\begin{acknowledgments}
This work is supported by the Air Force Office of Scientific Research (AFOSR) under Grant FA9550-20-1-0129. Stephen Warren-Smith is supported by an Australian Research Council (ARC) Future Fellowship (FT200100154). We thank Heike Ebendorff-Heidepriem, David J. Ottaway, Ori Henderson-Sapir, Shuen Wei at University of Adelaide, Linh V. Nguyen at University of South Australia, and Peter Rakich and Owen D.  Miller at Yale university for helpful technical discussions. We also thank Yale University and University of Adelaide for providing computational resources.
\end{acknowledgments}

\begin{appendices}
\appendix
\section{Elastically Isotropic Fibers} \label{appendixA}

\subsection{Acousto-Optic Interaction}
\renewcommand\thefigure{A.\arabic{figure}}    
\setcounter{figure}{0} 
\begin{figure*}[t!]
		\centering\includegraphics[width=\textwidth]{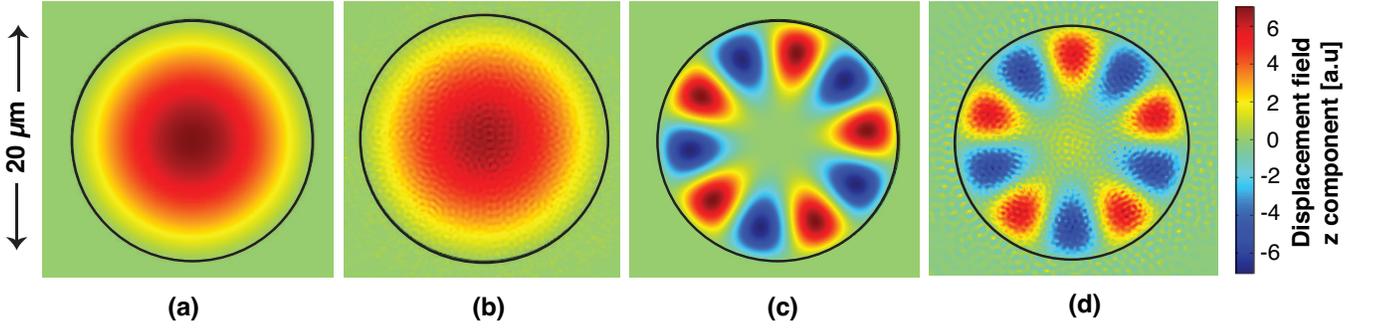}
		\caption{Mode profiles for the fundamental (a,b) and a higher order (c,d) longitudinal acoustic acoustic modes, calculated without shear terms (a,c) and with shear-longitudinal coupling (b,d), for a Ge doped circular step index silica fiber. This fiber is elastically isotropic. When shear is neglected, each acoustic mode is characterized by two indices $(i,j)$ and it varies in the radial direction as $i^{\rm th}$ order Bessel function (of the first kind) with $j-1$ zeros in the core, and as a cosine or a sine with $i$ nodal lines in the azimuthal direction. Inclusion of shear-longitudinal coupling leads to a rapidly varying (small feature size) perturbation in addition to the dominant Bessel-like behaviour. }
        \label{Fig:acmodes}
\end{figure*}
In this work we have described Acousto-optic interaction with a photoelastic tensor. We have defined a scaled version of usual photoelastic tensor, $\stackrel{\leftrightarrow}{\pi}=\epsilon_0 \epsilon_r^2\stackrel{\leftrightarrow}{p}$, where $\stackrel{\leftrightarrow}{p}$ is the usual fourth rank photoelastic tensor. As mentioned in \ref{sec:III}.B, for elastically isotropic fibers the full photoelastic tensor can be described by just two independent constants~\cite{nelson1971theory,feldman1975relations,biegelsen1974photoelastic}, resulting in dramatic simplification in the form of the overlap integral (Eq.\ref{Eq:overlap-simple}). In this section, we review the tensor theory in an elastically isotropic medium and provide a detailed derivation of the relevant overlap integral. The acoustic modes which appear in this  integral consist of a longitudinal and a shear term; we carefully evaluate the contribution to the integral of the shear term, finding it to be quite small, and hence we omit it for the calculations in the main text.

The form of $\stackrel{\leftrightarrow}{\pi}$ for an isotropic medium, can be written in the index notation as follows:
\begin{equation}
    \pi_{ijkl}=\pi_{1122}\delta_{ij}\delta_{kl}+\pi_{1212}(\delta_{ik}\delta_{jl}+\delta_{il}\delta_{jk}), 
    \label{Eq:pi-tensor}
\end{equation}
where $\pi_{1122}$ and $\pi_{1212}$ are the two independent parameters describing the entire photoelastic tensor, and $\delta_{ij}$ is the Kronecker delta function for indices $i$ and $j$. A useful way to visualize $\stackrel{\leftrightarrow}{\pi}$ in this case is by employing the Voigt notation~\cite{voight1928lehrbuch,slezak2020tensor}. We introduce a set of six new labels for the three diagonal and the three independent off-diagonal elements:
$\{1,2,3,4,5,6\}\equiv \{11,22,33,12,13,23\}$. With such notation, the fourth-rank tensor $\stackrel{\leftrightarrow}{\pi}$ can be written as the following $6\times6$ matrix~\cite{slezak2020tensor}:
\begin{equation}
\stackrel{\leftrightarrow}{\Pi}=
\begin{bmatrix}
\pi_{12}+2\pi_{44} & \pi_{12} & \pi_{12} & 0 & 0 & 0 \\
\pi_{12} & \pi_{12}+2\pi_{44} &  \pi_{12} & 0 & 0 & 0 \\
\pi_{12} &  \pi_{12} & \pi_{12}+2\pi_{44}  & 0 & 0 & 0 \\
0 & 0 & 0 & \pi_{44} & 0 & 0 \\
0 & 0 & 0 & 0 & \pi_{44} &  0 \\
0 & 0 & 0 &  0 & 0 & \pi_{44}.
\end{bmatrix}
\label{Eq:pi-voight}
\end{equation}
The $\pi_{12}$ component ($\pi_{1122}$ in original notation) is directly related to the electrostriction constant, $\gamma_e$. The formula for
$\stackrel{\leftrightarrow}{\pi}$ in Eq.~\ref{Eq:pi-tensor} can be substituted in Eq.~\ref{Eq:overlap} to simplify the overlap integral for the elastically isotropic materials. We contract $\stackrel{\leftrightarrow}{\pi}$ and  $\vec{f}^{(1)}_l\otimes \vec{f}^{(2)^*}_m$ to obtain: 
\begin{equation}
\begin{aligned}
O_{mlk}=  &\pi_{12}\langle\vec{\nabla}(\vec{f}^{(1)}_l\cdot \vec{f}^{(2)^*}_m)\cdot \vec{u}^{\:ml*}_{k} \rangle+ \\ & 2\pi_{44} \langle\vec{\nabla}\cdot(\vec{f}^{(1)}_l\otimes \vec{f}^{(2)^*}_m)\cdot \vec{u}^{\:ml*}_{k} \rangle.
\end{aligned}
\end{equation}
\noindent
The overlap integral contains two terms, one for each of the two independent components of $\stackrel{\leftrightarrow}{\pi}$.  Here, $\vec{\nabla}$ is equal to $\vec{\nabla}_T-iq_{mn}\hat{z}$. We can further simplify the overlap integral by using integration by parts:

\begin{equation}
\begin{aligned}
O_{mlk}= & \pi_{12}\langle\vec{f}^{(1)}_l\cdot \vec{f}^{(2)^*}_m\vec{\nabla}\cdot \vec{u}^{\:ml*}_{k}\rangle\\ &+  2\pi_{44} \langle(\vec{\nabla}\cdot\vec{f}^{(1)}_l)\vec{f}^{(2)^*}_m\cdot \vec{u}^{\:ml*}_{k} \rangle\\ & +2\pi_{44} \langle(\vec{f}^{(1)}_l\cdot\vec{\nabla})\vec{f}^{(2)^*}_m\cdot \vec{u}^{\:ml*}_{k} \rangle.
\label{Eq:overlap-isotropic}
\end{aligned}
\end{equation}
\noindent
The first term in the overlap integral (we call it the direct interaction term) is proportional to $\pi_{12}$ and the dot product between the Stokes and the signal mode profiles multiplied with the divergence of the displacement field profile. The other two terms (we call them cross interaction terms) consist of dot products between the optical mode and displacement field profiles multiplied with the derivative of the remaining optical mode profile. Typically, the direct interaction term is significantly larger than the cross interaction terms. This is due to the predominantly transverse nature of the optical modes~\cite{okamoto2021fundamentals,snyder1978modes} and longitudinal nature of the acoustic modes~\cite{dong2010formulation,waldron1969some,auld1973acoustic}, which leads to an extremely small dot product between optical and acoustic mode profiles. Therefore, we can ignore the cross interaction terms leading to:

\begin{equation}
O_{mlk} \approx  \pi_{12}\langle\vec{f}^{(1)}_l\cdot \vec{f}^{(2)^*}_m\vec{\nabla}\cdot \vec{u}^{\:ml*}_{k}\rangle.
\label{Eq:overlap-simple-A}
\end{equation}
\noindent
 
It should be noted that for specialty optical fibers or sufficiently higher order modes the primarily longitudinal and transverse character of acoustic and optical modes can break down leading to non-trivial contribution from the cross interactions, in which case Eq.~\ref{Eq:overlap-isotropic} should be used for accurate calculations. 

\subsection{Acoustic Modal Equation}

In the previous subsection, we showed that if the fiber is elastically isotropic the overlap integrals describing the acousto-optic interaction simplify dramatically due to the form of the photoelastic tensor. Similarly, elastic isotropy can be used to substantially  simplify the acoustic modal equation, Eq.~\ref{Eq:umode}. For isotropic materials, the elasticity tensor, $\stackrel{\leftrightarrow}{C}$, can be described by just two independent constants and takes the following form~\cite{auld1973acoustic,slaughter2012linearized}:
\begin{equation}
    C_{ijkl}=\lambda\delta_{ij}\delta_{kl}+\mu (\delta_{ik}\delta_{jl}+\delta_{il}\delta_{jk}), 
    \label{Eq:C-tensor}
\end{equation}
where $\lambda$ and $\mu$ are the well known Lam\'e parameters~\cite{slaughter2012linearized,auld1973acoustic} and are related to the longitudinal velocity $v_{\tiny L}$ and the shear velocity  $v_{\tiny s} $ for the acoustic waves; $\lambda=v^2_{\tiny L} \rho_0$ and $\mu=v^2_{\tiny s} \rho_0$, Where $\rho_0$ is the average density of the material.  $\delta_{ij}$ is Kronecker delta function for indices $i$ and $j$. This form of the elasticity tensor, $\stackrel{\leftrightarrow}{C}$, can be directly substituted in Eq.~\ref{Eq:umode} to obtain a simplified acoustic modal equation for isotropic fibers~\cite{dragic2010accurate,dong2010formulation,waldron1969some,auld1973acoustic,mccurdy2005modeling}:

\begin{equation}
  v_L^2 \vec{\nabla}(\vec{\nabla}.\vec{u}^{\:ij}_{k})+v_s^2 \vec{\nabla} \times \vec{\nabla} \times \vec{u}^{\:ij}_{k} + \Omega^2_{ijk} \vec{u}^{\:ij}_{k}=0.
  \label{Eq:isotropicaceq}
\end{equation}

Here, the first term is given by the gradient of the divergence of the displacement field and is proportional to the longitudinal acoustic velocity squared. This term is related directly to the density fluctuations, $\delta \rho = \rho_0\vec{\nabla}\cdot\vec{u}$, hence it maps onto the $\nabla_T^2$ term in the scalar acoustic wave equation~\cite{ke2014stimulated,agrawal2000nonlinear,boyd2020nonlinear,kobyakov2010stimulated}. The second term is given by the curl of curl of displacement field and captures the role of the shear forces, parameterized by the shear velocity, $v_s$. This term has no analog in the scalar acoustic wave equation. Typically in silica fibers the shear acoustic velocity is much smaller than the longitudinal acoustic velocity~\cite{dragic2009estimating,smith2016metamaterial}; this results in primarily longitudinal acoustic modes. However, the non-zero shear velocity does produce an observable effect even for these primarily longitudinal acoustic modes, because of the shear-longitudinal coupling due to the boundary conditions. We calculated the longitudinal acoustic modes for a circular step-index fiber, with germanium-doped silica core and pure silica cladding, with (Figs. \ref{Fig:acmodes}b and \ref{Fig:acmodes}d) and without the shear term (Figs. \ref{Fig:acmodes}a and \ref{Fig:acmodes}c). The details of the fiber parameters are given in Table~\ref{tab:my_label}. Without the shear term each acoustic mode is characterized by two indices $(i,j)$ and it varies in the radial direction as an $i^{\rm th}$ order Bessel function (of the first kind) with $j-1$ zeros in the core, and as a cosine or a sine function with $i$ nodal lines in the azimuthal direction. Including the shear--longitudinal coupling leads to a rapidly varying perturbation (small feature size) in addition to the dominant Bessel-like behaviour~\cite{dong2010formulation,waldron1969some}.  These fast variations are understood \cite{dong2010formulation, waldron1969some}, and are a result of higher shear propagation constants, due to lower shear velocity in the core compared to the longitudinal velocity. Additionally, the effect of the shear term is higher in the higher-order acoustic mode (Fig.~\ref{Fig:acmodes}d), compared to the fundamental acoustic mode (Fig.~\ref{Fig:acmodes}b). These variations can, in principle, substantially affect the calculation of the overlap integrals for the acousto-optic interaction, especially for materials with relatively high shear velocities. However, in practice we have found that for typical materials used in fibers the effect of fast variations due to the shear velocity term is averaged out in the overlap integrals and does not change the Brillouin gain spectra significantly (1--5\% error). For this reason, we conclude that using the scalar theory to evaluate the relevant acoustic modes is a useful and accurate approximation. In that case, to calculate the overlap integral in Eq.~\ref{Eq:overlap-simple-A}, we make the substitution $\frac{1}{q_{ml}}\vec{\nabla} \cdot \vec{u}^{\: ml}_{k}=\delta\rho^{\: ml}_{k}$,
where $\delta \rho^{\: lm}_{k}$ is $k^{\rm th}$ eigenmode (for optical mode pair $\{i,j\}$) of the scalar density fluctuation equation:
\begin{equation}
    [\nabla_T^2+(\frac{\Omega_{mlk}^2}{v_L^2}-q_{ml}^2)]{\delta \rho^{\: ml}_{k}}=0.
    \label{Eq:scalaraceq}
\end{equation}
Here, $\nabla_T^2$ is the transverse Laplacian, $\Omega_{mlk}$ is the modal eigenfrequency, and $q_{ml}$ is the acoustic propagation constant in the $z$ direction and is given by $q_{ml}=\beta_l+\gamma_m$.  The scalar acoustic eigenmodes are appropriately scaled so that they are normalized similar to the original vector acoustic modes. Equation~\ref{Eq:scalaraceq} is obtained immediately upon taking the divergence of Eq.~\ref{Eq:isotropicaceq}, which causes the curl of curl term to vanish. When the scalar acoustic modes are considered, the acousto-optic overlap integral is given by:
\begin{equation}
O_{mlk} \approx  \pi_{12}q_{ml}\langle\vec{f}^{(1)}_l\cdot \vec{f}^{(2)^*}_m \delta\rho^{ml}_{k}\rangle.
\label{Eq:overlap-ac-scalar-A}
\end{equation}
As stated in Sec.~\ref{sec:III}.B, this form of the overlap integral is accurate enough for fibers with relatively low shear velocity. It should be noted that ignoring the shear velocity contribution can lead to substantial errors especially if the shear velocity is comparable or even higher than the longitudinal velocity. In such a case, Eq.~\ref{Eq:overlap-simple-A} should be used for acoustic mode calculation to evaluate the overlap integrals.

\renewcommand\thefigure{B.\arabic{figure}}    
\setcounter{figure}{0} 
\begin{figure*}[t]
		\centering\includegraphics[width=0.9\textwidth]{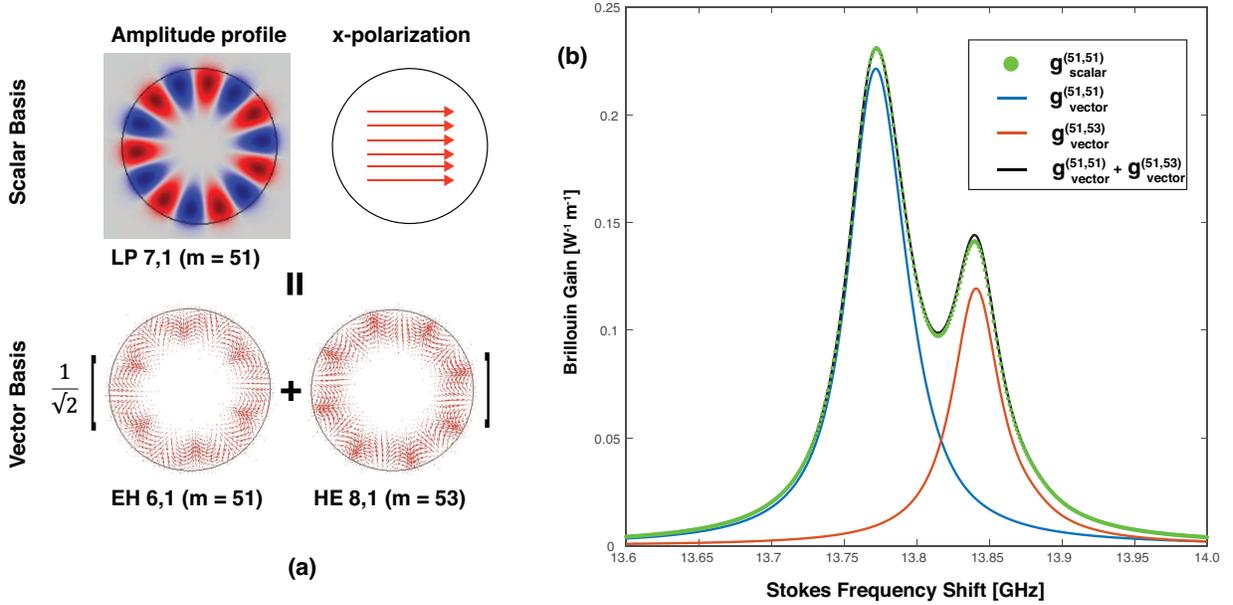}
		\caption{(a) An example of decomposition of scalar LP modes into exact fiber vector modes. An x-polarized LP 7,1 ($m=51$) mode is an equally weighted linear combination of two nearly degenerate vector modes EH 6,1 ($m=51$) and HE 8,1 ($m=53$). (b) Brillouin gain spectrum (BGS) for various mode combinations calculated using vector and scalar SBS formalism. It shows that intramodal BGS for scalar LP modes (here LP 7,1) is equal to the sum of intramodal and intermodal BGS for nearly degenerate vector modes (here EH 6,1 and HE 8,1) validating the consistency condition in Eq.~\ref{Eq:consistent}. }
        \label{Fig:vec-sca}
\end{figure*}

\section{Comparison to Scalar Theory} \label{appendixB}

In this section, we compare our phase-matched vector multimode SBS theory, simplified for elastically isotropic fibers, to the scalar multimode SBS theory presented in Ke. et al.~\cite{ke2014stimulated} (henceforth referred to as `the scalar theory').  We note that much of the work on SBS in SMFs and in other contexts also uses similar scalar approximations. Ke. et al were able to obtain the power growth equations for multimode Stokes growth similar to Eqs. \ref{Eq:powergrowth} and \ref{Eq:Brillouin}, which are capable of capturing important physics of SBS in multimode fibers. These authors did not discuss the efficacy of multimode excitation in suppressing SBS, nor did they apply it to calculate the full gain matrix for realistic MMFs, as we do in the current work. In the scalar theory, SBS growth equations were obtained by solving scalar optical and acoustic wave equations by expanding in terms of scalar linearly polarized (LP) fiber modes and eigenmodes of scalar density fluctuation equation respectively. This formulation is inadequate for non-isotropic fibers and waveguides, which can have non-negligible contributions from cross components of the photoelastic tensor. Even for isotropic fibers, there can be significant errors due to shear--longitudinal coupling in acoustic modes. However, most importantly, the use of uniformly polarized LP modes instead of exact vector fiber modes can lead to overestimation of intermodal gain due to neglect of space-dependent polarization variations in exact fiber modes, which can strongly affect the relevant overlap integrals.

There are special cases when the scalar theory is reasonably accurate. An important case is the SBS coupling between forward- (signal) and backward- (Stokes) propagating fundamental modes of the circular step-index fibers. This is by default the case in most SBS studies focusing on single-mode fibers. In this instance, because the vector fundamental mode has constant polarization in space, the scalar multiplication of the amplitudes is the same as the vector dot product between the mode profiles.  Another case when scalar theory is accurate is when the cross-section of the fiber has two well-defined polarization axes and therefore supports uniformly (in space) polarized modes, such as in the case of an elliptical or D-shaped cross-section (Fig.~\ref{Fig:optmodes}c,d). 

Interestingly, even circular step-index fibers can support uniformly polarized modes, commonly known as linearly polarized (LP) modes. However, these are only approximately the eigenmodes of the fiber, obtained for weakly guiding fibers. Exactly speaking, each LP mode (designated as LP$_{i,j}$, where $i$ is the azimuthal index and $j$ is the radial index) is a linear combination of two nearly degenerate vector modes [EH$_{i-1,n}$] and [HE$_{i+1,n}$]  for $i\geq1$ \cite{okamoto2021fundamentals},.
A representative example is shown in Fig.~\ref{Fig:vec-sca}(a). The x-polarized 
[LP 7,1] mode is equal to ${1}/{\sqrt{2}}$ times the sum of exact vector modes [EH 6,1] and [HE 8,1]. Because the EH and HE modes are not exactly degenerate there is a small difference, $\Delta \beta$, in their propagation constants.
Over short enough length scales, ($L \ll {2\pi}/{\Delta \beta}$), the effect of the difference in propagation constants is negligible, and the LP modes form a good basis. In this limit, the scalar theory utilizing the uniformly polarized LP modes should be reasonably accurate. Thus, the effective SBS gain calculated from the scalar and the vector theories is expected to match closely in this limit. This leads to a specific consistency condition on the BGS calculated by the two theories. If all the power is sent in an $x$-polarized LP mode with mode number $m$ (say, $m=51$), the effective BGS in that mode according to the scalar theory is given by $g^{m,m}_{\rm scalar}(\Omega)$. Generically  $g^{(m,l)}_{\rm scalar}(\Omega)$ denotes the BGS for Stokes-signal mode pair ($m,l$) in the scalar theory. In the vector basis, exciting the $x$-polarized LP mode with mode number $m$ is equivalent to exciting two vector modes with mode number $m$ and $m'$ with half the power in each mode. Here, $m'$ is the nearly degenerate partner of mode $m$.(for $m=51, m'=53$). Therefore, the effective BGS is given by $0.5[g^{(m,m)}_{\rm vector}(\Omega) + g^{(m,m')}_{\rm vector}(\Omega)]$. In addition, since in this limit, the vector modes are effectively degenerate, there are additional ``non-trivially phase-matched terms" (see Section~\ref{sec:III}) equal in number to the trivially phase-matched terms, which appear on the off-diagonals of the SBS coupling matrix (Eq.~\ref{Eq:coupmat}). This causes the maximum eigenvalue of the matrix to increase by a factor of two and minimum eigenvalue to go to zero, with the trace preserved. This is the well-known effect of eigenvalue repulsion in Hermitian matrices due to the off-diagonal elements. Thus the effective SBS gain in the vector theory is $g^{(m,m)}_{\rm vector} + g^{(m,m')}_{\rm vector}(\Omega)$. Hence the consistency requires:
\begin{equation}
    g^{(m,m)}_{\rm scalar}(\Omega) = g^{(m,m)}_{\rm vector}(\Omega) + g^{(m,m')}_{\rm vector}(\Omega)
    \label{Eq:consistent}
\end{equation}
We verify the validity of this relation by explicitly calculating the BGS using both our  vector formalism and the scalar theory. As an example, we have shown the results for $m=51$ (which gives $m'=53$) in Fig.~\ref{Fig:vec-sca}b. The individual BGS describing the self and cross interaction between $m$ and $m'$ calculated using the vector theory are shown in blue and red color respectively. The sum of these curves is given by the black curve. The green dots represent the BGS calculation using the scalar theory which closely matches the sum of the BGS from vector theory (black curve), verifying the relation in Eq.~\ref{Eq:consistent}. 

Note that since scalar theory implicitly assumes perfect degeneracy between the exact vector fiber modes, it assumes phase matching between some of the mode pairs which becomes invalid when degeneracy is sufficiently lifted. This can be due to a long-enough length of the fiber or due to the presence of disorder, leading to a suppression of the off-diagonal terms. Therefore, the scalar theory will typically overestimate the SBS gain, especially for intermodal SBS couplings. There is evidence for this result in Table (I) presented by Ke et al., where the values of the SBS coupling calculated from the scalar theory are compared with the experimental values provided in ref.~\cite{song2013characterization}. Although the intramodal gain values match closely with the experiments, the predicted values of intermodal gain are consistently higher than the experimental values. In such a case, the vector formalism presented in this work, with properly formulated phase-matching conditions, should lead to more accurate calculations.
\vspace{ 3mm}

\section{Derivation Steps for Coupled Mode Equations} \label{appendixC}

In this section, we present some useful intermediate steps in the derivation omitted in the main text of the paper. Once the solution to the acoustic amplitudes is obtained (Eq.~\ref{Eq:ucoeff}) in terms of optical amplitudes, it can be used to obtain the source terms in Eq.~\ref{Eq:Efield}:

\begin{equation}
\begin{aligned}
    {\rm RHS}&=\mu_0 \frac{\partial^2}{\partial t^2}\sum_{i,j}\sum_{k}  \frac{1}{2\rho_0}\frac{O_1}{\Omega^2_{ijk}-\Omega^2+ i\Omega\Gamma_{ijk}}A_i B^{*}_j \\
    & {\stackrel{\leftrightarrow}{\pi}}:\nabla \otimes \vec{u}^{ij}_{k} e^{i\Omega t}e^{i(\beta_i -\gamma_j)z} \cdot \vec{E}
\end{aligned}
\label{Eq:source}
\end{equation}

Next we substitute the ansatz for $\vec{E}$ from Eq.~\ref{Eq:Eansatz} and evaluate the LHS and RHS of Eq.~\ref{Eq:Efield}:
\begin{equation}
    {\rm LHS} = \sum_{m} 2\gamma_m \vec{f}^{(2)}_m \frac{d B_m}{dz} e^{i(\omega_2 t +\gamma_m z)} + \omega_1 \; \rm {terms}.
\end{equation}
Here we have used the modal equations (Eq.~\ref{Eq:Emode}) to simplify the LHS. Also, used slowly varying envelope approximation to neglect terms with second order z derivative. The terms with time derivative of amplitudes were set to zero to study the steady-state condition. The RHS now becomes:
\begin{equation}
\begin{aligned}
    {\rm RHS}&=\mu_0 \omega_2^2 \sum_{l}\sum_{i,j}\sum_{k}  \frac{1}{2\rho_0}\frac{O_1}{\Omega^2_{ijk}-\Omega^2+ i\Omega\Gamma_{ijk}}A_i B^{*}_j A_l  \\
    & {\stackrel{\leftrightarrow}{\pi}}:\nabla \otimes \vec{u}^{ij}_{k} \cdot \vec{f}^{(1)}_l  e^{i\omega_2 t}e^{i(\beta_i -\gamma_j+\beta_l)z} + \omega_1 \: \rm {terms}.
\end{aligned}
\label{Eq:source1}
\end{equation}

Finally, we match terms oscillating at the same frequencies on both LHS and RHS and isolate terms for a particular mode $m$ by taking the dot product with $\vec{f}^{(2)}_m$. Upon taking the integral over the fiber cross-section, we use the orthogonality of the fiber modes to separate out the terms of the LHS. This leads to the coupled mode equations given in Eq.~\ref{Eq:CAEs1} and Eq.~\ref{Eq:CAEs} with coupling coefficients in Eq.~\ref{Eq:coupcoeff}.

\end{appendices}

\bibliography{myref}

%apsrev4-2.bst 2019-01-14 (MD) hand-edited version of apsrev4-1.bst
%Control: key (0)
%Control: author (8) initials jnrlst
%Control: editor formatted (1) identically to author
%Control: production of article title (0) allowed
%Control: page (0) single
%Control: year (1) truncated
%Control: production of eprint (0) enabled
\begin{thebibliography}{97}%
\makeatletter
\providecommand \@ifxundefined [1]{%
 \@ifx{#1\undefined}
}%
\providecommand \@ifnum [1]{%
 \ifnum #1\expandafter \@firstoftwo
 \else \expandafter \@secondoftwo
 \fi
}%
\providecommand \@ifx [1]{%
 \ifx #1\expandafter \@firstoftwo
 \else \expandafter \@secondoftwo
 \fi
}%
\providecommand \natexlab [1]{#1}%
\providecommand \enquote  [1]{``#1''}%
\providecommand \bibnamefont  [1]{#1}%
\providecommand \bibfnamefont [1]{#1}%
\providecommand \citenamefont [1]{#1}%
\providecommand \href@noop [0]{\@secondoftwo}%
\providecommand \href [0]{\begingroup \@sanitize@url \@href}%
\providecommand \@href[1]{\@@startlink{#1}\@@href}%
\providecommand \@@href[1]{\endgroup#1\@@endlink}%
\providecommand \@sanitize@url [0]{\catcode `\\12\catcode `\$12\catcode
  `\&12\catcode `\#12\catcode `\^12\catcode `\_12\catcode `\%12\relax}%
\providecommand \@@startlink[1]{}%
\providecommand \@@endlink[0]{}%
\providecommand \url  [0]{\begingroup\@sanitize@url \@url }%
\providecommand \@url [1]{\endgroup\@href {#1}{\urlprefix }}%
\providecommand \urlprefix  [0]{URL }%
\providecommand \Eprint [0]{\href }%
\providecommand \doibase [0]{https://doi.org/}%
\providecommand \selectlanguage [0]{\@gobble}%
\providecommand \bibinfo  [0]{\@secondoftwo}%
\providecommand \bibfield  [0]{\@secondoftwo}%
\providecommand \translation [1]{[#1]}%
\providecommand \BibitemOpen [0]{}%
\providecommand \bibitemStop [0]{}%
\providecommand \bibitemNoStop [0]{.\EOS\space}%
\providecommand \EOS [0]{\spacefactor3000\relax}%
\providecommand \BibitemShut  [1]{\csname bibitem#1\endcsname}%
\let\auto@bib@innerbib\@empty
%</preamble>
\bibitem [{\citenamefont {Boyd}(2020)}]{boyd2020nonlinear}%
  \BibitemOpen
  \bibfield  {author} {\bibinfo {author} {\bibfnamefont {R.~W.}\ \bibnamefont
  {Boyd}},\ }\href@noop {} {\emph {\bibinfo {title} {Nonlinear optics}}}\
  (\bibinfo  {publisher} {Academic press},\ \bibinfo {year} {2020})\BibitemShut
  {NoStop}%
\bibitem [{\citenamefont {Agrawal}(2000)}]{agrawal2000nonlinear}%
  \BibitemOpen
  \bibfield  {author} {\bibinfo {author} {\bibfnamefont {G.~P.}\ \bibnamefont
  {Agrawal}},\ }\bibfield  {title} {\bibinfo {title} {Nonlinear fiber optics},\
  }in\ \href@noop {} {\emph {\bibinfo {booktitle} {Nonlinear Science at the
  Dawn of the 21st Century}}}\ (\bibinfo  {publisher} {Springer},\ \bibinfo
  {year} {2000})\ pp.\ \bibinfo {pages} {195--211}\BibitemShut {NoStop}%
\bibitem [{\citenamefont {Kobyakov}\ \emph {et~al.}(2010)\citenamefont
  {Kobyakov}, \citenamefont {Sauer},\ and\ \citenamefont
  {Chowdhury}}]{kobyakov2010stimulated}%
  \BibitemOpen
  \bibfield  {author} {\bibinfo {author} {\bibfnamefont {A.}~\bibnamefont
  {Kobyakov}}, \bibinfo {author} {\bibfnamefont {M.}~\bibnamefont {Sauer}},\
  and\ \bibinfo {author} {\bibfnamefont {D.}~\bibnamefont {Chowdhury}},\
  }\bibfield  {title} {\bibinfo {title} {Stimulated brillouin scattering in
  optical fibers},\ }\href@noop {} {\bibfield  {journal} {\bibinfo  {journal}
  {Advances in optics and photonics}\ }\textbf {\bibinfo {volume} {2}},\
  \bibinfo {pages} {1} (\bibinfo {year} {2010})}\BibitemShut {NoStop}%
\bibitem [{\citenamefont {Bai}\ \emph {et~al.}(2018)\citenamefont {Bai},
  \citenamefont {Yuan}, \citenamefont {Liu}, \citenamefont {Xu}, \citenamefont
  {Gao}, \citenamefont {Williams}, \citenamefont {Kitzler}, \citenamefont
  {Mildren}, \citenamefont {Wang},\ and\ \citenamefont
  {Lu}}]{bai2018stimulated}%
  \BibitemOpen
  \bibfield  {author} {\bibinfo {author} {\bibfnamefont {Z.}~\bibnamefont
  {Bai}}, \bibinfo {author} {\bibfnamefont {H.}~\bibnamefont {Yuan}}, \bibinfo
  {author} {\bibfnamefont {Z.}~\bibnamefont {Liu}}, \bibinfo {author}
  {\bibfnamefont {P.}~\bibnamefont {Xu}}, \bibinfo {author} {\bibfnamefont
  {Q.}~\bibnamefont {Gao}}, \bibinfo {author} {\bibfnamefont {R.~J.}\
  \bibnamefont {Williams}}, \bibinfo {author} {\bibfnamefont {O.}~\bibnamefont
  {Kitzler}}, \bibinfo {author} {\bibfnamefont {R.~P.}\ \bibnamefont
  {Mildren}}, \bibinfo {author} {\bibfnamefont {Y.}~\bibnamefont {Wang}},\ and\
  \bibinfo {author} {\bibfnamefont {Z.}~\bibnamefont {Lu}},\ }\bibfield
  {title} {\bibinfo {title} {Stimulated brillouin scattering materials,
  experimental design and applications: A review},\ }\href@noop {} {\bibfield
  {journal} {\bibinfo  {journal} {Optical Materials}\ }\textbf {\bibinfo
  {volume} {75}},\ \bibinfo {pages} {626} (\bibinfo {year} {2018})}\BibitemShut
  {NoStop}%
\bibitem [{\citenamefont {Wolff}\ \emph {et~al.}(2021)\citenamefont {Wolff},
  \citenamefont {Smith}, \citenamefont {Stiller},\ and\ \citenamefont
  {Poulton}}]{wolff2021brillouin}%
  \BibitemOpen
  \bibfield  {author} {\bibinfo {author} {\bibfnamefont {C.}~\bibnamefont
  {Wolff}}, \bibinfo {author} {\bibfnamefont {M.}~\bibnamefont {Smith}},
  \bibinfo {author} {\bibfnamefont {B.}~\bibnamefont {Stiller}},\ and\ \bibinfo
  {author} {\bibfnamefont {C.}~\bibnamefont {Poulton}},\ }\bibfield  {title}
  {\bibinfo {title} {Brillouin scattering—theory and experiment: tutorial},\
  }\href@noop {} {\bibfield  {journal} {\bibinfo  {journal} {JOSA B}\ }\textbf
  {\bibinfo {volume} {38}},\ \bibinfo {pages} {1243} (\bibinfo {year}
  {2021})}\BibitemShut {NoStop}%
\bibitem [{\citenamefont {Brillouin}(1922)}]{brillouin1922diffusion}%
  \BibitemOpen
  \bibfield  {author} {\bibinfo {author} {\bibfnamefont {L.}~\bibnamefont
  {Brillouin}},\ }\bibfield  {title} {\bibinfo {title} {Diffusion of light and
  x-rays by a homogeneous transparent body},\ }in\ \href@noop {} {\emph
  {\bibinfo {booktitle} {Annals of Physics}}},\ Vol.~\bibinfo {volume} {9}\
  (\bibinfo {year} {1922})\ pp.\ \bibinfo {pages} {88--122}\BibitemShut
  {NoStop}%
\bibitem [{\citenamefont {Chiao}\ \emph {et~al.}(1964)\citenamefont {Chiao},
  \citenamefont {Townes},\ and\ \citenamefont
  {Stoicheff}}]{chiao1964stimulated}%
  \BibitemOpen
  \bibfield  {author} {\bibinfo {author} {\bibfnamefont {R.}~\bibnamefont
  {Chiao}}, \bibinfo {author} {\bibfnamefont {C.~H.}\ \bibnamefont {Townes}},\
  and\ \bibinfo {author} {\bibfnamefont {B.}~\bibnamefont {Stoicheff}},\
  }\bibfield  {title} {\bibinfo {title} {Stimulated brillouin scattering and
  coherent generation of intense hypersonic waves},\ }\href@noop {} {\bibfield
  {journal} {\bibinfo  {journal} {Physical Review Letters}\ }\textbf {\bibinfo
  {volume} {12}},\ \bibinfo {pages} {592} (\bibinfo {year} {1964})}\BibitemShut
  {NoStop}%
\bibitem [{\citenamefont {Ippen}\ and\ \citenamefont
  {Stolen}(1972)}]{ippen1972stimulated}%
  \BibitemOpen
  \bibfield  {author} {\bibinfo {author} {\bibfnamefont {E.}~\bibnamefont
  {Ippen}}\ and\ \bibinfo {author} {\bibfnamefont {R.}~\bibnamefont {Stolen}},\
  }\bibfield  {title} {\bibinfo {title} {Stimulated brillouin scattering in
  optical fibers},\ }\href@noop {} {\bibfield  {journal} {\bibinfo  {journal}
  {Applied Physics Letters}\ }\textbf {\bibinfo {volume} {21}},\ \bibinfo
  {pages} {539} (\bibinfo {year} {1972})}\BibitemShut {NoStop}%
\bibitem [{\citenamefont {Th{\'e}venaz}(2008)}]{thevenaz2008slow}%
  \BibitemOpen
  \bibfield  {author} {\bibinfo {author} {\bibfnamefont {L.}~\bibnamefont
  {Th{\'e}venaz}},\ }\bibfield  {title} {\bibinfo {title} {Slow and fast light
  in optical fibres},\ }\href@noop {} {\bibfield  {journal} {\bibinfo
  {journal} {Nature photonics}\ }\textbf {\bibinfo {volume} {2}},\ \bibinfo
  {pages} {474} (\bibinfo {year} {2008})}\BibitemShut {NoStop}%
\bibitem [{\citenamefont {Kim}\ \emph {et~al.}(2015)\citenamefont {Kim},
  \citenamefont {Kuzyk}, \citenamefont {Han}, \citenamefont {Wang},\ and\
  \citenamefont {Bahl}}]{kim2015non}%
  \BibitemOpen
  \bibfield  {author} {\bibinfo {author} {\bibfnamefont {J.}~\bibnamefont
  {Kim}}, \bibinfo {author} {\bibfnamefont {M.~C.}\ \bibnamefont {Kuzyk}},
  \bibinfo {author} {\bibfnamefont {K.}~\bibnamefont {Han}}, \bibinfo {author}
  {\bibfnamefont {H.}~\bibnamefont {Wang}},\ and\ \bibinfo {author}
  {\bibfnamefont {G.}~\bibnamefont {Bahl}},\ }\bibfield  {title} {\bibinfo
  {title} {Non-reciprocal brillouin scattering induced transparency},\
  }\href@noop {} {\bibfield  {journal} {\bibinfo  {journal} {Nature Physics}\
  }\textbf {\bibinfo {volume} {11}},\ \bibinfo {pages} {275} (\bibinfo {year}
  {2015})}\BibitemShut {NoStop}%
\bibitem [{\citenamefont {Horiguchi}\ \emph {et~al.}(1995)\citenamefont
  {Horiguchi}, \citenamefont {Shimizu}, \citenamefont {Kurashima},
  \citenamefont {Tateda},\ and\ \citenamefont
  {Koyamada}}]{horiguchi1995development}%
  \BibitemOpen
  \bibfield  {author} {\bibinfo {author} {\bibfnamefont {T.}~\bibnamefont
  {Horiguchi}}, \bibinfo {author} {\bibfnamefont {K.}~\bibnamefont {Shimizu}},
  \bibinfo {author} {\bibfnamefont {T.}~\bibnamefont {Kurashima}}, \bibinfo
  {author} {\bibfnamefont {M.}~\bibnamefont {Tateda}},\ and\ \bibinfo {author}
  {\bibfnamefont {Y.}~\bibnamefont {Koyamada}},\ }\bibfield  {title} {\bibinfo
  {title} {Development of a distributed sensing technique using brillouin
  scattering},\ }\href@noop {} {\bibfield  {journal} {\bibinfo  {journal}
  {Journal of lightwave technology}\ }\textbf {\bibinfo {volume} {13}},\
  \bibinfo {pages} {1296} (\bibinfo {year} {1995})}\BibitemShut {NoStop}%
\bibitem [{\citenamefont {Ballmann}\ \emph {et~al.}(2015)\citenamefont
  {Ballmann}, \citenamefont {Thompson}, \citenamefont {Traverso}, \citenamefont
  {Meng}, \citenamefont {Scully},\ and\ \citenamefont
  {Yakovlev}}]{ballmann2015stimulated}%
  \BibitemOpen
  \bibfield  {author} {\bibinfo {author} {\bibfnamefont {C.~W.}\ \bibnamefont
  {Ballmann}}, \bibinfo {author} {\bibfnamefont {J.~V.}\ \bibnamefont
  {Thompson}}, \bibinfo {author} {\bibfnamefont {A.~J.}\ \bibnamefont
  {Traverso}}, \bibinfo {author} {\bibfnamefont {Z.}~\bibnamefont {Meng}},
  \bibinfo {author} {\bibfnamefont {M.~O.}\ \bibnamefont {Scully}},\ and\
  \bibinfo {author} {\bibfnamefont {V.~V.}\ \bibnamefont {Yakovlev}},\
  }\bibfield  {title} {\bibinfo {title} {Stimulated brillouin scattering
  microscopic imaging},\ }\href@noop {} {\bibfield  {journal} {\bibinfo
  {journal} {Scientific Reports}\ }\textbf {\bibinfo {volume} {5}},\ \bibinfo
  {pages} {18139} (\bibinfo {year} {2015})}\BibitemShut {NoStop}%
\bibitem [{\citenamefont {Rakich}\ \emph {et~al.}(2012)\citenamefont {Rakich},
  \citenamefont {Reinke}, \citenamefont {Camacho}, \citenamefont {Davids},\
  and\ \citenamefont {Wang}}]{rakich2012giant}%
  \BibitemOpen
  \bibfield  {author} {\bibinfo {author} {\bibfnamefont {P.~T.}\ \bibnamefont
  {Rakich}}, \bibinfo {author} {\bibfnamefont {C.}~\bibnamefont {Reinke}},
  \bibinfo {author} {\bibfnamefont {R.}~\bibnamefont {Camacho}}, \bibinfo
  {author} {\bibfnamefont {P.}~\bibnamefont {Davids}},\ and\ \bibinfo {author}
  {\bibfnamefont {Z.}~\bibnamefont {Wang}},\ }\bibfield  {title} {\bibinfo
  {title} {Giant enhancement of stimulated brillouin scattering in the
  subwavelength limit},\ }\href@noop {} {\bibfield  {journal} {\bibinfo
  {journal} {Physical Review X}\ }\textbf {\bibinfo {volume} {2}},\ \bibinfo
  {pages} {011008} (\bibinfo {year} {2012})}\BibitemShut {NoStop}%
\bibitem [{\citenamefont {Eggleton}\ \emph {et~al.}(2013)\citenamefont
  {Eggleton}, \citenamefont {Poulton},\ and\ \citenamefont
  {Pant}}]{eggleton2013inducing}%
  \BibitemOpen
  \bibfield  {author} {\bibinfo {author} {\bibfnamefont {B.~J.}\ \bibnamefont
  {Eggleton}}, \bibinfo {author} {\bibfnamefont {C.~G.}\ \bibnamefont
  {Poulton}},\ and\ \bibinfo {author} {\bibfnamefont {R.}~\bibnamefont
  {Pant}},\ }\bibfield  {title} {\bibinfo {title} {Inducing and harnessing
  stimulated brillouin scattering in photonic integrated circuits},\
  }\href@noop {} {\bibfield  {journal} {\bibinfo  {journal} {Advances in Optics
  and Photonics}\ }\textbf {\bibinfo {volume} {5}},\ \bibinfo {pages} {536}
  (\bibinfo {year} {2013})}\BibitemShut {NoStop}%
\bibitem [{\citenamefont {Shin}\ \emph {et~al.}(2013)\citenamefont {Shin},
  \citenamefont {Qiu}, \citenamefont {Jarecki}, \citenamefont {Cox},
  \citenamefont {Olsson~III}, \citenamefont {Starbuck}, \citenamefont {Wang},\
  and\ \citenamefont {Rakich}}]{shin2013tailorable}%
  \BibitemOpen
  \bibfield  {author} {\bibinfo {author} {\bibfnamefont {H.}~\bibnamefont
  {Shin}}, \bibinfo {author} {\bibfnamefont {W.}~\bibnamefont {Qiu}}, \bibinfo
  {author} {\bibfnamefont {R.}~\bibnamefont {Jarecki}}, \bibinfo {author}
  {\bibfnamefont {J.~A.}\ \bibnamefont {Cox}}, \bibinfo {author} {\bibfnamefont
  {R.~H.}\ \bibnamefont {Olsson~III}}, \bibinfo {author} {\bibfnamefont
  {A.}~\bibnamefont {Starbuck}}, \bibinfo {author} {\bibfnamefont
  {Z.}~\bibnamefont {Wang}},\ and\ \bibinfo {author} {\bibfnamefont {P.~T.}\
  \bibnamefont {Rakich}},\ }\bibfield  {title} {\bibinfo {title} {Tailorable
  stimulated brillouin scattering in nanoscale silicon waveguides},\
  }\href@noop {} {\bibfield  {journal} {\bibinfo  {journal} {Nature
  communications}\ }\textbf {\bibinfo {volume} {4}},\ \bibinfo {pages} {1944}
  (\bibinfo {year} {2013})}\BibitemShut {NoStop}%
\bibitem [{\citenamefont {Otterstrom}\ \emph {et~al.}(2018)\citenamefont
  {Otterstrom}, \citenamefont {Behunin}, \citenamefont {Kittlaus},
  \citenamefont {Wang},\ and\ \citenamefont {Rakich}}]{otterstrom2018silicon}%
  \BibitemOpen
  \bibfield  {author} {\bibinfo {author} {\bibfnamefont {N.~T.}\ \bibnamefont
  {Otterstrom}}, \bibinfo {author} {\bibfnamefont {R.~O.}\ \bibnamefont
  {Behunin}}, \bibinfo {author} {\bibfnamefont {E.~A.}\ \bibnamefont
  {Kittlaus}}, \bibinfo {author} {\bibfnamefont {Z.}~\bibnamefont {Wang}},\
  and\ \bibinfo {author} {\bibfnamefont {P.~T.}\ \bibnamefont {Rakich}},\
  }\bibfield  {title} {\bibinfo {title} {A silicon brillouin laser},\
  }\href@noop {} {\bibfield  {journal} {\bibinfo  {journal} {Science}\ }\textbf
  {\bibinfo {volume} {360}},\ \bibinfo {pages} {1113} (\bibinfo {year}
  {2018})}\BibitemShut {NoStop}%
\bibitem [{\citenamefont {Eggleton}\ \emph {et~al.}(2019)\citenamefont
  {Eggleton}, \citenamefont {Poulton}, \citenamefont {Rakich}, \citenamefont
  {Steel},\ and\ \citenamefont {Bahl}}]{eggleton2019brillouin}%
  \BibitemOpen
  \bibfield  {author} {\bibinfo {author} {\bibfnamefont {B.~J.}\ \bibnamefont
  {Eggleton}}, \bibinfo {author} {\bibfnamefont {C.~G.}\ \bibnamefont
  {Poulton}}, \bibinfo {author} {\bibfnamefont {P.~T.}\ \bibnamefont {Rakich}},
  \bibinfo {author} {\bibfnamefont {M.~J.}\ \bibnamefont {Steel}},\ and\
  \bibinfo {author} {\bibfnamefont {G.}~\bibnamefont {Bahl}},\ }\bibfield
  {title} {\bibinfo {title} {Brillouin integrated photonics},\ }\href@noop {}
  {\bibfield  {journal} {\bibinfo  {journal} {Nature Photonics}\ }\textbf
  {\bibinfo {volume} {13}},\ \bibinfo {pages} {664} (\bibinfo {year}
  {2019})}\BibitemShut {NoStop}%
\bibitem [{\citenamefont {Richardson}\ \emph {et~al.}(2010)\citenamefont
  {Richardson}, \citenamefont {Nilsson},\ and\ \citenamefont
  {Clarkson}}]{richardson2010high}%
  \BibitemOpen
  \bibfield  {author} {\bibinfo {author} {\bibfnamefont {D.~J.}\ \bibnamefont
  {Richardson}}, \bibinfo {author} {\bibfnamefont {J.}~\bibnamefont
  {Nilsson}},\ and\ \bibinfo {author} {\bibfnamefont {W.~A.}\ \bibnamefont
  {Clarkson}},\ }\bibfield  {title} {\bibinfo {title} {High power fiber lasers:
  current status and future perspectives},\ }\href@noop {} {\bibfield
  {journal} {\bibinfo  {journal} {JOSA B}\ }\textbf {\bibinfo {volume} {27}},\
  \bibinfo {pages} {B63} (\bibinfo {year} {2010})}\BibitemShut {NoStop}%
\bibitem [{\citenamefont {Zervas}\ and\ \citenamefont
  {Codemard}(2014)}]{zervas2014high}%
  \BibitemOpen
  \bibfield  {author} {\bibinfo {author} {\bibfnamefont {M.~N.}\ \bibnamefont
  {Zervas}}\ and\ \bibinfo {author} {\bibfnamefont {C.~A.}\ \bibnamefont
  {Codemard}},\ }\bibfield  {title} {\bibinfo {title} {High power fiber lasers:
  a review},\ }\href@noop {} {\bibfield  {journal} {\bibinfo  {journal} {IEEE
  Journal of selected topics in Quantum Electronics}\ }\textbf {\bibinfo
  {volume} {20}},\ \bibinfo {pages} {219} (\bibinfo {year} {2014})}\BibitemShut
  {NoStop}%
\bibitem [{\citenamefont {Fu}\ \emph {et~al.}(2017)\citenamefont {Fu},
  \citenamefont {Shi}, \citenamefont {Feng}, \citenamefont {Zhang},
  \citenamefont {Yang}, \citenamefont {Xu}, \citenamefont {Zhu}, \citenamefont
  {Norwood},\ and\ \citenamefont {Peyghambarian}}]{fu2017review}%
  \BibitemOpen
  \bibfield  {author} {\bibinfo {author} {\bibfnamefont {S.}~\bibnamefont
  {Fu}}, \bibinfo {author} {\bibfnamefont {W.}~\bibnamefont {Shi}}, \bibinfo
  {author} {\bibfnamefont {Y.}~\bibnamefont {Feng}}, \bibinfo {author}
  {\bibfnamefont {L.}~\bibnamefont {Zhang}}, \bibinfo {author} {\bibfnamefont
  {Z.}~\bibnamefont {Yang}}, \bibinfo {author} {\bibfnamefont {S.}~\bibnamefont
  {Xu}}, \bibinfo {author} {\bibfnamefont {X.}~\bibnamefont {Zhu}}, \bibinfo
  {author} {\bibfnamefont {R.}~\bibnamefont {Norwood}},\ and\ \bibinfo {author}
  {\bibfnamefont {N.}~\bibnamefont {Peyghambarian}},\ }\bibfield  {title}
  {\bibinfo {title} {Review of recent progress on single-frequency fiber
  lasers},\ }\href@noop {} {\bibfield  {journal} {\bibinfo  {journal} {JOSA B}\
  }\textbf {\bibinfo {volume} {34}},\ \bibinfo {pages} {A49} (\bibinfo {year}
  {2017})}\BibitemShut {NoStop}%
\bibitem [{\citenamefont {Pannell}\ \emph {et~al.}(1993)\citenamefont
  {Pannell}, \citenamefont {Russell},\ and\ \citenamefont
  {Newson}}]{pannell1993stimulated}%
  \BibitemOpen
  \bibfield  {author} {\bibinfo {author} {\bibfnamefont {C.}~\bibnamefont
  {Pannell}}, \bibinfo {author} {\bibfnamefont {P.~S.~J.}\ \bibnamefont
  {Russell}},\ and\ \bibinfo {author} {\bibfnamefont {T.}~\bibnamefont
  {Newson}},\ }\bibfield  {title} {\bibinfo {title} {Stimulated brillouin
  scattering in optical fibers: the effects of optical amplification},\
  }\href@noop {} {\bibfield  {journal} {\bibinfo  {journal} {JOSA B}\ }\textbf
  {\bibinfo {volume} {10}},\ \bibinfo {pages} {684} (\bibinfo {year}
  {1993})}\BibitemShut {NoStop}%
\bibitem [{\citenamefont {Panbhiharwala}\ \emph {et~al.}(2018)\citenamefont
  {Panbhiharwala}, \citenamefont {Harish}, \citenamefont {Venkitesh},
  \citenamefont {Nilsson},\ and\ \citenamefont
  {Srinivasan}}]{panbhiharwala2018investigation}%
  \BibitemOpen
  \bibfield  {author} {\bibinfo {author} {\bibfnamefont {Y.}~\bibnamefont
  {Panbhiharwala}}, \bibinfo {author} {\bibfnamefont {A.~V.}\ \bibnamefont
  {Harish}}, \bibinfo {author} {\bibfnamefont {D.}~\bibnamefont {Venkitesh}},
  \bibinfo {author} {\bibfnamefont {J.}~\bibnamefont {Nilsson}},\ and\ \bibinfo
  {author} {\bibfnamefont {B.}~\bibnamefont {Srinivasan}},\ }\bibfield  {title}
  {\bibinfo {title} {Investigation of temporal dynamics due to stimulated
  brillouin scattering using statistical correlation in a narrow-linewidth cw
  high power fiber amplifier},\ }\href@noop {} {\bibfield  {journal} {\bibinfo
  {journal} {Optics express}\ }\textbf {\bibinfo {volume} {26}},\ \bibinfo
  {pages} {33409} (\bibinfo {year} {2018})}\BibitemShut {NoStop}%
\bibitem [{\citenamefont {Wisal}\ \emph {et~al.}(2022)\citenamefont {Wisal},
  \citenamefont {Warren-Smith}, \citenamefont {Chen}, \citenamefont {Behunin},
  \citenamefont {Cao},\ and\ \citenamefont {Stone}}]{wisal2022generalized}%
  \BibitemOpen
  \bibfield  {author} {\bibinfo {author} {\bibfnamefont {K.}~\bibnamefont
  {Wisal}}, \bibinfo {author} {\bibfnamefont {S.~C.}\ \bibnamefont
  {Warren-Smith}}, \bibinfo {author} {\bibfnamefont {C.-W.}\ \bibnamefont
  {Chen}}, \bibinfo {author} {\bibfnamefont {R.}~\bibnamefont {Behunin}},
  \bibinfo {author} {\bibfnamefont {H.}~\bibnamefont {Cao}},\ and\ \bibinfo
  {author} {\bibfnamefont {A.~D.}\ \bibnamefont {Stone}},\ }\bibfield  {title}
  {\bibinfo {title} {Generalized theory of sbs in multimode fiber amplifiers},\
  }in\ \href@noop {} {\emph {\bibinfo {booktitle} {Physics and Simulation of
  Optoelectronic Devices XXX}}}\ (\bibinfo {organization} {SPIE},\ \bibinfo
  {year} {2022})\ p.\ \bibinfo {pages} {PC1199504}\BibitemShut {NoStop}%
\bibitem [{\citenamefont {Chen}\ \emph
  {et~al.}(2022{\natexlab{a}})\citenamefont {Chen}, \citenamefont {Wisal},
  \citenamefont {Ahmadi}, \citenamefont {Stone},\ and\ \citenamefont
  {Cao}}]{chen2022suppressingSBS}%
  \BibitemOpen
  \bibfield  {author} {\bibinfo {author} {\bibfnamefont {C.-W.}\ \bibnamefont
  {Chen}}, \bibinfo {author} {\bibfnamefont {K.}~\bibnamefont {Wisal}},
  \bibinfo {author} {\bibfnamefont {P.}~\bibnamefont {Ahmadi}}, \bibinfo
  {author} {\bibfnamefont {A.~D.}\ \bibnamefont {Stone}},\ and\ \bibinfo
  {author} {\bibfnamefont {H.}~\bibnamefont {Cao}},\ }\bibfield  {title}
  {\bibinfo {title} {Suppressing stimulated brillouin scattering by selective
  mode excitation in multimode fibers},\ }in\ \href@noop {} {\emph {\bibinfo
  {booktitle} {CLEO: QELS\_Fundamental Science}}}\ (\bibinfo {organization}
  {Optica Publishing Group},\ \bibinfo {year} {2022})\ pp.\ \bibinfo {pages}
  {FF2L--3}\BibitemShut {NoStop}%
\bibitem [{\citenamefont {Chen}\ \emph {et~al.}(2023)\citenamefont {Chen},
  \citenamefont {Nguyen}, \citenamefont {Wisal}, \citenamefont {Wei},
  \citenamefont {Warren-Smith}, \citenamefont {Henderson-Sapir}, \citenamefont
  {Ahmadi}, \citenamefont {Ottaway}, , \citenamefont {Stone}, \citenamefont
  {Ebendorff-Heidepriem},\ and\ \citenamefont {Cao}}]{SBSexpinPrep}%
  \BibitemOpen
  \bibfield  {author} {\bibinfo {author} {\bibfnamefont {C.-W.}\ \bibnamefont
  {Chen}}, \bibinfo {author} {\bibfnamefont {L.~V.}\ \bibnamefont {Nguyen}},
  \bibinfo {author} {\bibfnamefont {K.}~\bibnamefont {Wisal}}, \bibinfo
  {author} {\bibfnamefont {S.}~\bibnamefont {Wei}}, \bibinfo {author}
  {\bibfnamefont {S.}~\bibnamefont {Warren-Smith}}, \bibinfo {author}
  {\bibfnamefont {O.}~\bibnamefont {Henderson-Sapir}}, \bibinfo {author}
  {\bibfnamefont {P.}~\bibnamefont {Ahmadi}}, \bibinfo {author} {\bibfnamefont
  {D.~J.}\ \bibnamefont {Ottaway}}, , \bibinfo {author} {\bibfnamefont {A.~D.}\
  \bibnamefont {Stone}}, \bibinfo {author} {\bibfnamefont {H.}~\bibnamefont
  {Ebendorff-Heidepriem}},\ and\ \bibinfo {author} {\bibfnamefont
  {H.}~\bibnamefont {Cao}},\ }\bibfield  {title} {\bibinfo {title} {Suppression
  of stimulated brillouin scattering in optical fibers using wavefront
  shaping}} (\bibinfo {year} {2023}),\ \bibinfo {note} {\textit{in
  preparation}}\BibitemShut {NoStop}%
\bibitem [{\citenamefont {Ke}\ \emph {et~al.}(2014)\citenamefont {Ke},
  \citenamefont {Wang},\ and\ \citenamefont {Tang}}]{ke2014stimulated}%
  \BibitemOpen
  \bibfield  {author} {\bibinfo {author} {\bibfnamefont {W.-W.}\ \bibnamefont
  {Ke}}, \bibinfo {author} {\bibfnamefont {X.-J.}\ \bibnamefont {Wang}},\ and\
  \bibinfo {author} {\bibfnamefont {X.}~\bibnamefont {Tang}},\ }\bibfield
  {title} {\bibinfo {title} {Stimulated brillouin scattering model in
  multi-mode fiber lasers},\ }\href@noop {} {\bibfield  {journal} {\bibinfo
  {journal} {IEEE Journal of Selected Topics in Quantum Electronics}\ }\textbf
  {\bibinfo {volume} {20}},\ \bibinfo {pages} {305} (\bibinfo {year}
  {2014})}\BibitemShut {NoStop}%
\bibitem [{\citenamefont {Poulton}\ \emph {et~al.}(2013)\citenamefont
  {Poulton}, \citenamefont {Pant},\ and\ \citenamefont
  {Eggleton}}]{poulton2013acoustic}%
  \BibitemOpen
  \bibfield  {author} {\bibinfo {author} {\bibfnamefont {C.~G.}\ \bibnamefont
  {Poulton}}, \bibinfo {author} {\bibfnamefont {R.}~\bibnamefont {Pant}},\ and\
  \bibinfo {author} {\bibfnamefont {B.~J.}\ \bibnamefont {Eggleton}},\
  }\bibfield  {title} {\bibinfo {title} {Acoustic confinement and stimulated
  brillouin scattering in integrated optical waveguides},\ }\href@noop {}
  {\bibfield  {journal} {\bibinfo  {journal} {JOSA B}\ }\textbf {\bibinfo
  {volume} {30}},\ \bibinfo {pages} {2657} (\bibinfo {year}
  {2013})}\BibitemShut {NoStop}%
\bibitem [{\citenamefont {Dong}(2010)}]{dong2010formulation}%
  \BibitemOpen
  \bibfield  {author} {\bibinfo {author} {\bibfnamefont {L.}~\bibnamefont
  {Dong}},\ }\bibfield  {title} {\bibinfo {title} {Formulation of a complex
  mode solver for arbitrary circular acoustic waveguides},\ }\href@noop {}
  {\bibfield  {journal} {\bibinfo  {journal} {Journal of lightwave technology}\
  }\textbf {\bibinfo {volume} {28}},\ \bibinfo {pages} {3162} (\bibinfo {year}
  {2010})}\BibitemShut {NoStop}%
\bibitem [{\citenamefont {Supradeepa}(2013)}]{supradeepa2013stimulated}%
  \BibitemOpen
  \bibfield  {author} {\bibinfo {author} {\bibfnamefont {V.}~\bibnamefont
  {Supradeepa}},\ }\bibfield  {title} {\bibinfo {title} {Stimulated brillouin
  scattering thresholds in optical fibers for lasers linewidth broadened with
  noise},\ }\href@noop {} {\bibfield  {journal} {\bibinfo  {journal} {Optics
  express}\ }\textbf {\bibinfo {volume} {21}},\ \bibinfo {pages} {4677}
  (\bibinfo {year} {2013})}\BibitemShut {NoStop}%
\bibitem [{\citenamefont {Coles}\ \emph {et~al.}(2010)\citenamefont {Coles},
  \citenamefont {Kuo}, \citenamefont {Alic}, \citenamefont {Moro},
  \citenamefont {Bres}, \citenamefont {Boggio}, \citenamefont {Andrekson},
  \citenamefont {Karlsson},\ and\ \citenamefont {Radic}}]{coles2010bandwidth}%
  \BibitemOpen
  \bibfield  {author} {\bibinfo {author} {\bibfnamefont {J.~B.}\ \bibnamefont
  {Coles}}, \bibinfo {author} {\bibfnamefont {B.-P.}\ \bibnamefont {Kuo}},
  \bibinfo {author} {\bibfnamefont {N.}~\bibnamefont {Alic}}, \bibinfo {author}
  {\bibfnamefont {S.}~\bibnamefont {Moro}}, \bibinfo {author} {\bibfnamefont
  {C.-S.}\ \bibnamefont {Bres}}, \bibinfo {author} {\bibfnamefont {J.~C.}\
  \bibnamefont {Boggio}}, \bibinfo {author} {\bibfnamefont {P.}~\bibnamefont
  {Andrekson}}, \bibinfo {author} {\bibfnamefont {M.}~\bibnamefont
  {Karlsson}},\ and\ \bibinfo {author} {\bibfnamefont {S.}~\bibnamefont
  {Radic}},\ }\bibfield  {title} {\bibinfo {title} {Bandwidth-efficient phase
  modulation techniques for stimulated brillouin scattering suppression in
  fiber optic parametric amplifiers},\ }\href@noop {} {\bibfield  {journal}
  {\bibinfo  {journal} {Optics Express}\ }\textbf {\bibinfo {volume} {18}},\
  \bibinfo {pages} {18138} (\bibinfo {year} {2010})}\BibitemShut {NoStop}%
\bibitem [{\citenamefont {Liu}\ \emph {et~al.}(2009)\citenamefont {Liu},
  \citenamefont {Lv}, \citenamefont {Dong},\ and\ \citenamefont
  {Li}}]{liu2009research}%
  \BibitemOpen
  \bibfield  {author} {\bibinfo {author} {\bibfnamefont {Y.}~\bibnamefont
  {Liu}}, \bibinfo {author} {\bibfnamefont {Z.}~\bibnamefont {Lv}}, \bibinfo
  {author} {\bibfnamefont {Y.}~\bibnamefont {Dong}},\ and\ \bibinfo {author}
  {\bibfnamefont {Q.}~\bibnamefont {Li}},\ }\bibfield  {title} {\bibinfo
  {title} {Research on stimulated brillouin scattering suppression based on
  multi-frequency phase modulation},\ }\href@noop {} {\bibfield  {journal}
  {\bibinfo  {journal} {Chinese Optics Letters}\ }\textbf {\bibinfo {volume}
  {7}},\ \bibinfo {pages} {29} (\bibinfo {year} {2009})}\BibitemShut {NoStop}%
\bibitem [{\citenamefont {Yoshizawa}\ and\ \citenamefont
  {Imai}(1993)}]{yoshizawa1993stimulated}%
  \BibitemOpen
  \bibfield  {author} {\bibinfo {author} {\bibfnamefont {N.}~\bibnamefont
  {Yoshizawa}}\ and\ \bibinfo {author} {\bibfnamefont {T.}~\bibnamefont
  {Imai}},\ }\bibfield  {title} {\bibinfo {title} {Stimulated brillouin
  scattering suppression by means of applying strain distribution to fiber with
  cabling},\ }\href@noop {} {\bibfield  {journal} {\bibinfo  {journal} {Journal
  of Lightwave Technology}\ }\textbf {\bibinfo {volume} {11}},\ \bibinfo
  {pages} {1518} (\bibinfo {year} {1993})}\BibitemShut {NoStop}%
\bibitem [{\citenamefont {Liu}(2007)}]{liu2007suppressing}%
  \BibitemOpen
  \bibfield  {author} {\bibinfo {author} {\bibfnamefont {A.}~\bibnamefont
  {Liu}},\ }\bibfield  {title} {\bibinfo {title} {Suppressing stimulated
  brillouin scattering in fiber amplifiers using nonuniform fiber and
  temperature gradient},\ }\href@noop {} {\bibfield  {journal} {\bibinfo
  {journal} {Optics express}\ }\textbf {\bibinfo {volume} {15}},\ \bibinfo
  {pages} {977} (\bibinfo {year} {2007})}\BibitemShut {NoStop}%
\bibitem [{\citenamefont {Kobyakov}\ \emph {et~al.}(2005)\citenamefont
  {Kobyakov}, \citenamefont {Kumar}, \citenamefont {Chowdhury}, \citenamefont
  {Ruffin}, \citenamefont {Sauer}, \citenamefont {Bickham},\ and\ \citenamefont
  {Mishra}}]{kobyakov2005design}%
  \BibitemOpen
  \bibfield  {author} {\bibinfo {author} {\bibfnamefont {A.}~\bibnamefont
  {Kobyakov}}, \bibinfo {author} {\bibfnamefont {S.}~\bibnamefont {Kumar}},
  \bibinfo {author} {\bibfnamefont {D.~Q.}\ \bibnamefont {Chowdhury}}, \bibinfo
  {author} {\bibfnamefont {A.~B.}\ \bibnamefont {Ruffin}}, \bibinfo {author}
  {\bibfnamefont {M.}~\bibnamefont {Sauer}}, \bibinfo {author} {\bibfnamefont
  {S.~R.}\ \bibnamefont {Bickham}},\ and\ \bibinfo {author} {\bibfnamefont
  {R.}~\bibnamefont {Mishra}},\ }\bibfield  {title} {\bibinfo {title} {Design
  concept for optical fibers with enhanced sbs threshold},\ }\href@noop {}
  {\bibfield  {journal} {\bibinfo  {journal} {Optics Express}\ }\textbf
  {\bibinfo {volume} {13}},\ \bibinfo {pages} {5338} (\bibinfo {year}
  {2005})}\BibitemShut {NoStop}%
\bibitem [{\citenamefont {Dragic}\ \emph {et~al.}(2005)\citenamefont {Dragic},
  \citenamefont {Liu}, \citenamefont {Papen},\ and\ \citenamefont
  {Galvanauskas}}]{dragic2005optical}%
  \BibitemOpen
  \bibfield  {author} {\bibinfo {author} {\bibfnamefont {P.~D.}\ \bibnamefont
  {Dragic}}, \bibinfo {author} {\bibfnamefont {C.-H.}\ \bibnamefont {Liu}},
  \bibinfo {author} {\bibfnamefont {G.~C.}\ \bibnamefont {Papen}},\ and\
  \bibinfo {author} {\bibfnamefont {A.}~\bibnamefont {Galvanauskas}},\
  }\bibfield  {title} {\bibinfo {title} {Optical fiber with an acoustic guiding
  layer for stimulated brillouin scattering suppression},\ }in\ \href@noop {}
  {\emph {\bibinfo {booktitle} {(CLEO). Conference on Lasers and
  Electro-Optics, 2005.}}},\ Vol.~\bibinfo {volume} {3}\ (\bibinfo
  {organization} {IEEE},\ \bibinfo {year} {2005})\ pp.\ \bibinfo {pages}
  {1984--1986}\BibitemShut {NoStop}%
\bibitem [{\citenamefont {Li}\ \emph {et~al.}(2007)\citenamefont {Li},
  \citenamefont {Chen}, \citenamefont {Wang}, \citenamefont {Gray},
  \citenamefont {Liu}, \citenamefont {Demeritt}, \citenamefont {Ruffin},
  \citenamefont {Crowley}, \citenamefont {Walton},\ and\ \citenamefont
  {Zenteno}}]{li2007ge}%
  \BibitemOpen
  \bibfield  {author} {\bibinfo {author} {\bibfnamefont {M.-J.}\ \bibnamefont
  {Li}}, \bibinfo {author} {\bibfnamefont {X.}~\bibnamefont {Chen}}, \bibinfo
  {author} {\bibfnamefont {J.}~\bibnamefont {Wang}}, \bibinfo {author}
  {\bibfnamefont {S.}~\bibnamefont {Gray}}, \bibinfo {author} {\bibfnamefont
  {A.}~\bibnamefont {Liu}}, \bibinfo {author} {\bibfnamefont {J.~A.}\
  \bibnamefont {Demeritt}}, \bibinfo {author} {\bibfnamefont {A.~B.}\
  \bibnamefont {Ruffin}}, \bibinfo {author} {\bibfnamefont {A.~M.}\
  \bibnamefont {Crowley}}, \bibinfo {author} {\bibfnamefont {D.~T.}\
  \bibnamefont {Walton}},\ and\ \bibinfo {author} {\bibfnamefont {L.~A.}\
  \bibnamefont {Zenteno}},\ }\bibfield  {title} {\bibinfo {title} {Al/ge
  co-doped large mode area fiber with high sbs threshold},\ }\href@noop {}
  {\bibfield  {journal} {\bibinfo  {journal} {Optics Express}\ }\textbf
  {\bibinfo {volume} {15}},\ \bibinfo {pages} {8290} (\bibinfo {year}
  {2007})}\BibitemShut {NoStop}%
\bibitem [{\citenamefont {Hawkins}\ \emph {et~al.}(2021)\citenamefont
  {Hawkins}, \citenamefont {Dragic}, \citenamefont {Yu}, \citenamefont
  {Flores}, \citenamefont {Engholm},\ and\ \citenamefont
  {Ballato}}]{hawkins2021kilowatt}%
  \BibitemOpen
  \bibfield  {author} {\bibinfo {author} {\bibfnamefont {T.}~\bibnamefont
  {Hawkins}}, \bibinfo {author} {\bibfnamefont {P.}~\bibnamefont {Dragic}},
  \bibinfo {author} {\bibfnamefont {N.}~\bibnamefont {Yu}}, \bibinfo {author}
  {\bibfnamefont {A.}~\bibnamefont {Flores}}, \bibinfo {author} {\bibfnamefont
  {M.}~\bibnamefont {Engholm}},\ and\ \bibinfo {author} {\bibfnamefont
  {J.}~\bibnamefont {Ballato}},\ }\bibfield  {title} {\bibinfo {title}
  {Kilowatt power scaling of an intrinsically low brillouin and thermo-optic
  yb-doped silica fiber},\ }\href@noop {} {\bibfield  {journal} {\bibinfo
  {journal} {JOSA B}\ }\textbf {\bibinfo {volume} {38}},\ \bibinfo {pages}
  {F38} (\bibinfo {year} {2021})}\BibitemShut {NoStop}%
\bibitem [{\citenamefont {Shiraki}\ \emph {et~al.}(1995)\citenamefont
  {Shiraki}, \citenamefont {Ohashi},\ and\ \citenamefont
  {Tateda}}]{shiraki1995suppression}%
  \BibitemOpen
  \bibfield  {author} {\bibinfo {author} {\bibfnamefont {K.}~\bibnamefont
  {Shiraki}}, \bibinfo {author} {\bibfnamefont {M.}~\bibnamefont {Ohashi}},\
  and\ \bibinfo {author} {\bibfnamefont {M.}~\bibnamefont {Tateda}},\
  }\bibfield  {title} {\bibinfo {title} {Suppression of stimulated brillouin
  scattering in a fibre by changing the core radius},\ }\href@noop {}
  {\bibfield  {journal} {\bibinfo  {journal} {Electronics letters}\ }\textbf
  {\bibinfo {volume} {31}},\ \bibinfo {pages} {668} (\bibinfo {year}
  {1995})}\BibitemShut {NoStop}%
\bibitem [{\citenamefont {Robin}\ and\ \citenamefont
  {Dajani}(2011)}]{robin2011acoustically}%
  \BibitemOpen
  \bibfield  {author} {\bibinfo {author} {\bibfnamefont {C.}~\bibnamefont
  {Robin}}\ and\ \bibinfo {author} {\bibfnamefont {I.}~\bibnamefont {Dajani}},\
  }\bibfield  {title} {\bibinfo {title} {Acoustically segmented photonic
  crystal fiber for single-frequency high-power laser applications},\
  }\href@noop {} {\bibfield  {journal} {\bibinfo  {journal} {Optics letters}\
  }\textbf {\bibinfo {volume} {36}},\ \bibinfo {pages} {2641} (\bibinfo {year}
  {2011})}\BibitemShut {NoStop}%
\bibitem [{\citenamefont {Augst}\ \emph {et~al.}(2007)\citenamefont {Augst},
  \citenamefont {Ranka}, \citenamefont {Fan},\ and\ \citenamefont
  {Sanchez}}]{augst2007beam}%
  \BibitemOpen
  \bibfield  {author} {\bibinfo {author} {\bibfnamefont {S.~J.}\ \bibnamefont
  {Augst}}, \bibinfo {author} {\bibfnamefont {J.~K.}\ \bibnamefont {Ranka}},
  \bibinfo {author} {\bibfnamefont {T.}~\bibnamefont {Fan}},\ and\ \bibinfo
  {author} {\bibfnamefont {A.}~\bibnamefont {Sanchez}},\ }\bibfield  {title}
  {\bibinfo {title} {Beam combining of ytterbium fiber amplifiers},\
  }\href@noop {} {\bibfield  {journal} {\bibinfo  {journal} {JOSA B}\ }\textbf
  {\bibinfo {volume} {24}},\ \bibinfo {pages} {1707} (\bibinfo {year}
  {2007})}\BibitemShut {NoStop}%
\bibitem [{\citenamefont {Loftus}\ \emph {et~al.}(2007)\citenamefont {Loftus},
  \citenamefont {Thomas}, \citenamefont {Hoffman}, \citenamefont {Norsen},
  \citenamefont {Royse}, \citenamefont {Liu},\ and\ \citenamefont
  {Honea}}]{loftus2007spectrally}%
  \BibitemOpen
  \bibfield  {author} {\bibinfo {author} {\bibfnamefont {T.~H.}\ \bibnamefont
  {Loftus}}, \bibinfo {author} {\bibfnamefont {A.~M.}\ \bibnamefont {Thomas}},
  \bibinfo {author} {\bibfnamefont {P.~R.}\ \bibnamefont {Hoffman}}, \bibinfo
  {author} {\bibfnamefont {M.}~\bibnamefont {Norsen}}, \bibinfo {author}
  {\bibfnamefont {R.}~\bibnamefont {Royse}}, \bibinfo {author} {\bibfnamefont
  {A.}~\bibnamefont {Liu}},\ and\ \bibinfo {author} {\bibfnamefont {E.~C.}\
  \bibnamefont {Honea}},\ }\bibfield  {title} {\bibinfo {title} {Spectrally
  beam-combined fiber lasers for high-average-power applications},\ }\href@noop
  {} {\bibfield  {journal} {\bibinfo  {journal} {IEEE journal of selected
  topics in quantum electronics}\ }\textbf {\bibinfo {volume} {13}},\ \bibinfo
  {pages} {487} (\bibinfo {year} {2007})}\BibitemShut {NoStop}%
\bibitem [{\citenamefont {Buikema}\ \emph {et~al.}(2019)\citenamefont
  {Buikema}, \citenamefont {Jose}, \citenamefont {Augst}, \citenamefont
  {Fritschel},\ and\ \citenamefont {Mavalvala}}]{buikema2019narrow}%
  \BibitemOpen
  \bibfield  {author} {\bibinfo {author} {\bibfnamefont {A.}~\bibnamefont
  {Buikema}}, \bibinfo {author} {\bibfnamefont {F.}~\bibnamefont {Jose}},
  \bibinfo {author} {\bibfnamefont {S.~J.}\ \bibnamefont {Augst}}, \bibinfo
  {author} {\bibfnamefont {P.}~\bibnamefont {Fritschel}},\ and\ \bibinfo
  {author} {\bibfnamefont {N.}~\bibnamefont {Mavalvala}},\ }\bibfield  {title}
  {\bibinfo {title} {Narrow-linewidth fiber amplifier for gravitational-wave
  detectors},\ }\href@noop {} {\bibfield  {journal} {\bibinfo  {journal}
  {Optics Letters}\ }\textbf {\bibinfo {volume} {44}},\ \bibinfo {pages} {3833}
  (\bibinfo {year} {2019})}\BibitemShut {NoStop}%
\bibitem [{\citenamefont {Ward}\ \emph {et~al.}(2012)\citenamefont {Ward},
  \citenamefont {Robin},\ and\ \citenamefont {Dajani}}]{ward2012origin}%
  \BibitemOpen
  \bibfield  {author} {\bibinfo {author} {\bibfnamefont {B.}~\bibnamefont
  {Ward}}, \bibinfo {author} {\bibfnamefont {C.}~\bibnamefont {Robin}},\ and\
  \bibinfo {author} {\bibfnamefont {I.}~\bibnamefont {Dajani}},\ }\bibfield
  {title} {\bibinfo {title} {Origin of thermal modal instabilities in large
  mode area fiber amplifiers},\ }\href@noop {} {\bibfield  {journal} {\bibinfo
  {journal} {Optics Express}\ }\textbf {\bibinfo {volume} {20}},\ \bibinfo
  {pages} {11407} (\bibinfo {year} {2012})}\BibitemShut {NoStop}%
\bibitem [{\citenamefont {Tzang}\ \emph {et~al.}(2018)\citenamefont {Tzang},
  \citenamefont {Caravaca-Aguirre}, \citenamefont {Wagner},\ and\ \citenamefont
  {Piestun}}]{tzang2018adaptive}%
  \BibitemOpen
  \bibfield  {author} {\bibinfo {author} {\bibfnamefont {O.}~\bibnamefont
  {Tzang}}, \bibinfo {author} {\bibfnamefont {A.~M.}\ \bibnamefont
  {Caravaca-Aguirre}}, \bibinfo {author} {\bibfnamefont {K.}~\bibnamefont
  {Wagner}},\ and\ \bibinfo {author} {\bibfnamefont {R.}~\bibnamefont
  {Piestun}},\ }\bibfield  {title} {\bibinfo {title} {Adaptive wavefront
  shaping for controlling nonlinear multimode interactions in optical fibres},\
  }\href@noop {} {\bibfield  {journal} {\bibinfo  {journal} {Nature Photonics}\
  }\textbf {\bibinfo {volume} {12}},\ \bibinfo {pages} {368} (\bibinfo {year}
  {2018})}\BibitemShut {NoStop}%
\bibitem [{\citenamefont {Deliancourt}\ \emph {et~al.}(2019)\citenamefont
  {Deliancourt}, \citenamefont {Fabert}, \citenamefont {Tonello}, \citenamefont
  {Krupa}, \citenamefont {Desfarges-Berthelemot}, \citenamefont {Kermene},
  \citenamefont {Millot}, \citenamefont {Barth{\'e}l{\'e}my}, \citenamefont
  {Wabnitz},\ and\ \citenamefont {Couderc}}]{deliancourt2019wavefront}%
  \BibitemOpen
  \bibfield  {author} {\bibinfo {author} {\bibfnamefont {E.}~\bibnamefont
  {Deliancourt}}, \bibinfo {author} {\bibfnamefont {M.}~\bibnamefont {Fabert}},
  \bibinfo {author} {\bibfnamefont {A.}~\bibnamefont {Tonello}}, \bibinfo
  {author} {\bibfnamefont {K.}~\bibnamefont {Krupa}}, \bibinfo {author}
  {\bibfnamefont {A.}~\bibnamefont {Desfarges-Berthelemot}}, \bibinfo {author}
  {\bibfnamefont {V.}~\bibnamefont {Kermene}}, \bibinfo {author} {\bibfnamefont
  {G.}~\bibnamefont {Millot}}, \bibinfo {author} {\bibfnamefont
  {A.}~\bibnamefont {Barth{\'e}l{\'e}my}}, \bibinfo {author} {\bibfnamefont
  {S.}~\bibnamefont {Wabnitz}},\ and\ \bibinfo {author} {\bibfnamefont
  {V.}~\bibnamefont {Couderc}},\ }\bibfield  {title} {\bibinfo {title}
  {Wavefront shaping for optimized many-mode kerr beam self-cleaning in
  graded-index multimode fiber},\ }\href@noop {} {\bibfield  {journal}
  {\bibinfo  {journal} {Optics Express}\ }\textbf {\bibinfo {volume} {27}},\
  \bibinfo {pages} {17311} (\bibinfo {year} {2019})}\BibitemShut {NoStop}%
\bibitem [{\citenamefont {Shutova}\ \emph {et~al.}(2019)\citenamefont
  {Shutova}, \citenamefont {Shutov}, \citenamefont {Zhdanova}, \citenamefont
  {Thompson},\ and\ \citenamefont {Sokolov}}]{shutova2019coherent}%
  \BibitemOpen
  \bibfield  {author} {\bibinfo {author} {\bibfnamefont {M.}~\bibnamefont
  {Shutova}}, \bibinfo {author} {\bibfnamefont {A.~D.}\ \bibnamefont {Shutov}},
  \bibinfo {author} {\bibfnamefont {A.~A.}\ \bibnamefont {Zhdanova}}, \bibinfo
  {author} {\bibfnamefont {J.~V.}\ \bibnamefont {Thompson}},\ and\ \bibinfo
  {author} {\bibfnamefont {A.~V.}\ \bibnamefont {Sokolov}},\ }\bibfield
  {title} {\bibinfo {title} {Coherent raman generation controlled by wavefront
  shaping},\ }\href@noop {} {\bibfield  {journal} {\bibinfo  {journal}
  {Scientific reports}\ }\textbf {\bibinfo {volume} {9}},\ \bibinfo {pages}
  {1565} (\bibinfo {year} {2019})}\BibitemShut {NoStop}%
\bibitem [{\citenamefont {Te{\u{g}}in}\ \emph {et~al.}(2020)\citenamefont
  {Te{\u{g}}in}, \citenamefont {Rahmani}, \citenamefont {Kakkava},
  \citenamefont {Borhani}, \citenamefont {Moser},\ and\ \citenamefont
  {Psaltis}}]{teugin2020controlling}%
  \BibitemOpen
  \bibfield  {author} {\bibinfo {author} {\bibfnamefont {U.}~\bibnamefont
  {Te{\u{g}}in}}, \bibinfo {author} {\bibfnamefont {B.}~\bibnamefont
  {Rahmani}}, \bibinfo {author} {\bibfnamefont {E.}~\bibnamefont {Kakkava}},
  \bibinfo {author} {\bibfnamefont {N.}~\bibnamefont {Borhani}}, \bibinfo
  {author} {\bibfnamefont {C.}~\bibnamefont {Moser}},\ and\ \bibinfo {author}
  {\bibfnamefont {D.}~\bibnamefont {Psaltis}},\ }\bibfield  {title} {\bibinfo
  {title} {Controlling spatiotemporal nonlinearities in multimode fibers with
  deep neural networks},\ }\href@noop {} {\bibfield  {journal} {\bibinfo
  {journal} {Apl Photonics}\ }\textbf {\bibinfo {volume} {5}},\ \bibinfo
  {pages} {030804} (\bibinfo {year} {2020})}\BibitemShut {NoStop}%
\bibitem [{\citenamefont {Chen}\ \emph
  {et~al.}(2022{\natexlab{b}})\citenamefont {Chen}, \citenamefont {Wisal},
  \citenamefont {Eliezer}, \citenamefont {Stone},\ and\ \citenamefont
  {Cao}}]{chen2022suppressing}%
  \BibitemOpen
  \bibfield  {author} {\bibinfo {author} {\bibfnamefont {C.-W.}\ \bibnamefont
  {Chen}}, \bibinfo {author} {\bibfnamefont {K.}~\bibnamefont {Wisal}},
  \bibinfo {author} {\bibfnamefont {Y.}~\bibnamefont {Eliezer}}, \bibinfo
  {author} {\bibfnamefont {A.~D.}\ \bibnamefont {Stone}},\ and\ \bibinfo
  {author} {\bibfnamefont {H.}~\bibnamefont {Cao}},\ }\bibfield  {title}
  {\bibinfo {title} {Suppressing transverse mode instability through multimode
  excitation in a fiber amplifier},\ }\href@noop {} {\bibfield  {journal}
  {\bibinfo  {journal} {arXiv preprint arXiv:2206.15438}\ } (\bibinfo {year}
  {2022}{\natexlab{b}})}\BibitemShut {NoStop}%
\bibitem [{\citenamefont {Pl{\"o}schner}\ \emph {et~al.}(2015)\citenamefont
  {Pl{\"o}schner}, \citenamefont {Tyc},\ and\ \citenamefont
  {{\v{C}}i{\v{z}}m{\'a}r}}]{ploschner2015seeing}%
  \BibitemOpen
  \bibfield  {author} {\bibinfo {author} {\bibfnamefont {M.}~\bibnamefont
  {Pl{\"o}schner}}, \bibinfo {author} {\bibfnamefont {T.}~\bibnamefont {Tyc}},\
  and\ \bibinfo {author} {\bibfnamefont {T.}~\bibnamefont
  {{\v{C}}i{\v{z}}m{\'a}r}},\ }\bibfield  {title} {\bibinfo {title} {Seeing
  through chaos in multimode fibres},\ }\href@noop {} {\bibfield  {journal}
  {\bibinfo  {journal} {Nature Photonics}\ }\textbf {\bibinfo {volume} {9}},\
  \bibinfo {pages} {529} (\bibinfo {year} {2015})}\BibitemShut {NoStop}%
\bibitem [{\citenamefont {Xiong}\ \emph {et~al.}(2016)\citenamefont {Xiong},
  \citenamefont {Ambichl}, \citenamefont {Bromberg}, \citenamefont {Redding},
  \citenamefont {Rotter},\ and\ \citenamefont {Cao}}]{xiong2016spatiotemporal}%
  \BibitemOpen
  \bibfield  {author} {\bibinfo {author} {\bibfnamefont {W.}~\bibnamefont
  {Xiong}}, \bibinfo {author} {\bibfnamefont {P.}~\bibnamefont {Ambichl}},
  \bibinfo {author} {\bibfnamefont {Y.}~\bibnamefont {Bromberg}}, \bibinfo
  {author} {\bibfnamefont {B.}~\bibnamefont {Redding}}, \bibinfo {author}
  {\bibfnamefont {S.}~\bibnamefont {Rotter}},\ and\ \bibinfo {author}
  {\bibfnamefont {H.}~\bibnamefont {Cao}},\ }\bibfield  {title} {\bibinfo
  {title} {Spatiotemporal control of light transmission through a multimode
  fiber with strong mode coupling},\ }\href@noop {} {\bibfield  {journal}
  {\bibinfo  {journal} {Physical review letters}\ }\textbf {\bibinfo {volume}
  {117}},\ \bibinfo {pages} {053901} (\bibinfo {year} {2016})}\BibitemShut
  {NoStop}%
\bibitem [{\citenamefont {Florentin}\ \emph {et~al.}(2017)\citenamefont
  {Florentin}, \citenamefont {Kermene}, \citenamefont {Benoist}, \citenamefont
  {Desfarges-Berthelemot}, \citenamefont {Pagnoux}, \citenamefont
  {Barth{\'e}l{\'e}my},\ and\ \citenamefont {Huignard}}]{florentin2017shaping}%
  \BibitemOpen
  \bibfield  {author} {\bibinfo {author} {\bibfnamefont {R.}~\bibnamefont
  {Florentin}}, \bibinfo {author} {\bibfnamefont {V.}~\bibnamefont {Kermene}},
  \bibinfo {author} {\bibfnamefont {J.}~\bibnamefont {Benoist}}, \bibinfo
  {author} {\bibfnamefont {A.}~\bibnamefont {Desfarges-Berthelemot}}, \bibinfo
  {author} {\bibfnamefont {D.}~\bibnamefont {Pagnoux}}, \bibinfo {author}
  {\bibfnamefont {A.}~\bibnamefont {Barth{\'e}l{\'e}my}},\ and\ \bibinfo
  {author} {\bibfnamefont {J.-P.}\ \bibnamefont {Huignard}},\ }\bibfield
  {title} {\bibinfo {title} {Shaping the light amplified in a multimode
  fiber},\ }\href@noop {} {\bibfield  {journal} {\bibinfo  {journal} {Light:
  Science \& Applications}\ }\textbf {\bibinfo {volume} {6}},\ \bibinfo {pages}
  {e16208} (\bibinfo {year} {2017})}\BibitemShut {NoStop}%
\bibitem [{\citenamefont {Gomes}\ \emph {et~al.}(2022)\citenamefont {Gomes},
  \citenamefont {Turtaev}, \citenamefont {Du},\ and\ \citenamefont
  {{\v{C}}i{\v{z}}m{\'a}r}}]{gomes2022near}%
  \BibitemOpen
  \bibfield  {author} {\bibinfo {author} {\bibfnamefont {A.~D.}\ \bibnamefont
  {Gomes}}, \bibinfo {author} {\bibfnamefont {S.}~\bibnamefont {Turtaev}},
  \bibinfo {author} {\bibfnamefont {Y.}~\bibnamefont {Du}},\ and\ \bibinfo
  {author} {\bibfnamefont {T.}~\bibnamefont {{\v{C}}i{\v{z}}m{\'a}r}},\
  }\bibfield  {title} {\bibinfo {title} {Near perfect focusing through
  multimode fibres},\ }\href@noop {} {\bibfield  {journal} {\bibinfo  {journal}
  {Optics Express}\ }\textbf {\bibinfo {volume} {30}},\ \bibinfo {pages}
  {10645} (\bibinfo {year} {2022})}\BibitemShut {NoStop}%
\bibitem [{\citenamefont {Tei}\ \emph {et~al.}(2001)\citenamefont {Tei},
  \citenamefont {Tsuruoka}, \citenamefont {Uchiyama},\ and\ \citenamefont
  {Fujioka}}]{tei2001critical}%
  \BibitemOpen
  \bibfield  {author} {\bibinfo {author} {\bibfnamefont {K.}~\bibnamefont
  {Tei}}, \bibinfo {author} {\bibfnamefont {Y.}~\bibnamefont {Tsuruoka}},
  \bibinfo {author} {\bibfnamefont {T.}~\bibnamefont {Uchiyama}},\ and\
  \bibinfo {author} {\bibfnamefont {T.}~\bibnamefont {Fujioka}},\ }\bibfield
  {title} {\bibinfo {title} {Critical power of stimulated brillouin scattering
  in multimode optical fibers},\ }\href@noop {} {\bibfield  {journal} {\bibinfo
   {journal} {Japanese Journal of Applied Physics}\ }\textbf {\bibinfo {volume}
  {40}},\ \bibinfo {pages} {3191} (\bibinfo {year} {2001})}\BibitemShut
  {NoStop}%
\bibitem [{\citenamefont {Kovalev}\ and\ \citenamefont
  {Harrison}(2002)}]{kovalev2002waveguide}%
  \BibitemOpen
  \bibfield  {author} {\bibinfo {author} {\bibfnamefont {V.~I.}\ \bibnamefont
  {Kovalev}}\ and\ \bibinfo {author} {\bibfnamefont {R.~G.}\ \bibnamefont
  {Harrison}},\ }\bibfield  {title} {\bibinfo {title} {Waveguide-induced
  inhomogeneous spectral broadening of stimulated brillouin scattering in
  optical fiber},\ }\href@noop {} {\bibfield  {journal} {\bibinfo  {journal}
  {Optics letters}\ }\textbf {\bibinfo {volume} {27}},\ \bibinfo {pages} {2022}
  (\bibinfo {year} {2002})}\BibitemShut {NoStop}%
\bibitem [{\citenamefont {Sj{\"o}berg}\ \emph {et~al.}(2003)\citenamefont
  {Sj{\"o}berg}, \citenamefont {Quiroga-Teixeiro}, \citenamefont {Galt},\ and\
  \citenamefont {H{\aa}rd}}]{sjoberg2003dependence}%
  \BibitemOpen
  \bibfield  {author} {\bibinfo {author} {\bibfnamefont {M.}~\bibnamefont
  {Sj{\"o}berg}}, \bibinfo {author} {\bibfnamefont {M.~L.}\ \bibnamefont
  {Quiroga-Teixeiro}}, \bibinfo {author} {\bibfnamefont {S.}~\bibnamefont
  {Galt}},\ and\ \bibinfo {author} {\bibfnamefont {S.}~\bibnamefont
  {H{\aa}rd}},\ }\bibfield  {title} {\bibinfo {title} {Dependence of stimulated
  brillouin scattering in multimode fibers on beam quality, pulse duration, and
  coherence length},\ }\href@noop {} {\bibfield  {journal} {\bibinfo  {journal}
  {JOSA B}\ }\textbf {\bibinfo {volume} {20}},\ \bibinfo {pages} {434}
  (\bibinfo {year} {2003})}\BibitemShut {NoStop}%
\bibitem [{\citenamefont {Minardo}\ \emph {et~al.}(2014)\citenamefont
  {Minardo}, \citenamefont {Bernini},\ and\ \citenamefont
  {Zeni}}]{minardo2014experimental}%
  \BibitemOpen
  \bibfield  {author} {\bibinfo {author} {\bibfnamefont {A.}~\bibnamefont
  {Minardo}}, \bibinfo {author} {\bibfnamefont {R.}~\bibnamefont {Bernini}},\
  and\ \bibinfo {author} {\bibfnamefont {L.}~\bibnamefont {Zeni}},\ }\bibfield
  {title} {\bibinfo {title} {Experimental and numerical study on stimulated
  brillouin scattering in a graded-index multimode fiber},\ }\href@noop {}
  {\bibfield  {journal} {\bibinfo  {journal} {Optics express}\ }\textbf
  {\bibinfo {volume} {22}},\ \bibinfo {pages} {17480} (\bibinfo {year}
  {2014})}\BibitemShut {NoStop}%
\bibitem [{\citenamefont {Wang}\ \emph {et~al.}(2020)\citenamefont {Wang},
  \citenamefont {Alvarado-Zacarias}, \citenamefont {Habib}, \citenamefont
  {Wen}, \citenamefont {Antonio-Lopez}, \citenamefont {Sillard}, \citenamefont
  {Amezcua-Correa}, \citenamefont {Sch{\"u}lzgen}, \citenamefont
  {Amezcua-Correa},\ and\ \citenamefont {Li}}]{wang2020mode}%
  \BibitemOpen
  \bibfield  {author} {\bibinfo {author} {\bibfnamefont {N.}~\bibnamefont
  {Wang}}, \bibinfo {author} {\bibfnamefont {J.}~\bibnamefont
  {Alvarado-Zacarias}}, \bibinfo {author} {\bibfnamefont {M.~S.}\ \bibnamefont
  {Habib}}, \bibinfo {author} {\bibfnamefont {H.}~\bibnamefont {Wen}}, \bibinfo
  {author} {\bibfnamefont {J.}~\bibnamefont {Antonio-Lopez}}, \bibinfo {author}
  {\bibfnamefont {P.}~\bibnamefont {Sillard}}, \bibinfo {author} {\bibfnamefont
  {A.}~\bibnamefont {Amezcua-Correa}}, \bibinfo {author} {\bibfnamefont
  {A.}~\bibnamefont {Sch{\"u}lzgen}}, \bibinfo {author} {\bibfnamefont
  {R.}~\bibnamefont {Amezcua-Correa}},\ and\ \bibinfo {author} {\bibfnamefont
  {G.}~\bibnamefont {Li}},\ }\bibfield  {title} {\bibinfo {title}
  {Mode-selective few-mode brillouin fiber lasers based on intramodal and
  intermodal sbs},\ }\href@noop {} {\bibfield  {journal} {\bibinfo  {journal}
  {Optics Letters}\ }\textbf {\bibinfo {volume} {45}},\ \bibinfo {pages} {2323}
  (\bibinfo {year} {2020})}\BibitemShut {NoStop}%
\bibitem [{\citenamefont {Song}\ and\ \citenamefont
  {Kim}(2013)}]{song2013characterization}%
  \BibitemOpen
  \bibfield  {author} {\bibinfo {author} {\bibfnamefont {K.~Y.}\ \bibnamefont
  {Song}}\ and\ \bibinfo {author} {\bibfnamefont {Y.~H.}\ \bibnamefont {Kim}},\
  }\bibfield  {title} {\bibinfo {title} {Characterization of stimulated
  brillouin scattering in a few-mode fiber},\ }\href@noop {} {\bibfield
  {journal} {\bibinfo  {journal} {Optics letters}\ }\textbf {\bibinfo {volume}
  {38}},\ \bibinfo {pages} {4841} (\bibinfo {year} {2013})}\BibitemShut
  {NoStop}%
\bibitem [{\citenamefont {Srinivasan}\ \emph {et~al.}(2021)\citenamefont
  {Srinivasan}, \citenamefont {Agrawal}, \citenamefont {Venkitesh} \emph
  {et~al.}}]{srinivasan2021role}%
  \BibitemOpen
  \bibfield  {author} {\bibinfo {author} {\bibfnamefont {B.}~\bibnamefont
  {Srinivasan}}, \bibinfo {author} {\bibfnamefont {G.~P.}\ \bibnamefont
  {Agrawal}}, \bibinfo {author} {\bibfnamefont {D.}~\bibnamefont {Venkitesh}},
  \emph {et~al.},\ }\bibfield  {title} {\bibinfo {title} {Role of the modal
  composition of pump in the multi-peak brillouin gain spectrum in a few-mode
  fiber},\ }\href@noop {} {\bibfield  {journal} {\bibinfo  {journal} {Optics
  Communications}\ }\textbf {\bibinfo {volume} {494}},\ \bibinfo {pages}
  {127052} (\bibinfo {year} {2021})}\BibitemShut {NoStop}%
\bibitem [{\citenamefont {L{\"u}}\ \emph {et~al.}(2015)\citenamefont {L{\"u}},
  \citenamefont {Zhou}, \citenamefont {Wang},\ and\ \citenamefont
  {Jiang}}]{lu2015theoretical}%
  \BibitemOpen
  \bibfield  {author} {\bibinfo {author} {\bibfnamefont {H.}~\bibnamefont
  {L{\"u}}}, \bibinfo {author} {\bibfnamefont {P.}~\bibnamefont {Zhou}},
  \bibinfo {author} {\bibfnamefont {X.}~\bibnamefont {Wang}},\ and\ \bibinfo
  {author} {\bibfnamefont {Z.}~\bibnamefont {Jiang}},\ }\bibfield  {title}
  {\bibinfo {title} {Theoretical and numerical study of the threshold of
  stimulated brillouin scattering in multimode fibers},\ }\href@noop {}
  {\bibfield  {journal} {\bibinfo  {journal} {Journal of Lightwave Technology}\
  }\textbf {\bibinfo {volume} {33}},\ \bibinfo {pages} {4464} (\bibinfo {year}
  {2015})}\BibitemShut {NoStop}%
\bibitem [{\citenamefont {Cotter}(1983)}]{cotter1983stimulated}%
  \BibitemOpen
  \bibfield  {author} {\bibinfo {author} {\bibfnamefont {D.}~\bibnamefont
  {Cotter}},\ }\bibfield  {title} {\bibinfo {title} {Stimulated brillouin
  scattering in monomode optical fiber},\ }\href@noop {} {\bibfield  {journal}
  {\bibinfo  {journal} {Journal of Optical Communications}\ }\textbf {\bibinfo
  {volume} {4}},\ \bibinfo {pages} {10} (\bibinfo {year} {1983})}\BibitemShut
  {NoStop}%
\bibitem [{\citenamefont {Dragic}\ and\ \citenamefont
  {Ward}(2010)}]{dragic2010accurate}%
  \BibitemOpen
  \bibfield  {author} {\bibinfo {author} {\bibfnamefont {P.~D.}\ \bibnamefont
  {Dragic}}\ and\ \bibinfo {author} {\bibfnamefont {B.~G.}\ \bibnamefont
  {Ward}},\ }\bibfield  {title} {\bibinfo {title} {Accurate modeling of the
  intrinsic brillouin linewidth via finite-element analysis},\ }\href@noop {}
  {\bibfield  {journal} {\bibinfo  {journal} {IEEE Photonics Technology
  Letters}\ }\textbf {\bibinfo {volume} {22}},\ \bibinfo {pages} {1698}
  (\bibinfo {year} {2010})}\BibitemShut {NoStop}%
\bibitem [{\citenamefont {Suni}\ and\ \citenamefont
  {Falk}(1986)}]{suni1986theory}%
  \BibitemOpen
  \bibfield  {author} {\bibinfo {author} {\bibfnamefont {P.}~\bibnamefont
  {Suni}}\ and\ \bibinfo {author} {\bibfnamefont {J.}~\bibnamefont {Falk}},\
  }\bibfield  {title} {\bibinfo {title} {Theory of phase conjugation by
  stimulated brillouin scattering},\ }\href@noop {} {\bibfield  {journal}
  {\bibinfo  {journal} {JOSA B}\ }\textbf {\bibinfo {volume} {3}},\ \bibinfo
  {pages} {1681} (\bibinfo {year} {1986})}\BibitemShut {NoStop}%
\bibitem [{\citenamefont {Hu}\ \emph {et~al.}(1989)\citenamefont {Hu},
  \citenamefont {Goldstone},\ and\ \citenamefont {Ma}}]{hu1989theoretical}%
  \BibitemOpen
  \bibfield  {author} {\bibinfo {author} {\bibfnamefont {P.}~\bibnamefont
  {Hu}}, \bibinfo {author} {\bibfnamefont {J.}~\bibnamefont {Goldstone}},\ and\
  \bibinfo {author} {\bibfnamefont {S.}~\bibnamefont {Ma}},\ }\bibfield
  {title} {\bibinfo {title} {Theoretical study of phase conjugation in
  stimulated brillouin scattering},\ }\href@noop {} {\bibfield  {journal}
  {\bibinfo  {journal} {JOSA B}\ }\textbf {\bibinfo {volume} {6}},\ \bibinfo
  {pages} {1813} (\bibinfo {year} {1989})}\BibitemShut {NoStop}%
\bibitem [{\citenamefont {Lombard}\ \emph {et~al.}(2006)\citenamefont
  {Lombard}, \citenamefont {Brignon}, \citenamefont {Huignard}, \citenamefont
  {Lallier},\ and\ \citenamefont {Georges}}]{lombard2006beam}%
  \BibitemOpen
  \bibfield  {author} {\bibinfo {author} {\bibfnamefont {L.}~\bibnamefont
  {Lombard}}, \bibinfo {author} {\bibfnamefont {A.}~\bibnamefont {Brignon}},
  \bibinfo {author} {\bibfnamefont {J.-P.}\ \bibnamefont {Huignard}}, \bibinfo
  {author} {\bibfnamefont {E.}~\bibnamefont {Lallier}},\ and\ \bibinfo {author}
  {\bibfnamefont {P.}~\bibnamefont {Georges}},\ }\bibfield  {title} {\bibinfo
  {title} {Beam cleanup in a self-aligned gradient-index brillouin cavity for
  high-power multimode fiber amplifiers},\ }\href@noop {} {\bibfield  {journal}
  {\bibinfo  {journal} {Optics Letters}\ }\textbf {\bibinfo {volume} {31}},\
  \bibinfo {pages} {158} (\bibinfo {year} {2006})}\BibitemShut {NoStop}%
\bibitem [{\citenamefont {Qiu}\ \emph {et~al.}(2013)\citenamefont {Qiu},
  \citenamefont {Rakich}, \citenamefont {Shin}, \citenamefont {Dong},
  \citenamefont {Solja{\v{c}}i{\'c}},\ and\ \citenamefont
  {Wang}}]{qiu2013stimulated}%
  \BibitemOpen
  \bibfield  {author} {\bibinfo {author} {\bibfnamefont {W.}~\bibnamefont
  {Qiu}}, \bibinfo {author} {\bibfnamefont {P.~T.}\ \bibnamefont {Rakich}},
  \bibinfo {author} {\bibfnamefont {H.}~\bibnamefont {Shin}}, \bibinfo {author}
  {\bibfnamefont {H.}~\bibnamefont {Dong}}, \bibinfo {author} {\bibfnamefont
  {M.}~\bibnamefont {Solja{\v{c}}i{\'c}}},\ and\ \bibinfo {author}
  {\bibfnamefont {Z.}~\bibnamefont {Wang}},\ }\bibfield  {title} {\bibinfo
  {title} {Stimulated brillouin scattering in nanoscale silicon step-index
  waveguides: a general framework of selection rules and calculating sbs
  gain},\ }\href@noop {} {\bibfield  {journal} {\bibinfo  {journal} {Optics
  express}\ }\textbf {\bibinfo {volume} {21}},\ \bibinfo {pages} {31402}
  (\bibinfo {year} {2013})}\BibitemShut {NoStop}%
\bibitem [{\citenamefont {Wolff}\ \emph {et~al.}(2015)\citenamefont {Wolff},
  \citenamefont {Steel}, \citenamefont {Eggleton},\ and\ \citenamefont
  {Poulton}}]{wolff2015stimulated}%
  \BibitemOpen
  \bibfield  {author} {\bibinfo {author} {\bibfnamefont {C.}~\bibnamefont
  {Wolff}}, \bibinfo {author} {\bibfnamefont {M.~J.}\ \bibnamefont {Steel}},
  \bibinfo {author} {\bibfnamefont {B.~J.}\ \bibnamefont {Eggleton}},\ and\
  \bibinfo {author} {\bibfnamefont {C.~G.}\ \bibnamefont {Poulton}},\
  }\bibfield  {title} {\bibinfo {title} {Stimulated brillouin scattering in
  integrated photonic waveguides: Forces, scattering mechanisms, and
  coupled-mode analysis},\ }\href@noop {} {\bibfield  {journal} {\bibinfo
  {journal} {Physical Review A}\ }\textbf {\bibinfo {volume} {92}},\ \bibinfo
  {pages} {013836} (\bibinfo {year} {2015})}\BibitemShut {NoStop}%
\bibitem [{\citenamefont {Shi}\ \emph {et~al.}(2017)\citenamefont {Shi},
  \citenamefont {Cerjan},\ and\ \citenamefont {Fan}}]{shi2017invited}%
  \BibitemOpen
  \bibfield  {author} {\bibinfo {author} {\bibfnamefont {Y.}~\bibnamefont
  {Shi}}, \bibinfo {author} {\bibfnamefont {A.}~\bibnamefont {Cerjan}},\ and\
  \bibinfo {author} {\bibfnamefont {S.}~\bibnamefont {Fan}},\ }\bibfield
  {title} {\bibinfo {title} {Invited article: Acousto-optic finite-difference
  frequency-domain algorithm for first-principles simulations of on-chip
  acousto-optic devices},\ }\href@noop {} {\bibfield  {journal} {\bibinfo
  {journal} {APL Photonics}\ }\textbf {\bibinfo {volume} {2}},\ \bibinfo
  {pages} {020801} (\bibinfo {year} {2017})}\BibitemShut {NoStop}%
\bibitem [{\citenamefont {Jackson}(1999)}]{jackson1999classical}%
  \BibitemOpen
  \bibfield  {author} {\bibinfo {author} {\bibfnamefont {J.~D.}\ \bibnamefont
  {Jackson}},\ }\href@noop {} {\bibinfo {title} {Classical electrodynamics}}
  (\bibinfo {year} {1999})\BibitemShut {NoStop}%
\bibitem [{\citenamefont {Nelson}\ and\ \citenamefont
  {Lax}(1971)}]{nelson1971theory}%
  \BibitemOpen
  \bibfield  {author} {\bibinfo {author} {\bibfnamefont {D.}~\bibnamefont
  {Nelson}}\ and\ \bibinfo {author} {\bibfnamefont {M.}~\bibnamefont {Lax}},\
  }\bibfield  {title} {\bibinfo {title} {Theory of the photoelastic
  interaction},\ }\href@noop {} {\bibfield  {journal} {\bibinfo  {journal}
  {Physical Review B}\ }\textbf {\bibinfo {volume} {3}},\ \bibinfo {pages}
  {2778} (\bibinfo {year} {1971})}\BibitemShut {NoStop}%
\bibitem [{\citenamefont {Auld}(1973)}]{auld1973acoustic}%
  \BibitemOpen
  \bibfield  {author} {\bibinfo {author} {\bibfnamefont {B.~A.}\ \bibnamefont
  {Auld}},\ }\href@noop {} {\emph {\bibinfo {title} {Acoustic fields and waves
  in solids}}}\ (\bibinfo {year} {1973})\BibitemShut {NoStop}%
\bibitem [{\citenamefont {Landau}\ \emph {et~al.}(1986)\citenamefont {Landau},
  \citenamefont {Lif{\v{s}}ic}, \citenamefont {Lifshitz}, \citenamefont
  {Kosevich},\ and\ \citenamefont {Pitaevskii}}]{landau1986theory}%
  \BibitemOpen
  \bibfield  {author} {\bibinfo {author} {\bibfnamefont {L.~D.}\ \bibnamefont
  {Landau}}, \bibinfo {author} {\bibfnamefont {E.~M.}\ \bibnamefont
  {Lif{\v{s}}ic}}, \bibinfo {author} {\bibfnamefont {E.~M.}\ \bibnamefont
  {Lifshitz}}, \bibinfo {author} {\bibfnamefont {A.~M.}\ \bibnamefont
  {Kosevich}},\ and\ \bibinfo {author} {\bibfnamefont {L.~P.}\ \bibnamefont
  {Pitaevskii}},\ }\href@noop {} {\emph {\bibinfo {title} {Theory of
  elasticity: volume 7}}},\ Vol.~\bibinfo {volume} {7}\ (\bibinfo  {publisher}
  {Elsevier},\ \bibinfo {year} {1986})\BibitemShut {NoStop}%
\bibitem [{\citenamefont {Feldman}(1975)}]{feldman1975relations}%
  \BibitemOpen
  \bibfield  {author} {\bibinfo {author} {\bibfnamefont {A.}~\bibnamefont
  {Feldman}},\ }\bibfield  {title} {\bibinfo {title} {Relations between
  electrostriction and the stress-optical effect},\ }\href@noop {} {\bibfield
  {journal} {\bibinfo  {journal} {Physical Review B}\ }\textbf {\bibinfo
  {volume} {11}},\ \bibinfo {pages} {5112} (\bibinfo {year}
  {1975})}\BibitemShut {NoStop}%
\bibitem [{\citenamefont {Waldron}(1969)}]{waldron1969some}%
  \BibitemOpen
  \bibfield  {author} {\bibinfo {author} {\bibfnamefont {R.}~\bibnamefont
  {Waldron}},\ }\bibfield  {title} {\bibinfo {title} {Some problems in the
  theory of guided microsonic waves},\ }\href@noop {} {\bibfield  {journal}
  {\bibinfo  {journal} {IEEE Transactions on Microwave Theory and Techniques}\
  }\textbf {\bibinfo {volume} {17}},\ \bibinfo {pages} {893} (\bibinfo {year}
  {1969})}\BibitemShut {NoStop}%
\bibitem [{\citenamefont {Ndagano}\ \emph {et~al.}(2015)\citenamefont
  {Ndagano}, \citenamefont {Br{\"u}ning}, \citenamefont {McLaren},
  \citenamefont {Duparr{\'e}},\ and\ \citenamefont
  {Forbes}}]{ndagano2015fiber}%
  \BibitemOpen
  \bibfield  {author} {\bibinfo {author} {\bibfnamefont {B.}~\bibnamefont
  {Ndagano}}, \bibinfo {author} {\bibfnamefont {R.}~\bibnamefont
  {Br{\"u}ning}}, \bibinfo {author} {\bibfnamefont {M.}~\bibnamefont
  {McLaren}}, \bibinfo {author} {\bibfnamefont {M.}~\bibnamefont
  {Duparr{\'e}}},\ and\ \bibinfo {author} {\bibfnamefont {A.}~\bibnamefont
  {Forbes}},\ }\bibfield  {title} {\bibinfo {title} {Fiber propagation of
  vector modes},\ }\href@noop {} {\bibfield  {journal} {\bibinfo  {journal}
  {Optics express}\ }\textbf {\bibinfo {volume} {23}},\ \bibinfo {pages}
  {17330} (\bibinfo {year} {2015})}\BibitemShut {NoStop}%
\bibitem [{\citenamefont {Okamoto}(2021)}]{okamoto2021fundamentals}%
  \BibitemOpen
  \bibfield  {author} {\bibinfo {author} {\bibfnamefont {K.}~\bibnamefont
  {Okamoto}},\ }\href@noop {} {\emph {\bibinfo {title} {Fundamentals of optical
  waveguides}}}\ (\bibinfo  {publisher} {Elsevier},\ \bibinfo {year}
  {2021})\BibitemShut {NoStop}%
\bibitem [{\citenamefont {Snyder}\ and\ \citenamefont
  {Young}(1978)}]{snyder1978modes}%
  \BibitemOpen
  \bibfield  {author} {\bibinfo {author} {\bibfnamefont {A.~W.}\ \bibnamefont
  {Snyder}}\ and\ \bibinfo {author} {\bibfnamefont {W.~R.}\ \bibnamefont
  {Young}},\ }\bibfield  {title} {\bibinfo {title} {Modes of optical
  waveguides},\ }\href@noop {} {\bibfield  {journal} {\bibinfo  {journal}
  {JOSA}\ }\textbf {\bibinfo {volume} {68}},\ \bibinfo {pages} {297} (\bibinfo
  {year} {1978})}\BibitemShut {NoStop}%
\bibitem [{com()}]{comsol}%
  \BibitemOpen
  \href@noop {} {\bibinfo {title} {Comsol multiphysics v. 5.5. www.comsol.com.
  comsol ab, stockholm, sweden}}\BibitemShut {NoStop}%
\bibitem [{\citenamefont {Bittner}\ \emph {et~al.}(2018)\citenamefont
  {Bittner}, \citenamefont {Guazzotti}, \citenamefont {Zeng}, \citenamefont
  {Hu}, \citenamefont {Y{\i}lmaz}, \citenamefont {Kim}, \citenamefont {Oh},
  \citenamefont {Wang}, \citenamefont {Hess},\ and\ \citenamefont
  {Cao}}]{bittner2018suppressing}%
  \BibitemOpen
  \bibfield  {author} {\bibinfo {author} {\bibfnamefont {S.}~\bibnamefont
  {Bittner}}, \bibinfo {author} {\bibfnamefont {S.}~\bibnamefont {Guazzotti}},
  \bibinfo {author} {\bibfnamefont {Y.}~\bibnamefont {Zeng}}, \bibinfo {author}
  {\bibfnamefont {X.}~\bibnamefont {Hu}}, \bibinfo {author} {\bibfnamefont
  {H.}~\bibnamefont {Y{\i}lmaz}}, \bibinfo {author} {\bibfnamefont
  {K.}~\bibnamefont {Kim}}, \bibinfo {author} {\bibfnamefont {S.~S.}\
  \bibnamefont {Oh}}, \bibinfo {author} {\bibfnamefont {Q.~J.}\ \bibnamefont
  {Wang}}, \bibinfo {author} {\bibfnamefont {O.}~\bibnamefont {Hess}},\ and\
  \bibinfo {author} {\bibfnamefont {H.}~\bibnamefont {Cao}},\ }\bibfield
  {title} {\bibinfo {title} {Suppressing spatiotemporal lasing instabilities
  with wave-chaotic microcavities},\ }\href@noop {} {\bibfield  {journal}
  {\bibinfo  {journal} {Science}\ }\textbf {\bibinfo {volume} {361}},\ \bibinfo
  {pages} {1225} (\bibinfo {year} {2018})}\BibitemShut {NoStop}%
\bibitem [{\citenamefont {Manley}\ and\ \citenamefont
  {Rowe}(1959)}]{manley1959general}%
  \BibitemOpen
  \bibfield  {author} {\bibinfo {author} {\bibfnamefont {J.}~\bibnamefont
  {Manley}}\ and\ \bibinfo {author} {\bibfnamefont {H.}~\bibnamefont {Rowe}},\
  }\bibfield  {title} {\bibinfo {title} {General energy relations in nonlinear
  reactances},\ }\href@noop {} {\bibfield  {journal} {\bibinfo  {journal}
  {PROCEEDINGS OF THE INSTITUTE OF RADIO ENGINEERS}\ }\textbf {\bibinfo
  {volume} {47}},\ \bibinfo {pages} {2115} (\bibinfo {year}
  {1959})}\BibitemShut {NoStop}%
\bibitem [{\citenamefont {Brizard}\ and\ \citenamefont
  {Kaufman}(1995)}]{brizard1995local}%
  \BibitemOpen
  \bibfield  {author} {\bibinfo {author} {\bibfnamefont {A.}~\bibnamefont
  {Brizard}}\ and\ \bibinfo {author} {\bibfnamefont {A.}~\bibnamefont
  {Kaufman}},\ }\bibfield  {title} {\bibinfo {title} {Local manley-rowe
  relations for noneikonal wave fields},\ }\href@noop {} {\bibfield  {journal}
  {\bibinfo  {journal} {Physical review letters}\ }\textbf {\bibinfo {volume}
  {74}},\ \bibinfo {pages} {4567} (\bibinfo {year} {1995})}\BibitemShut
  {NoStop}%
\bibitem [{\citenamefont {Giscard}\ \emph {et~al.}(2015)\citenamefont
  {Giscard}, \citenamefont {Lui}, \citenamefont {Thwaite},\ and\ \citenamefont
  {Jaksch}}]{giscard2015exact}%
  \BibitemOpen
  \bibfield  {author} {\bibinfo {author} {\bibfnamefont {P.-L.}\ \bibnamefont
  {Giscard}}, \bibinfo {author} {\bibfnamefont {K.}~\bibnamefont {Lui}},
  \bibinfo {author} {\bibfnamefont {S.}~\bibnamefont {Thwaite}},\ and\ \bibinfo
  {author} {\bibfnamefont {D.}~\bibnamefont {Jaksch}},\ }\bibfield  {title}
  {\bibinfo {title} {An exact formulation of the time-ordered exponential using
  path-sums},\ }\href@noop {} {\bibfield  {journal} {\bibinfo  {journal}
  {Journal of Mathematical Physics}\ }\textbf {\bibinfo {volume} {56}},\
  \bibinfo {pages} {053503} (\bibinfo {year} {2015})}\BibitemShut {NoStop}%
\bibitem [{\citenamefont {Biegelsen}(1974)}]{biegelsen1974photoelastic}%
  \BibitemOpen
  \bibfield  {author} {\bibinfo {author} {\bibfnamefont {D.~K.}\ \bibnamefont
  {Biegelsen}},\ }\bibfield  {title} {\bibinfo {title} {Photoelastic tensor of
  silicon and the volume dependence of the average gap},\ }\href@noop {}
  {\bibfield  {journal} {\bibinfo  {journal} {Physical Review Letters}\
  }\textbf {\bibinfo {volume} {32}},\ \bibinfo {pages} {1196} (\bibinfo {year}
  {1974})}\BibitemShut {NoStop}%
\bibitem [{\citenamefont {Mermelstein}\ \emph {et~al.}(2007)\citenamefont
  {Mermelstein}, \citenamefont {Ramachandran}, \citenamefont {Fini},\ and\
  \citenamefont {Ghalmi}}]{mermelstein2007sbs}%
  \BibitemOpen
  \bibfield  {author} {\bibinfo {author} {\bibfnamefont {M.}~\bibnamefont
  {Mermelstein}}, \bibinfo {author} {\bibfnamefont {S.}~\bibnamefont
  {Ramachandran}}, \bibinfo {author} {\bibfnamefont {J.}~\bibnamefont {Fini}},\
  and\ \bibinfo {author} {\bibfnamefont {S.}~\bibnamefont {Ghalmi}},\
  }\bibfield  {title} {\bibinfo {title} {Sbs gain efficiency measurements and
  modeling in a 1714 $\mu$m 2 effective area lp 08 higher-order mode optical
  fiber},\ }\href@noop {} {\bibfield  {journal} {\bibinfo  {journal} {Optics
  Express}\ }\textbf {\bibinfo {volume} {15}},\ \bibinfo {pages} {15952}
  (\bibinfo {year} {2007})}\BibitemShut {NoStop}%
\bibitem [{\citenamefont {Van~Deventer}\ and\ \citenamefont
  {Boot}(1994)}]{van1994polarization}%
  \BibitemOpen
  \bibfield  {author} {\bibinfo {author} {\bibfnamefont {M.~O.}\ \bibnamefont
  {Van~Deventer}}\ and\ \bibinfo {author} {\bibfnamefont {A.~J.}\ \bibnamefont
  {Boot}},\ }\bibfield  {title} {\bibinfo {title} {Polarization properties of
  stimulated brillouin scattering in single-mode fibers},\ }\href@noop {}
  {\bibfield  {journal} {\bibinfo  {journal} {Journal of Lightwave Technology}\
  }\textbf {\bibinfo {volume} {12}},\ \bibinfo {pages} {585} (\bibinfo {year}
  {1994})}\BibitemShut {NoStop}%
\bibitem [{\citenamefont {Jauregui}\ \emph {et~al.}(2021)\citenamefont
  {Jauregui}, \citenamefont {Stihler}, \citenamefont {Kholaif}, \citenamefont
  {Tu},\ and\ \citenamefont {Limpert}}]{jauregui2021mitigation}%
  \BibitemOpen
  \bibfield  {author} {\bibinfo {author} {\bibfnamefont {C.}~\bibnamefont
  {Jauregui}}, \bibinfo {author} {\bibfnamefont {C.}~\bibnamefont {Stihler}},
  \bibinfo {author} {\bibfnamefont {S.}~\bibnamefont {Kholaif}}, \bibinfo
  {author} {\bibfnamefont {Y.}~\bibnamefont {Tu}},\ and\ \bibinfo {author}
  {\bibfnamefont {J.}~\bibnamefont {Limpert}},\ }\bibfield  {title} {\bibinfo
  {title} {Mitigation of transverse mode instability in polarization
  maintaining, high-power fiber amplifiers},\ }in\ \href@noop {} {\emph
  {\bibinfo {booktitle} {Fiber Lasers XVIII: Technology and Systems}}},\ Vol.\
  \bibinfo {volume} {11665}\ (\bibinfo {organization} {SPIE},\ \bibinfo {year}
  {2021})\ pp.\ \bibinfo {pages} {147--152}\BibitemShut {NoStop}%
\bibitem [{\citenamefont {Wisal}\ \emph {et~al.}(2023)\citenamefont {Wisal},
  \citenamefont {Chen}, \citenamefont {Kuang}, \citenamefont {Warren-Smith},
  \citenamefont {Miller}, \citenamefont {Cao},\ and\ \citenamefont
  {Stone}}]{SBSoptinPrep}%
  \BibitemOpen
  \bibfield  {author} {\bibinfo {author} {\bibfnamefont {K.}~\bibnamefont
  {Wisal}}, \bibinfo {author} {\bibfnamefont {C.-W.}\ \bibnamefont {Chen}},
  \bibinfo {author} {\bibfnamefont {Z.}~\bibnamefont {Kuang}}, \bibinfo
  {author} {\bibfnamefont {S.}~\bibnamefont {Warren-Smith}}, \bibinfo {author}
  {\bibfnamefont {O.~D.}\ \bibnamefont {Miller}}, \bibinfo {author}
  {\bibfnamefont {H.}~\bibnamefont {Cao}},\ and\ \bibinfo {author}
  {\bibfnamefont {A.~D.}\ \bibnamefont {Stone}},\ }\bibfield  {title} {\bibinfo
  {title} {Optimal input excitation for sbs suppression in multimode fibers}}
  (\bibinfo {year} {2023}),\ \bibinfo {note} {\textit{in
  preparation}}\BibitemShut {NoStop}%
\bibitem [{\citenamefont {Ho}(2012)}]{ho2012exact}%
  \BibitemOpen
  \bibfield  {author} {\bibinfo {author} {\bibfnamefont {K.-P.}\ \bibnamefont
  {Ho}},\ }\bibfield  {title} {\bibinfo {title} {Exact model for mode-dependent
  gains and losses in multimode fiber},\ }\href@noop {} {\bibfield  {journal}
  {\bibinfo  {journal} {Journal of lightwave technology}\ }\textbf {\bibinfo
  {volume} {30}},\ \bibinfo {pages} {3603} (\bibinfo {year}
  {2012})}\BibitemShut {NoStop}%
\bibitem [{\citenamefont {Paschotta}\ \emph {et~al.}(1997)\citenamefont
  {Paschotta}, \citenamefont {Nilsson}, \citenamefont {Tropper},\ and\
  \citenamefont {Hanna}}]{paschotta1997ytterbium}%
  \BibitemOpen
  \bibfield  {author} {\bibinfo {author} {\bibfnamefont {R.}~\bibnamefont
  {Paschotta}}, \bibinfo {author} {\bibfnamefont {J.}~\bibnamefont {Nilsson}},
  \bibinfo {author} {\bibfnamefont {A.~C.}\ \bibnamefont {Tropper}},\ and\
  \bibinfo {author} {\bibfnamefont {D.~C.}\ \bibnamefont {Hanna}},\ }\bibfield
  {title} {\bibinfo {title} {Ytterbium-doped fiber amplifiers},\ }\href@noop {}
  {\bibfield  {journal} {\bibinfo  {journal} {IEEE Journal of quantum
  electronics}\ }\textbf {\bibinfo {volume} {33}},\ \bibinfo {pages} {1049}
  (\bibinfo {year} {1997})}\BibitemShut {NoStop}%
\bibitem [{\citenamefont {Iezzi}\ \emph {et~al.}(2011)\citenamefont {Iezzi},
  \citenamefont {Loranger}, \citenamefont {Harhira}, \citenamefont {Kashyap},
  \citenamefont {Saad}, \citenamefont {Gomes},\ and\ \citenamefont
  {Rehman}}]{iezzi2011stimulated}%
  \BibitemOpen
  \bibfield  {author} {\bibinfo {author} {\bibfnamefont {V.~L.}\ \bibnamefont
  {Iezzi}}, \bibinfo {author} {\bibfnamefont {S.}~\bibnamefont {Loranger}},
  \bibinfo {author} {\bibfnamefont {A.}~\bibnamefont {Harhira}}, \bibinfo
  {author} {\bibfnamefont {R.}~\bibnamefont {Kashyap}}, \bibinfo {author}
  {\bibfnamefont {M.}~\bibnamefont {Saad}}, \bibinfo {author} {\bibfnamefont
  {A.}~\bibnamefont {Gomes}},\ and\ \bibinfo {author} {\bibfnamefont
  {S.}~\bibnamefont {Rehman}},\ }\bibfield  {title} {\bibinfo {title}
  {Stimulated brillouin scattering in multi-mode fiber for sensing
  applications},\ }in\ \href@noop {} {\emph {\bibinfo {booktitle} {Workshop on
  Fibre and Optical Passive Components}}}\ (\bibinfo {organization} {IEEE},\
  \bibinfo {year} {2011})\ pp.\ \bibinfo {pages} {1--4}\BibitemShut {NoStop}%
\bibitem [{\citenamefont {Liu}\ \emph {et~al.}(2016)\citenamefont {Liu},
  \citenamefont {Wright}, \citenamefont {Christodoulides},\ and\ \citenamefont
  {Wise}}]{liu2016kerr}%
  \BibitemOpen
  \bibfield  {author} {\bibinfo {author} {\bibfnamefont {Z.}~\bibnamefont
  {Liu}}, \bibinfo {author} {\bibfnamefont {L.~G.}\ \bibnamefont {Wright}},
  \bibinfo {author} {\bibfnamefont {D.~N.}\ \bibnamefont {Christodoulides}},\
  and\ \bibinfo {author} {\bibfnamefont {F.~W.}\ \bibnamefont {Wise}},\
  }\bibfield  {title} {\bibinfo {title} {Kerr self-cleaning of
  femtosecond-pulsed beams in graded-index multimode fiber},\ }\href@noop {}
  {\bibfield  {journal} {\bibinfo  {journal} {Optics letters}\ }\textbf
  {\bibinfo {volume} {41}},\ \bibinfo {pages} {3675} (\bibinfo {year}
  {2016})}\BibitemShut {NoStop}%
\bibitem [{\citenamefont {Voight}(1928)}]{voight1928lehrbuch}%
  \BibitemOpen
  \bibfield  {author} {\bibinfo {author} {\bibfnamefont {W.}~\bibnamefont
  {Voight}},\ }\bibfield  {title} {\bibinfo {title} {Lehrbuch der
  kristallphysik},\ }\href@noop {} {\bibfield  {journal} {\bibinfo  {journal}
  {Teubner, Leipzig}\ } (\bibinfo {year} {1928})}\BibitemShut {NoStop}%
\bibitem [{\citenamefont {Slez{\'a}k}\ \emph {et~al.}(2020)\citenamefont
  {Slez{\'a}k}, \citenamefont {Lucianetti},\ and\ \citenamefont
  {Mocek}}]{slezak2020tensor}%
  \BibitemOpen
  \bibfield  {author} {\bibinfo {author} {\bibfnamefont {O.}~\bibnamefont
  {Slez{\'a}k}}, \bibinfo {author} {\bibfnamefont {A.}~\bibnamefont
  {Lucianetti}},\ and\ \bibinfo {author} {\bibfnamefont {T.}~\bibnamefont
  {Mocek}},\ }\bibfield  {title} {\bibinfo {title} {Tensor-to-matrix mapping in
  elasto-optics},\ }\href@noop {} {\bibfield  {journal} {\bibinfo  {journal}
  {JOSA B}\ }\textbf {\bibinfo {volume} {37}},\ \bibinfo {pages} {1090}
  (\bibinfo {year} {2020})}\BibitemShut {NoStop}%
\bibitem [{\citenamefont {Slaughter}(2012)}]{slaughter2012linearized}%
  \BibitemOpen
  \bibfield  {author} {\bibinfo {author} {\bibfnamefont {W.~S.}\ \bibnamefont
  {Slaughter}},\ }\href@noop {} {\emph {\bibinfo {title} {The linearized theory
  of elasticity}}}\ (\bibinfo  {publisher} {Springer Science \& Business
  Media},\ \bibinfo {year} {2012})\BibitemShut {NoStop}%
\bibitem [{\citenamefont {McCurdy}(2005)}]{mccurdy2005modeling}%
  \BibitemOpen
  \bibfield  {author} {\bibinfo {author} {\bibfnamefont {A.~H.}\ \bibnamefont
  {McCurdy}},\ }\bibfield  {title} {\bibinfo {title} {Modeling of stimulated
  brillouin scattering in optical fibers with arbitrary radial index profile},\
  }\href@noop {} {\bibfield  {journal} {\bibinfo  {journal} {Journal of
  lightwave technology}\ }\textbf {\bibinfo {volume} {23}},\ \bibinfo {pages}
  {3509} (\bibinfo {year} {2005})}\BibitemShut {NoStop}%
\bibitem [{\citenamefont {Dragic}(2009)}]{dragic2009estimating}%
  \BibitemOpen
  \bibfield  {author} {\bibinfo {author} {\bibfnamefont {P.~D.}\ \bibnamefont
  {Dragic}},\ }\bibfield  {title} {\bibinfo {title} {Estimating the effect of
  ge doping on the acoustic damping coefficient via a highly ge-doped mcvd
  silica fiber},\ }\href@noop {} {\bibfield  {journal} {\bibinfo  {journal}
  {JOSA B}\ }\textbf {\bibinfo {volume} {26}},\ \bibinfo {pages} {1614}
  (\bibinfo {year} {2009})}\BibitemShut {NoStop}%
\bibitem [{\citenamefont {Smith}\ \emph {et~al.}(2016)\citenamefont {Smith},
  \citenamefont {Kuhlmey}, \citenamefont {de~Sterke}, \citenamefont {Wolff},
  \citenamefont {Lapine},\ and\ \citenamefont
  {Poulton}}]{smith2016metamaterial}%
  \BibitemOpen
  \bibfield  {author} {\bibinfo {author} {\bibfnamefont {M.}~\bibnamefont
  {Smith}}, \bibinfo {author} {\bibfnamefont {B.}~\bibnamefont {Kuhlmey}},
  \bibinfo {author} {\bibfnamefont {C.~M.}\ \bibnamefont {de~Sterke}}, \bibinfo
  {author} {\bibfnamefont {C.}~\bibnamefont {Wolff}}, \bibinfo {author}
  {\bibfnamefont {M.}~\bibnamefont {Lapine}},\ and\ \bibinfo {author}
  {\bibfnamefont {C.}~\bibnamefont {Poulton}},\ }\bibfield  {title} {\bibinfo
  {title} {Metamaterial control of stimulated brillouin scattering},\
  }\href@noop {} {\bibfield  {journal} {\bibinfo  {journal} {Optics letters}\
  }\textbf {\bibinfo {volume} {41}},\ \bibinfo {pages} {2338} (\bibinfo {year}
  {2016})}\BibitemShut {NoStop}%
\end{thebibliography}%

\end{document}